%% file: FlowThesis.tex
\documentclass[chap,10pt, oneside]{book}

\usepackage{latexsym}

\newcommand{\M}{{\cal M}}
\newcommand{\G}{{\cal G}}
\newcommand{\C}{{\cal C}}
\newcommand\be{\begin{equation}}
\newcommand\ee{\end{equation}}
\newcommand{\Qd}{Q_0^{\dagger}}

\usepackage{epsfig}
\usepackage{graphicx}

\if@twoside                  
\else                        
\fi
\usepackage{amsfonts}

\begin{document}

\begin{titlepage}
\vspace*{50mm}
\begin{center}
 {\bf FLOW EQUATIONS FOR HAMILTONIANS FROM CONTINUOUS UNITARY
 TRANSFORMATIONS}
 \begin{center} By \end{center}
 \begin{center} Bruce Bartlett \end{center}
\vspace{80mm}
\begin{center} Thesis presented in partial fulfilment of the
requirements for the degree of \end{center}
\begin{center} MASTER OF SCIENCE at the University of
Stellenbosch. \end{center}

\vspace{5mm} Supervisors : Professor F.G. Scholtz

\hspace{18.5mm} Professor H.B. Geyer

\hspace{-5mm}April 2003

\end{center}

\end{titlepage}

\frontmatter
\include{Declaration}
\addcontentsline{toc}{chapter}{Abstract - Opsomming}
\include{Abstract}
 \addcontentsline{toc}{chapter}{Acknowledgements}
\include{Acknowledgements}
 \addcontentsline{toc}{chapter}{List of Figures}
 \listoffigures
\include{Intro}

\tableofcontents
\mainmatter
\include{Chapter1}
\include{Chapter2}
\include{Chapter3}
\include{Chapter4}
\include{Conclusion}
\include{Appendices}

\include{Bib}
\end{document}

%% file: Declaration.tex
    \hbox{ }
    \vspace{5cm}
    \begin{center}
        \large\uppercase{Declaration} \\ 
        I, the undersigned, hereby declare that the work contained
        in this thesis is my own original work and that I have not
        previously in its entirety or in part submitted it at any
        university for a degree.
        \vskip 2cm
        \leftline{\rule[15pt]{5cm}{.3pt}
                  \hspace{-5cm}
                  Signature
                  \hspace{8cm}
                  \rule[15pt]{4cm}{.3pt}
                  \hspace{-4cm}
                  Date
                  \hfil}
    \end{center}

%% file: Abstract.tex
\begin{center} {\bf ABSTRACT} \end{center}

This thesis presents an overview of the flow equations recently
introduced by Wegner. The little known mathematical framework is
established in the initial chapter and used as a background for
the entire presentation. The application of flow equations to the
Foldy-Wouthuysen transformation and to the elimination of the
electron-phonon coupling in a solid is reviewed. Recent flow
equations approaches to the Lipkin model are examined thoroughly,
paying special attention to their utility near the phase change
boundary. We present more robust schemes by requiring that
expectation values be flow dependent; either through a variational
or self-consistent calculation. The similarity renormalization
group equations recently developed by Glazek and Wilson are also
reviewed. Their relationship to Wegner's flow equations is
investigated through the aid of an instructive model.

\vspace{1cm}

\begin{center} {\bf OPSOMMING} \end{center}

Hierdie tesis bied 'n oorsig van die vloeivergelykings soos dit
onlangs deur Wegner voorgestel is. Die betreklik onbekende
wiskundige raamwerk word in die eerste hoofstuk geskets en
deurgans as agtergrond gebruik. 'n Oorsig word gegee van die
aanwending van die vloeivergelyking vir die Foldy-Wouthuysen
transformasie en die eliminering van die elektron-fonon
wisselwerking in 'n vastestof. Onlangse benaderings tot die Lipkin
model, deur middel van vloeivergelykings, word ook deeglik
ondersoek. Besondere aandag word gegee aan hul aanwending naby
fasegrense. 'n Meer stewige skema word voorgestel deur te vereis
dat verwagtingswaardes vloei-afhanklik is; \'{o}f deur gevarieerde
\'{o}f self-konsistente berekenings. 'n Inleiding tot die
gelyksoortigheids renormerings groep vergelykings, soos onlangs
ontwikkel deur Glazek en Wilson, word ook aangebied. Hulle
verwantskap met die Wegner vloeivergelykings word bespreek aan die
hand van 'n instruktiewe voorbeeld.

%% file: Acknowledgements.tex
\begin{center} {\bf ACKNOWLEDGEMENTS} \end{center}

The realization of this thesis would not have been possible
without financial assistance from Stellenbosch University, the
National Research Foundation (NRF) and the Harry Crossley
Foundation. The financial assistance of the Department of Labour
(DoL) towards this research is hereby acknowledged. Opinions
expressed and conclusions derived at, are those of the author and
are not necessarily to be attributed to the DoL.

%% file: Intro.tex
\chapter*{Introduction}
 \vspace{-1cm}
 \setcounter{page}{6}
 \addcontentsline{toc}{chapter}{Introduction}
From the early days of Heisenberg's matrix mechanics, it became
clear that the language in which quantum physics described the
world was in terms of matrices and linear operators. Specifically,
the physical states of a system, and the values of any physical
observable, were intimately connected with the mathematical
problem of finding the eigenvectors and eigenvalues of a special
Hermitian operator known as the Hamiltonian of the system. Simply
put, the central problem of quantum mechanics is to diagonalize
large (mostly infinite) matrices.

Of course, this grand problem is severely and thoroughly
intractable, and sophisticated approximation schemes must be
employed to obtain a grip on the nature of the solution. Moreover,
the initial Hamiltonian is often expressed in terms of microscopic
variables in such a way that it obscures the larger scale dynamics
of the system. Before one attempts to solve for the energies and
eigenstates of the system, we should first understand {\em how}
the Hamiltonian works. In other words, one often requires an
equivalent description of the same physical system, expressed in
more familiar terms. This is expressed on the mathematical level
by finding a unitarily equivalent Hamiltonian which can be viewed
as a {\em renormalization} of the original theory, in the sense
that the constants appearing in the model have been modified in
order to accommodate the transformed nature of the Hamiltonian.

Such renormalization procedures have existed for a long time in
the context of both quantum mechanics and statistical physics, and
invariably result in a set of flow equations for the parameters
present in the Hamiltonian \cite{BDFN}. In statistical physics one
is normally interested in how the correlation between different
microscopic elements of the system behaves as the length scale
increases. Often this means ``integrating out'' the smaller length
scales so as to provide an effective theory on the larger scale.
In quantum field theory, one is interested in redefining coupling
constants so as to reconcile them with their physically measurable
counterparts.

Recently, Wegner \cite{Wegner1} and independently, Glazek and
Wilson \cite{GlazekWilson1, GlazekWilson2}, have developed a new
framework for flow equations. The authors have approached the
subject from different contexts; the former from condensed matter
physics and the latter from light-front quantum chromodynamics.
Both approaches, though, are similar in style and purpose. Their
distinction over previous methods is expressed in the title of the
present thesis. Specifically, the flow equations are written
directly in terms of the Hamiltonian, and do not involve the
Lagrangian framework with their associated path integral methods.
Secondly, the flow equations are continuous, as opposed to other
methods which take place in discrete steps. Thirdly, the flow
equations are unitary, so that no information about the system is
lost. The transformed Hamiltonian is completely equivalent to the
initial Hamiltonian at each point during the flow. Finally, the
flow equations are designed to diagonalize or block-diagonalize
the Hamiltonian (in Wegner's scheme), or to continuously eliminate
matrix elements involving large energy jumps, so as to render the
Hamiltonian more and more band diagonal (in Glazek and Wilson's
scheme).

The purpose of this study is to present an overview of this new
field, as well as to present new techniques which have proved
useful. The little known general mathematical framework of flow
equations is utilized extensively so as to provide a unified and
unique presentation of the subject. A specific model from nuclear
physics, the Lipkin model, is used as a central example against
which to test various approaches.

The material is organized in the following way. In Chapter
\ref{arb} Wegner's flow equation is introduced and solved
perturbatively. The mathematical framework behind the flow
equation is presented and the steepest descent nature of the flow
is revealed. The chapter concludes with considering two
modifications of Wegner's flow equation, namely Safonov's one step
scheme and block-diagonal flow equations. Chapter \ref{Chapter2}
discusses two pedagogical applications of flow equations to
familiar problems. The Foldy-Wouthuysen transformation of the
Dirac equation is derived using the new framework in a novel and
illuminating way. Flow equations are also used as a means of
renormalizing the electron-phonon interaction in solid-state
physics into an effective electron-electron attraction term. This
approach is compared with previous results using unitary
transformations such as that of Fr\"{o}hlich.

Chapter \ref{Chap3} introduces the Lipkin model and explains the
phase transition present in the model. As a background to further
discussion, numerical results are presented using brute force
diagonalization of the matrix and numerical solution of the flow
equations. This sets the stage for an overview of three separate
flow equations treatments of the model, all of which fail to
accommodate the second phase in a satisfying manner. Our own new
method is presented which revolves around tracking the ground
state during the flow, and proves to be of some use in treating
both phases in a consistent way. Two approaches are presented, the
first of which uses an external variational calculation while the
other uses a self-consistent approximation. Other possible
approaches are also considered, and the merits and drawbacks of
each scheme are enumerated. The chapter ends with a discussion and
summary of the important features of each approach.

 \markright{INTRODUCTION}

Chapter \ref{RenormChapter} addresses the important question of
how flow equations work relate to renormalization. It is here that
Glazek and Wilson's similarity renormalization group is presented,
and compared to Wegner's scheme. The renormalization properties of
the flow equations are elucidated by considering two examples, one
a toy model designed by Glazek and Wilson and the other the
familiar electron-phonon problem.

The philosophy behind this work has been to attempt to expound all
the finer details carefully, and some concepts are explained
repeatedly. The author has tried to follow the sound advice given
by Quintilian, 1900 years ago: {\quote \em One should not aim at
being possible to understand, but at being impossible to
misunderstand.}

%% file: Chapter1.tex
\chapter{Flow equations} \label{arb}

\section{Wegner's flow equation} \label{WegFE}

We intend to perform a continuous unitary transformation on an
initial Hamiltonian $H_0$ in such a way that the Hamiltonian flows
towards diagonal form. By a continuous unitary transformation we
mean that the Hamiltonian travels on a unitary path
 \be
 H(\ell) = U(\ell)H_0U^{\dagger}(\ell), \quad H(0) = H_0.
 \ee
The flow of the Hamiltonian can be expressed in an infinitesimal
form by computing the derivative with respect to $\ell$:
 \begin{eqnarray}
 \frac{dH}{d\ell} & = & \frac{dU}{d\ell}H(0)U^\dagger +
 UH(0)\frac{dU^\dagger}{d \ell} \\
 &=& \frac{dU}{d\ell}U^\dagger U H(0)U^\dagger - UH(0)U^\dagger
 \frac{dU}{d\ell}U^\dagger \\
 &=& \frac{dU}{d\ell}U^\dagger H(\ell) - H(\ell) \frac{dU}{d\ell}
 U^\dagger \\
 &=& [\eta(\ell), H(\ell)],
 \end{eqnarray}
where the derivative of $UU^{\dagger}=1$ has been employed in the
second line. Hence, the derivative of the Hamiltonian can be
expressed as the commutator of an anti-Hermitian generator
$\eta^\dagger = -\eta$ with the Hamiltonian,
 \be \label{roughe}
 \frac{dH}{d\ell} = [\eta(\ell), H(\ell)]  ,
 \ee
where $\eta(\ell) = \frac{dU}{d\ell}U^\dagger$. Instead of
concentrating on the unitary transformation $U(\ell)$, one may
instead shift interest onto the generator itself by recognizing
equation (\ref{roughe}) as the most general form of a unitary flow
on the Hamiltonian. The idea is to choose $\eta$ so as to solve
the problem, which in our case is diagonalizing $H_0$. The usual
Jacobi iterative numerical method for diagonalizing a matrix
performs a sequence of unitary transformations $U_{ij}$ on $H_0$:
 \be
 H' = U_{i_N j_N}\cdots U_{i_1 j_1}H_0U^{\dagger}_{i_1 j_1}\cdots U^{\dagger}_{i_N
 j_N}.
 \ee
Each transformation $U_{ij}$ is designed to eliminate the
off-diagonal term $H_{ij}$. In general the next unitary
transformation will cause $H_{ij}$ to reappear, but due to the
construction of the unitary transformations the new $H_{ij}$ has a
reduced magnitude. In this way one obtains a sequence of
Hamiltonians $H'^{(m)}$ which converge to diagonal form.

Wegner found a generator which will perform the diagonalization
continuously:
 \be \label{gc1}
 \eta(\ell) = [\mbox{Diag}(H(\ell)), H(\ell)] \rightarrow \quad \eta_{ij} =
 H_{ij}(\varepsilon_i - \varepsilon_j).
 \ee
Diag($H$) refers to the diagonal part of $H$. The $\varepsilon_i$
are simply the diagonal entries $H_{ii}$. To prove that this
choice diagonalizes the Hamiltonian, we substitute the generator
(\ref{gc1}) into the general flow equation (\ref{roughe}). With
the convention that $v_{ij}$ is the $\ell$-dependent off-diagonal
part of $H(\ell)$ (note that $v_{ii}=0$), we obtain the following
differential equations for the diagonal and off-diagonal matrix
elements,
  \begin{eqnarray}
 \dot{\varepsilon_i} &=& 2\sum_k (\varepsilon_i -
 \varepsilon_k)|v_{ik}|^2 \label{gc3} \\
 \dot{v_{ij}} &=& -(\varepsilon_i - \varepsilon_j)^2 v_{ij} + \sum_k
 (\varepsilon_i + \varepsilon_j - 2\varepsilon_k)v_{ik}v_{kj} ,
 \label{gc4}
 \end{eqnarray}
where the dot indicates differentiation with respect to $\ell$.
Eqs. (\ref{gc3}) and (\ref{gc4}) are Wegner's flow equations
written explicitly in matrix element form. The sum of the squares
of the diagonal matrix elements must increase:
 \begin{eqnarray}
 \frac{d}{d\ell} \sum_i \varepsilon_i^2 &=& 2\sum_i \varepsilon_i
 \frac{d \varepsilon_i}{d\ell} \nonumber \\
 &=& 2\sum_{i,k}2\varepsilon_i(\varepsilon_i -
 \varepsilon_k)|v_{ik}|^2  \\
 &=& 2 \sum_{i,k}(\varepsilon_i - \varepsilon_k)^2 |v_{ik}|^2 \geq 0.
 \label{gc5}
 \end{eqnarray}
Since the trace of a matrix remains invariant under unitary
transformations,
 \be \label{gc98}
 \frac{d}{d\ell} \mbox{Tr}(H^2(\ell)) = \frac{d}{d\ell} \left(
 \sum_i \varepsilon_i^2 + \sum_{ij}|v_{ij}|^2 \right)
 = 0,
 \ee
Eq. (\ref{gc5}) implies that the off-diagonal elements must
monotonically decrease until the only off-diagonal matrix elements
that can possibly remain are those between states with equal
diagonal matrix elements. We conclude that the choice of generator
(\ref{gc1}) has the remarkable property of causing the Hamiltonian
to flow towards diagonality.

\section{Perturbative solution} \label{WegPertS}

Let us now solve the flow equations (\ref{gc3}) and (\ref{gc4}),
written in terms of the matrix elements, perturbatively. We assume
that the initial Hamiltonian can be written as
 \be
 H_0 = D_0 + gV_0,
 \ee
with $D_0$ and $V_0$ diagonal and off-diagonal respectively, and
$g$ the bare coupling constant. We expand $\varepsilon_i(\ell)$
and $v_{ij}(\ell)$ in powers of $g$,
 \begin{eqnarray}
 \varepsilon_i(\ell) &=& \varepsilon_i^{(0)}(\ell) +
 g\varepsilon_i^{(1)}(\ell) + g^2\varepsilon_i^{(2)}(\ell) + \cdots \\
 v_{ij}(\ell) &=& gv_{ij}^{(1)} + g^2v_{ij}^{(2)} + \cdots
 \end{eqnarray}
and substitute into the flow equations (\ref{gc3}) and
(\ref{gc4}). The result up to second order in $g$ is
 \begin{eqnarray}
 \varepsilon_i(\ell) &=& E_i + g^2 \sum_k
 \frac{V_{ik}^2}{\Delta_{ik}}\left(1 - e^{-2\Delta_{ik}^2\ell}
 \right) + \cdots \label{pe} \\
 v_{ij}(\ell) &=& gV_{ij}e^{-\Delta_{ij}^2 \ell} + g^2
 \sum_{k}\frac{V_{ik}V_{kj}(\Delta_{ik} +
 \Delta_{jk})}{\Delta^2_{ik} + \Delta^2_{kj} -
 \Delta^2_{ij}}\left(1 - e^{-(\Delta^2_{ik} + \Delta^2_{kj} -
 \Delta^2_{ij})\ell}\right)e^{-\Delta^2_{ij}\ell} + \cdots
 \label{pv}
 \end{eqnarray}
where $E_i = \langle i |D_0| i\rangle$, $V_{ij} = \langle i| V_0 |
j\rangle$ and $\Delta_{ij} = E_i - E_j$. Eqs. (\ref{pe}) and
(\ref{pv}) should be seen as a generalization of ordinary
Rayleigh-Schr\"{o}dinger perturbation theory, which only
concentrates on the series for the final diagonal form of the
eigenvalues, $\varepsilon_i(\infty)$. The beauty of the flow
equations result is that we now have an idea of precisely {\em
how} the initial matrix $H_0$ is continuously diagonalized into
its final form. To second order in $g$, the diagonal elements
decay exponentially to the eigenvalues. The off-diagonal elements
decay exponentially to zero in the first order, but follow a
slightly more complicated route in second order.

 \begin{figure}
 \begin{center}
 \includegraphics[width=0.8\textwidth]{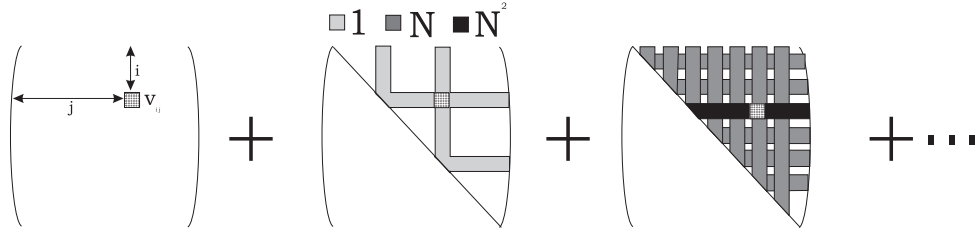}
 \end{center}
 \caption[Coupling diagram for perturbative solution of flow equations]{\label{MeshFig} Coupling diagram illustrating how the off-diagonal element $v_{ij}$ depends on
 the other off-diagonal elements, order for order. Each colour is a different power of $N$, the dimension
 of the matrix. Since the Hamiltonian is hermitian, it is only necessary to display the upper-half
 of the matrix. }
 \end{figure}

The perturbative solutions (\ref{pe}) and (\ref{pv}) also display
in a concrete fashion the {\em way} in which the matrix elements
are coupled and intermeshed to each other, by the constraint of
remaining unitarily equivalent to the initial Hamiltonian. It is
instructive to illustrate this by means of coupling diagrams, as
in Fig. \ref{MeshFig}. The orders in perturbation theory for the
off-diagonal elements are related schematically to the initial
off-diagonal elements $V_{ij}$ by
 \be
 v_{ij}(\ell) = gV_{ij}(\cdots) + g^2\sum_k{V_{ik}V_{kj}}(\cdots) +
 g^3\sum_{k,m}V_{ik}V_{im}V_{mk}(\cdots) + \cdots
 \ee
Fig. \ref{MeshFig} colours each matrix element according to the
number of times it is counted in the above sum. The diagram offers
a graphical illustration of the strength to which $v_{ij}$ is
coupled to the other matrix elements, order for order.

\section{Mathematics of flow equations} \label{MathChapter}

In this section the flow equations will be abstracted from their
physical setting and discussed from a purely mathematical point of
view.

Flow equations such as Wegner's were, in fact, already discussed
in the mathematical literature in 1983 \cite{Dei}, eleven years
before Wegner's paper. It is apparent from the physics literature
that this fact is unknown. Although there are few mathematical
papers that deal with the subject, there are some important
aspects worth mentioning that classify precisely the status of a
`flow equation'.

\subsection{Preliminaries} Let us first review some basic notions
of groups and manifolds. Let ${\cal G}$ be a group and ${\cal M}$
be a set. Then a map $\alpha : {\cal G} \times {\cal M}
\rightarrow {\cal M}$ is an {\em action} of ${\cal G}$ on ${\cal
M}$ if it satisfies the following conditions for all $g$, $h \in
{\cal G}$ and $m \in {\cal M}$:
\[ e\ast m=m \quad \mbox{and} \quad (gh)\ast m = g \ast (h \ast m), \]
where $e$ is the identity of the group $\cal G$. For $m \in \M$
let the {\em orbit} of $m$ be the action of the whole group $\G$
on $m$;
\[ \mbox{Orbit}(m) = \G \ast m = \{g \ast m : g \in \G\}. \]
 \setlength{\unitlength}{1mm}
 \begin{figure}
 \begin{center}
 \includegraphics[width=0.5\textwidth]{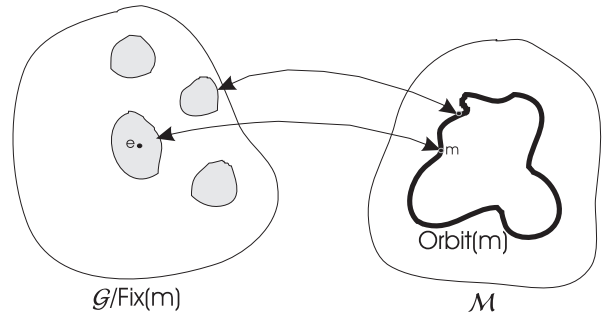}
 \caption {Orbit($m$) may be placed in one to one correspondence with $\G/\mbox{Fix}(m)$   }
 \end{center}
 \end{figure}
Consider the set of elements of $\G$ that leave $m \in \M$
invariant:
\[ \mbox{Fix}(m)=\{g \in \G : g \ast m = m\}. \]
It is easy to show that Fix($m$) is actually a normal subgroup of
$\G$, so we may construct the quotient group $\G /$ Fix($m$). Now,
every element in a given coset $g$Fix($m$) maps $m$ onto the same
point in Orbit($m$). This gives us the following important 1-1
mapping between the elements of $\G /$ Fix($m$) and the orbit of
$m$:
\[ \G / \mbox{Fix} (m) \rightarrow \mbox{Orbit}(m) \, : \,
 g\mbox{Fix}(m) \rightarrow g \ast m .\]
We are now in a position to state a well-known and important
result \cite{Var}.
\newline
Let $\M$ be a smooth manifold, and let $\G$ be a Lie group acting
smoothly on $\M$, and let $m \in \M$. Then the orbit $\G \ast m$
of $m$ is a smooth homogeneous manifold of $\M$ with the following
dimension: \be \label{MathA}
 \mbox{dim}(\G \ast m) \, = \, \mbox{dim}(\G) - \mbox{dim
 Fix}(m) .\ee

\subsection{The manifold of unitarily equivalent matrices}

 Let $H_0$ be a $n \times n$ Hermitian matrix and consider the
class $\C$ of all matrices unitarily equivalent to it:
\[ \C(H_0) = \{H \; : \; H \mbox{ is unitarily equivalent to }
 H_0\}. \]
Since two Hermitian matrices are unitarily equivalent if and only
if there is a unitary transformation that connects the two, $\C$
is clearly equivalent to the action of the group of unitary
matrices $U(n)$ on $\Lambda$, where $\Lambda$ is a diagonalized
form of $H_0$ (there are $n!$ such forms if $H_0$ is nondegenerate
since the eigenvalues may be shuffled down the diagonal) and the
action is defined by $Q \ast \Lambda \equiv Q \Lambda
Q^{\dagger}$. From the theorem (\ref{MathA}) we now conclude that
$\C(H_0)$ is a smooth manifold. Let us now compute Fix($\Lambda$);
that is, which unitary matrices $Q$ leave $Q\Lambda Q^{\dagger} =
\Lambda$? Consider the case where the eigenvalues $\lambda_i$ of
$\Lambda$ are all distinct. This is easy, since $Q \Lambda
Q^{\dagger} = \Lambda$ $\Rightarrow$ $[Q, \Lambda]=0$ which means
$Q$ must be diagonal. However, $QQ^{\dagger}=1$ thus
\[ \mbox{Fix}(\Lambda) \, = \, \{\mbox{diag}(\alpha_1, \alpha_2,
 ..., \alpha_n) \, : \, |\alpha_i | = 1 \} ,\]
which is $n$-dimensional. If the eigenvalues are not distinct, say
$\lambda_i=\lambda_j$, then care should be exercised since Q may
have off diagonal entries at $q_{ij}$ and $q_{ji}$.

Since dim $U(n)= n^2$ we conclude that, providing the eigenvalues
of $H_0$ are distinct, $\C(H_0)$ is a smooth, compact, homogeneous
manifold of dimension $n^2-n$.  $\Box$

Indeed, this result could easily be anticipated since two $n
\times n$ matrices $A$ and $B$ are unitarily equivalent if and
only if their moments up to order $n$ are equal \cite{Horn}:
\be\label{MathB} \mbox{tr}(A^i) \, =  \, \mbox{tr}(B^i), \quad
i=1..n \ee Since a complex Hermitian matrix has $n^2$ real degrees
of freedom, these $n$ constraints gives the dimension of the set
of matrices unitarily equivalent to $A$ as $n^2-n$. Unfortunately
the equations in (\ref{MathB}) do not help us to eliminate the
redundant variables directly since the expressions in the traces
are high order (up to $n$) polynomials in the matrix elements.

For those aficionados who think that all this effort to prove that
the class of unitarily equivalent matrices is a smooth manifold is
unnecessary, since just about all sets are manifolds anyway,
consider the following counterexample. The set of unitarily
equivalent $3 \times 3$ {\em tridiagonal} matrices, in which two
eigenvalues coincide, fails to be a manifold since it is
homeomorphic to a figure eight {\cite{Tomei}. \footnote{A figure
of eight fails the conditions of an analytic manifold at the
intersection point.}

Now that we have established that $\C(H_0)$ is a manifold, we may
go on to compute its Euler characteristic, its differential
geometry and so on. We shall not follow this route, but simply
mention two important examples:
\begin{itemize}
 \item For $3 \times 3$ real matrices the manifold is a sphere with 2 handles \cite{Drie5}.
 This is in contrast to the manifold of SO(3), a sphere with opposite points on the surface
 identified (Fig. \ref{Manifolds}(a)).

 \item For $3 \times 3$ real tridiagonal matrices
 \be
 \left( \begin{array}{ccc}
 a_1 & b_1 & 0 \\ b_1 & a_2 & b_2 \\ 0 & b_2 & a_3
 \end{array} \right),
 \ee
 with distinct eigenvalues $\lambda_1 > \lambda_2 > \lambda_3$ the manifold
 breaks into four closed components $++$, $+-$, $-+$, $--$,
 depending on the signs of $b_1$ and $b_2$ ($+$$-$, for example,
 means $b_1 \geq 0$, $b_2 \leq 0$). The set $++$ is homeomorphic
 to a hexagon \cite{Tomei} (Fig. \ref{Manifolds}(b)).
\end{itemize}

 \begin{figure}
 \begin{minipage}[t]{0.6\textwidth}
 \begin{center}
 \includegraphics[width=\textwidth]{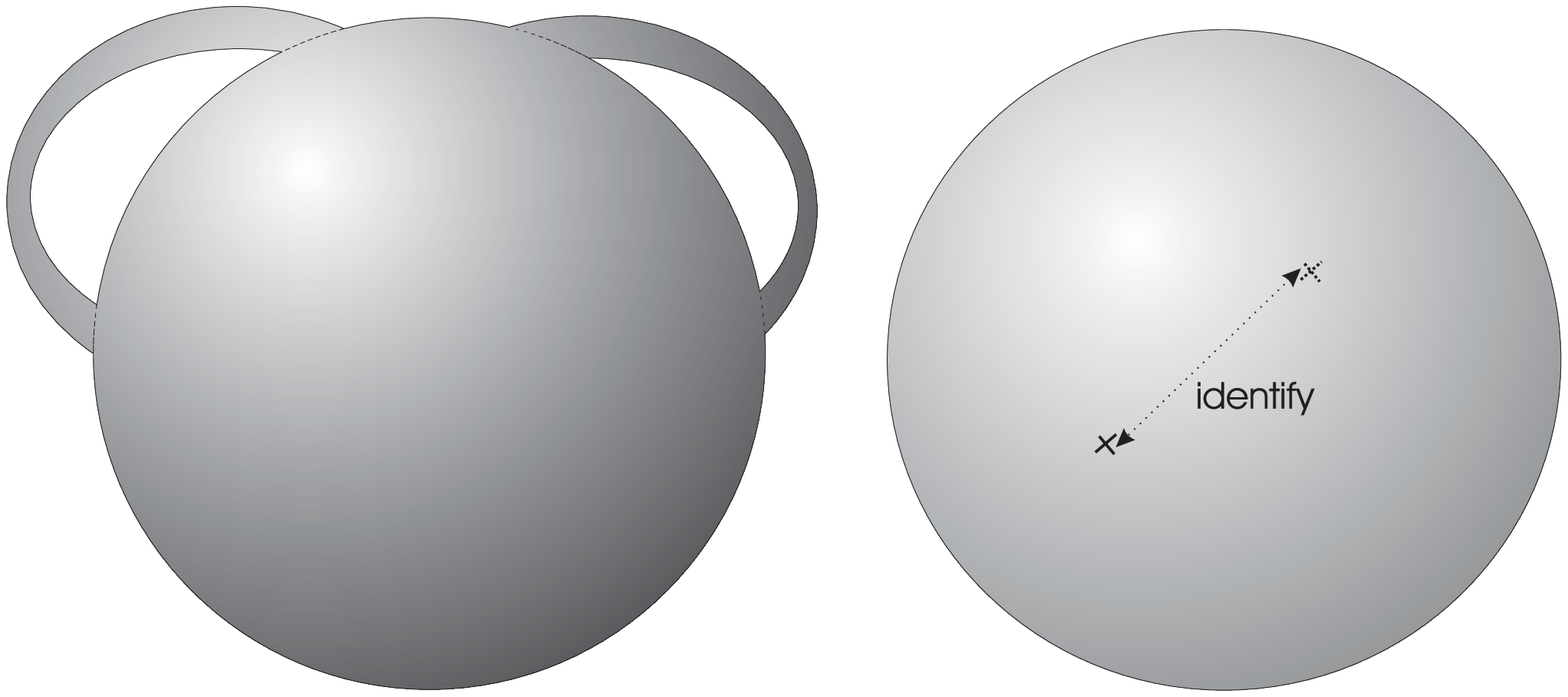}
 (a)
 \end{center}
 \end{minipage}
 \hfill
 \begin{minipage}[t]{0.3\textwidth}
 \begin{center}
 \includegraphics[width=\textwidth]{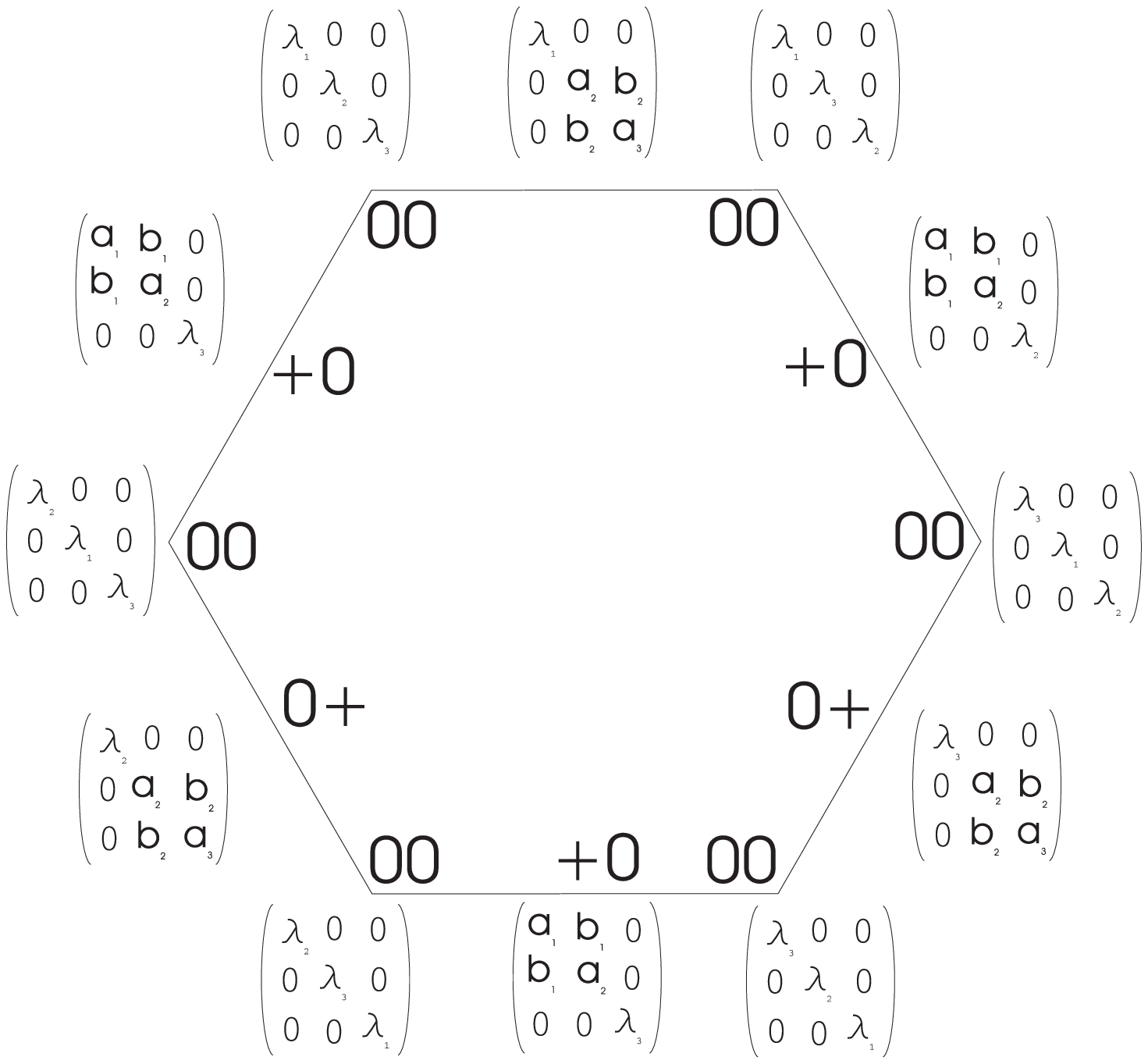}
 (b)
 \end{center}
 \end{minipage}
 \caption[Manifold structure for different matrix constraints] {\label{Manifolds} (a) The difference between the manifold of $3\times 3$ unitarily equivalent
 real matrices, and the manifold SO(3). (b) The $++$ section of $3\times 3$ real tridiagonal
 unitarily equivalent matrices is homeomorphic to a hexagon. }
 \end{figure}

\subsection{The steepest descent formulation} \label{sdf} Here we
follow the work of Refs. \cite{Drie3} and \cite{Brock}. The flow
equations can be placed into a more general framework based on the
following problem:

Find $X \in \C(H_0)$ that minimizes the function
 \be \label{OptFunc}
 F(X) = \|X - P(X)\|^2.
 \ee
Here $\|\cdot\|$ means the Frobenius norm defined as
 \be
 ||X|| = \sqrt{\left<X,X\right>},
 \ee
where the standard trace inner product is intended,  $
\left<A,B\right> = \mbox{tr}(A^\dagger B)$. $P(X)$ is a projection
of $X$ onto a subspace $\Phi$ of the space of $n \times n$
Hermitian matrices. In other words, we are trying to find the
closest possible matrix to $\Phi$ under the constraint that it be
unitarily equivalent to $H_0$. By choosing $\Phi$ appropriately
the problem may be adapted to mathematical problems such as the
inverse eigenvalue problem \cite{Chu2}, Toda flows \cite{Chu1,
Drie4} and displacement flows \cite{Drie4}. Physicists have yet to
discover this formulation but it does define a precise framework
for some problems, such as finding the Hamiltonian closest to
block diagonal form, in a renormalization type scheme. The idea is
simply to set up the steepest descent differential equations so as
to maximally decrease $F(X)$ in each step. The flow may be viewed
as taking place in the Hamiltonian space $\C(H_0)$ {\em and} in
the unitary matrix space SU($n$) simultaneously.

 \begin{figure}[h]
 \begin{center}
 \includegraphics[width=0.5\textwidth]{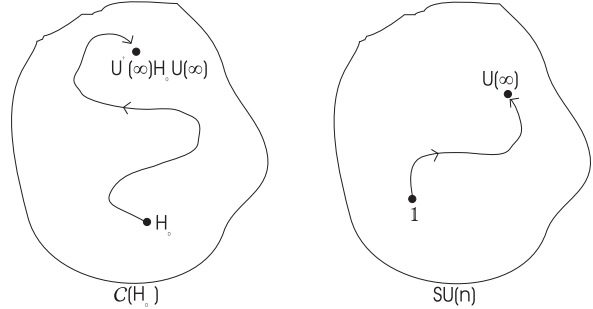}
 \end{center}
 \caption[Flow in $\C(H_0)$ and SU($n$)] {The flow may be viewed either on the level of the matrix elements themselves, or as a
 flow in the unitary matrices that accomplish the diagonalization.}
 \end{figure}
In $\C(H_0)$, the matrix begins at $H_0$ and ends when it is
closest to $\Phi$. In SU($n$), the flow begins at the identity
matrix and ends at the unitary matrix that accomplishes the job.
We may restrict our attention to SU($n$) as opposed to U($n$)
since any element of U($n$) can be written as $U=e^{i\phi}V$, with
$V \in \mbox{U}(n)$, and the phase $\phi$ does not affect the
transformed Hamiltonian.

It can easily be shown \cite{Drie3} that the only fixed point $X$
of the steepest descent of the objective function (\ref{OptFunc})
in $\C(H_0)$ occurs when $X$ commutes with its own projection:
\[ \left[ X, P(X) \right] = 0 \]
Thus to solve the eigenvalue problem, we may choose the subspace
$\Phi$ to be either: \newline
 (1) the space of all diagonal matrices :  $P(X)$ = Diag($X$), or   \newline
 (2) a fixed diagonal matrix $N$ :  $P(X)$ = $N$, \newline
where $N$ is some arbitrary diagonal matrix with distinct entries
chosen to `guide' the equations. These choices work since
 \be \label{commcond}
  [X,D]_{ij} = 0 \; \Leftrightarrow \; X_{ij}(D_{jj} - D_{ii}) =
  0,
 \ee
where $D$ is a diagonal matrix. For (2) this implies that $X$ must
be diagonal while in (1) $X_{ij}$ may still be non-zero for $i
\neq j$, providing that the diagonal entries $X_{ii}$ and $X_{jj}$
are equal.
\newline
\newline
For completeness, we shall now illustrate the steps leading to the
steepest descent flow equations for case (2). Let us then consider
the objective function $\bar{F}$ in SU($n$),
\begin{eqnarray}
 \bar{F}(Q) & = & \|Q^{\dagger} H_0 Q - N \|^2 \\
            & = & \mbox{tr} \left( (Q^{\dagger} H_0 Q - N)( Q^{\dagger} H_0 Q -
            N) \right) \\
            & = & \mbox{tr} (H_0^2) + \mbox{tr}(N^2) - 2\mbox{tr}(Q^{\dagger} H_0
            Q N).
\end{eqnarray}
Ignoring the terms which are independent of $Q$, the function we
wish to maximize is tr($Q^{\dagger} H_0 Q N)$. The unitary matrix
$Q$ will trace out a path $Q(\l)$. Consider a point $Q_0$ along
this path. We parametrize its neighbourhood as
 \be \label{parQ}
 Q(\Omega) = Q_0 \left(1 +
 \Omega + \frac{\Omega^2}{2!} + \cdots \, \right)
 \ee
with $\Omega$ anti-hermitian. To first order in $\Omega$ we have
\begin{eqnarray}
 \mbox{tr} \left( (1-\Omega) \Qd H_0 Q_0 (1 + \Omega) N \right) & = & \mbox{tr} \left(\Qd H_0
 Q_0 \right) - \mbox{tr}\left(\Omega \Qd H_0 Q_0 N \right) + \\
 & & \qquad \qquad \qquad \mbox{tr}\left(\Qd H_0 Q_0 \Omega.
 N\right)
\end{eqnarray}
Using the cyclic properties of the trace function gives
\begin{eqnarray}
\mbox{tr} \left( (1-\Omega) \Qd H_0 Q_0 (1 + \Omega) N \right)
-\mbox{tr} \left(\Qd H_0 Q_0 \right) & = & -\mbox{tr}(\Qd H_0 Q_0
N \Omega) + \mbox{tr}(N \Qd H_0 Q_0 \Omega) \\
& = & \mbox{tr}\left( [N, \Qd H_0 Q_0] \Omega \right) \\
& = & \mbox{tr}\left([\Qd H_0 Q_0, N]^{\dagger} \Omega \right) \\
& = & \left< [\Qd H_0 Q_0, N], \, \Omega \right>.
\end{eqnarray}
Since we have worked to first order in $\Omega$ we see that
$\left< [Q^{\dagger} H_0 Q, N], \, \cdot \, \right>$ represents
the gradient function at $Q$, in the sense that it accounts for
the infinitesimal behavior of $\bar{F}$ there. Following the
parameterization of the neighborhood (\ref{parQ}), we choose
$\dot{Q} = Q \Omega$, so now we can express the gradient flow as
\[ Q^{\dagger}\dot{Q} = [Q^{\dagger} H_0 Q, N] ,\]
or
 \be \label{OrthEqn}
 \dot{Q} = Q[Q^{\dagger} H_0 Q, N].
 \ee
This equation evolves in SU($n$) and is cubic in $Q$. Let us now
evaluate the projection of the flow equation in the space of
hermitian matrices. Let $H=Q^{\dagger}H_0Q$. Then
 \begin{eqnarray} \label{HermEqn}
 \dot{H} & = & [N, Q^{\dagger} H_0 Q]Q^{\dagger}H_0Q +
 Q^{\dagger}H_0Q[Q^{\dagger} H_0 Q, N]\\
  & = & -HNH + H^2N - HNH + NH^2 \\
  & = & \left[ \, [N, H]\, , \, H \,\right],
 \end{eqnarray}
which is the form of the flow equation we shall use for most of
this work. Wegner's original flow equation is recovered by
choosing $\mbox{P}(X)=\mbox{Diag}(X)$ as in case(1) above. The
analysis runs similarly to the one we have just presented. The
flow equation in SU($n$) is
\[ \dot{Q} = Q[Q^{\dagger} H_0 Q, \mbox{Diag}(Q^{\dagger} H_0 Q)] ,\]
while the corresponding flow in the Hermitian matrices is
\[ \dot{H} = \left[ [\mbox{Diag}(H), H] ,  H \right] ,\]
which is precisely Wegner's flow equation! To summarize, we see
that {\em Wegner's flow equations determine steepest descent
curves to diagonality in matrix space.} The difference between the
two cases presented above is that using $[N,H]$ as the generator
produces steepest flow towards $N$, while $[\mbox{Diag}(H), H]$
produces steepest descent flow to the diagonalized form of $H_0$.
Both generators cause $H$ to flow to diagonal form, with the
latter achieving this slightly quicker. The advantage of the first
formulation (using $N$) is that
 \begin{itemize}
 \item The equations are quadratic in $H$ and cubic in $Q$, while
 in the latter case they are cubic in $H$ and pentic in $Q$.
 \item By choosing an appropriate $N$ it may be possible to
 restrict the flow to some submanifold of $\C(H_0)$, for instance
 the tridiagonal matrices. This is what we shall do in Chapter \ref{Chap3}, in
 the context of the Lipkin model (see Eq. (\ref{genc})).
 \item Whereas the only stable fixed point of the flow equations using
 $[N, H$] as generator is the final diagonal matrix, it is
 possible for the flow equations using $[\mbox{Diag}(H), H]$ to
 end with remaining offdiagonal elements between states with equal
 diagonal elements (see Secion \ref{WegFE}). This is because these
 points are local minima in the objective function $F$, from which
 a steepest descent formulation will not escape.
 \item It can be shown \cite{Brock} that the final diagonal
 matrix to which the flow proceeds has its eigenvalues listed down
 the diagonal in the same order as in $N$.
 \end{itemize}
The last property follows from the fact that, if $H$ is diagonal,
\[ F(H)\,=\, \| H-N\|^2 = \sum_i (h_{ii} - n_{ii})^2, \]
so that $F$ is minimized when $H$ is similarly ordered to $N$.
This viewpoint clears up some of the confusion evident in the
physics literature about these topics. Some authors have expressed
reservations about using generators like $[N,H]$ instead of
Wegner's original choice \cite{PF}. Others, in order to gain the
advantages mentioned above, have set up new generators
\cite{Mielke1}. In the mathematical framework we have presented
here it has been shown that the flow equations are robust - quite
generally, the differential equation to minimize the distance of
$H$ to some subspace $P(H)$ is
\[ \dot{H} = [[P(H), H], H] .\]

\section{Other methods}
\subsection{One step continuous unitary transformations}
A simpler but less general diagonalization scheme is Safonov's
`one step' unitary transformation \cite{Saf1, Saf2, Saf3, Saf4,
Saf5}. The idea is to choose a fixed scalable generator instead of
a completely dynamic generator as used in the steepest descent
flow equations. We parametrize the transformations on the
Hamiltonian as
\[ H(\ell) = e^{\ell R}H_0e^{-\ell R}, \]
where $R$ is a fixed anti-hermitian generator, and $\ell$ a
parameter. This expression for $H(\ell)$ is a solution of the
differential equation
 \be \label{Safonov1}
 \frac{dH}{d \ell} = [R, H(\ell)].
 \ee
This should be compared to the more general steepest descent
formulation
 \be \label{Safonov2}
 \frac{dH}{dl} = [\eta (l), H(l)] \qquad
 \eta(l)=[\mbox{Diag}(H(l)),H(l)].
 \ee
It is clear that the steepest descent equation rotates $H$ around
$\eta$ at each step, where $\eta$ evolves dynamically with $H$.
Safonov's method restricts $H$ to be rotated around a fixed $R$ at
each step.

The method involves:
 \begin{itemize}
 \item parametrising the
 flow of $H$ in terms of operator combinations involving unknown
 $\ell$ dependent coefficients,
 \be \label{Safonov3}
 H(\ell) = a_1(\ell) H_1 + a_2(\ell) H_2 + \cdots
 a_N(\ell) H_N,
 \ee

\item Expanding the generator $R$ in terms of unknown ($\ell$
independent) anti-Hermitian terms,
 \be \label{Safonov4}
 R = b_1 R_1 + b_2 R_2 + \cdots b_M R_M.
 \ee
\end{itemize}
The two expansions (\ref{Safonov3}) and (\ref{Safonov4}) are then
substituted into the flow equation (\ref{Safonov1}), which results
in a set of {\em linear} differential equations for the $a_i$.
Solving these equations with the initial conditions provided by
the parameter values in the initial Hamiltonian, we obtain the
transformed Hamiltonian $H(\ell)$. In order to eliminate
`inconvenient' terms (i.e. the off-diagonal terms in a
diagonalization scheme) one needs to set their coefficients (for
example, for $\ell=1$) equal to zero; $a_p(1)=0$, for the unwanted
$a_p$. This gives a second set of equations which determines the
$b_i$ and hence $R$.

The advantage of Safonov's one step scheme is that the
differential equations to be solved are linear, since the
generator is $\ell$-independent. The difficulty is that extra
interaction terms are still generated, for almost any (useful)
choice of $R$. Providing the transformation involved is reasonably
small, these extra terms may be neglected. Another difficulty is
that the static nature of the generator $R$ does not allow for the
unitary transformation to change dynamically during the flow. This
is evident when Safonov's method is used to eliminate the
electron-phonon interaction terms in an interacting electron
system, as will be done in Section \ref{TEPI}. It gives the same
result as Fr\"{o}hlich's original expansion of the unitary
transformation by the BCH formula, which is to be expected since
the two approaches have much in common.

\subsection{Block-diagonal flow equations}
Perhaps it is asking too much to attempt to diagonalize the
Hamiltonian of the relevant problem completely. One may impose a
less grand requirement, by asking only that the Hamiltonian flows
to block-diagonal form. The individual blocks may then be analysed
separately. This manner of thinking lends itself to Hamiltonians
in relativistic field theory, which do not conserve the number of
particles. In this case one may attempt to find an effective
Hamiltonian which decouples the Fock spaces with different number
of particles from each other.

Wegner's flow equation can be extended in a straightforward
fashion for this purpose. The method is most clearly explained in
Ref. \cite{Pauli}, where it was applied to determine an effective
$q\bar{q}$-interaction in QCD. A good overall reference can be
found in Ref. \cite{Gubankova}, where it was applied to QED on the
light front.

One divides the Hilbert space into two pieces, the $P$ and the $Q$
space. We have only used two spaces here for clarity but of course
in a field theory one would divide it into an infinite number of
$P_i$ Fock spaces, where $i$ labels the number of particles in the
space. With a slight abuse of notation, let $P$ and $Q=1-P$
represent the projection operators onto the $P$ and $Q$ spaces
respectively. We intend to transform the Hamiltonian using flow
equations into block-diagonal form:
 \be
 \left(
 \begin{array}{cc}
 PHP & PHQ \\
 QHP & QHQ
 \end{array}
 \right) \rightarrow H(\infty) =
 \left(
 \begin{array}{cc}
 PH(\infty)P & 0 \\
 0 & QH(\infty)Q
 \end{array}
 \right).
 \ee
If we consider the $PHP$ and $QHQ$ blocks to be the ``diagonal''
part of the Hamiltonian $H_d$, and the $PHQ$ and $QHP$ blocks to
be the ``rest'' $H_r$, then the usual Wegner formula would dictate
that
 \be \label{bd1}
 \eta(\ell) = [H_d(\ell), H_r(\ell)]
 \ee
is the choice of generator to employ. Indeed, this choice shall be
proved shortly to propagate the Hamiltonian to block-diagonal
form. Firstly note that the generator is always off-diagonal (in
our block-diagonal sense of the term):
 \be
 P\eta(\ell)P = Q\eta(\ell)Q = 0.
 \ee
Evaluating (\ref{bd1}) in the two upper blocks gives
 \begin{eqnarray}
 \frac{d}{d\ell}(PHP) &=& P\eta QHP - PHQ \eta P \label{bd2} \\
 \frac{d}{d\ell}(PHQ) &=& P\eta QHQ - PHP\eta Q \label{bd3}.
 \end{eqnarray}
To show that the Hamiltonian flows to block-diagonal form, we use
a simple extension of the original Wegner matrix element proof
from Section \ref{WegFE} by defining a measure of off-diagonality
${\cal O} = ||PHQ||^2 = \mbox{Tr}(PHQHP)$. Using the expressions
for the flow in each block (\ref{bd2}) and (\ref{bd3}) we find its
derivative to be
 \begin{eqnarray}
 \frac{d {\cal O}}{d\ell} &=& \mbox{Tr}\left(P\eta Q(QHQHP -
 QHPHP)\right) + \mbox{Tr}\left((PHPHQ - PHQHQ)Q\eta P\right)\\
 &=& 2\mbox{Tr}(P\eta Q\eta P) = -2\mbox{Tr}((Q \eta P)^{\dagger}
 Q\eta P) = -2||Q\eta P||^2 \leq 0,
 \end{eqnarray}
which proves that $H_r$ must decrease during the flow, leading to
a block-diagonal Hamiltonian. This illustrates yet again how
flexible double commutator flow equations such as Wegner's can be.

It is important to compare the flow equations approach to
constructing a block-diagonal effective Hamiltonian, and other
previous approaches such as that of Lee and Suzuki
\cite{LeeSuzuki}. This latter approach writes the transformed
Hamiltonian as
 \be
 H' = e^{-\omega}He^{\omega}
 \ee
and then requires that $H'$ be block diagonal,
 \be \label{LeeS1}
 QH'P = 0
 \ee
and that the generator only has nonzero entries in the bottom-left
block,
 \be \label{LeeS2}
 \omega = Q\omega P
 \ee
This latter requirement is inconsistent with the primed
Hamiltonian $H'$ remaining unitarily equivalent to $H$, since the
generator $\omega$ is no longer anti-hermitian. The transformation
though is still a similarity transformation, so that the
eigenvalues of $H$ and $H'$ coincide. The requirements
(\ref{LeeS1}) and (\ref{LeeS2}) lead to a non-linear equation for
the generator $\omega$, which must be solved perturbatively. The
difference between this and the flow equations approach is that
the latter remains a strictly unitary transformation. Furthermore,
the differential flow equations constitute an approach whereby the
Hamiltonian is transformed continuously. In contrast, the method
of Lee and Suzuki attempts to find a one step transformation,
although this is normally computed in a discrete iterative
procedure.

%% file: Chapter2.tex
\chapter{Examples of flow equations} \label{Chapter2}

For the benefit of the reader who is anxious to discover if flow
equations have any merit with physical problems, we briefly review
here two recent treatments. They have been chosen out of the
myriad of other possibilities for pedagogical reasons, since the
first is rather simple and rapidly leads to an exact but
perturbative solution. The second is a good example of how flow
equations can be applied in a condensed matter context, in order
to find effective Hamiltonians. It illustrates the concepts of
renormalization (this will be returned to in Section
\ref{RenSec}), and the ${\cal L}$ ordering operation, as well as
dealing with unwanted newly-generated terms.

\section{Foldy-Wouthuysen transformation} \label{FWSec}

The Foldy-Wouthuysen transformation is a unitary transformation
that decouples the upper and lower pairs of components in the
Dirac equation. It is normally derived as an expansion in powers
of $1/m$ \cite{Gross}. We will derive it here using flow equations
following Ref. \cite{Bylev}, since it is a good example of a
non-trivial problem that can be solved in a perturbative but exact
treatment.

The initial Hamiltonian is (see eg. Ref. \cite{Gross})
\[ H_0 = \vec{\alpha} \cdot \left(\vec{p} - \vec{A}(\vec{x})\right) + \beta m
 + eA_0(\vec{x}) . \]
This Hamiltonian contains the $\vec{\alpha}$ terms which connect
the upper and lower components of the Dirac spinor. The objective
is to transform it to block-diagonal form. The key observation is
that the matrix $\beta$ has the special property that
 \begin{itemize}
 \item The most general form of the Hamiltonian during the flow
 can be written as a sum of even and odd terms, which commute and
 anticommute respectively with $\beta$ (This statement will be
 justified shortly).
 \begin{eqnarray}
 H(\ell) & = & \cal{E}(\ell) + \cal{O}( \ell ) \label{lline1} \\
 \left[ \cal{E}(\ell) , \beta \right] & = & 0 \label{lline2}\\
 \{ \cal{O} (\ell) , \beta \}  & = & 0. \label{lline3}
 \end{eqnarray}

 \item The required final Hamiltonian should commute with $\beta$
 in order to be block-diagonal
 \[ [H(\infty), \beta] = 0. \]
\end{itemize}

To ensure steepest descent to block-diagonality we choose the
generator as
 \be
  \eta(\ell) = [\beta m, H(\ell)] \label{lll2},
 \ee
where the mass $m$ appears in order to formulate a perturbative
solution in $1/m$. In this way we see that $\beta$ enters into
both the generator (\ref{lll2}) and into the form of the
Hamiltonian during the flow, via the parity relations
(\ref{lline2}) and (\ref{lline3}). Substituting the Hamiltonian
(\ref{lline1}) into the generator (\ref{lll2}) gives
\[ \label{line3} \eta(\ell) = 2m\beta {\cal O}(\ell). \]
The initial even and odd components are
\begin{eqnarray}
 {\cal E}(0) & = & \beta m + eA_0 \\
 {\cal O}(0) & = & \vec{\alpha} \cdot (\vec{p} - e \vec{A}).
\end{eqnarray}
By applying the commutation relations (\ref{lline2}) and
(\ref{lline3}) one sees that the flow equations can be written in
the following closed form
\begin{eqnarray} \label{closedeqs}
\frac{d {\cal E}(\ell)}{d \ell} & = & 4m\beta {\cal O}^2(\ell) \\
\frac{d {\cal O}(\ell)}{d \ell} & = & 2m\beta [{\cal O}(\ell),
{\cal E}(\ell)]
\end{eqnarray}
A word about such a closed form of equations is in order. ${\cal
E}(\ell)$ and ${\cal O}(\ell)$ are not simply scalar coefficients
but {\em operator-valued} functions of $\ell$. Nevertheless the
usual rules of calculus can be used to treat (\ref{closedeqs}) as
if it was a system of differential equations in scalar variables.
We now proceed to solve these equations perturbatively in $1/m$.
In order to conveniently distinguish different orders we introduce
the dimensionless\footnote{It is clear that $\ell$ has dimensions
$1/\mbox{(energy)}^2$ since the right hand side of the flow
equations has dimensions $\mbox{(energy)}^3$, while the left hand
side has dimensions $\mbox{(energy)}^1$.} flow parameter $s = m^2
\ell$. Now we express ${\cal E}(s)$ and ${\cal O}(s)$ in orders of
$1/m$. Since ${\cal E}(0) = \beta m + eA_0$ the expansion of
${\cal E}(s)/m$ contains terms starting with the zeroth order term
 \be \label{pertE}
 \frac{1}{m} {\cal
E}(s) = {\cal E}_0(s) + \frac{1}{m} {\cal E}_1(s) +
\frac{1}{m^2}{\cal E}_2(s) + \cdots
 \ee
while the expansion of ${\cal O}(s)$ starts with the first order
 \be \label{pertO}
 \frac{1}{m} {\cal O}
(s) = \frac{1}{m} {\cal O}_1(s) + \frac{1}{m^2}{\cal O}_2(s) +
\cdots
 \ee
Substituting the expansions (\ref{pertE}) and (\ref{pertO}) into
the flow equations (\ref{closedeqs}), equating terms of the same
order, and using the commutation relations (\ref{lline2}) and
(\ref{lline3}) gives
\begin{eqnarray}
\frac{d {\cal E}_n(s)}{d s} & = & 4\beta
\sum_{k=1}^{n-1} {\cal O}_k (s) {\cal O}_{n-k} (s) \\
\frac{d {\cal O}_n(s)}{d s} & = & -4{\cal O}_n (s) + 2\beta
\sum_{k=1}^{n-1} [{\cal O}_k (s), {\cal E}_{n-k} (s)].
\end{eqnarray}
These equations can be integrated to give the recursive type
solution
\begin{eqnarray}
{\cal E}_n (s) & = & {\cal E}_n (0) + 4\beta \int_0^{s} ds'
\sum_{k=1}^{n-1}{\cal O}_k (s')
{\cal O}_{n-k} (s')  \label {ESol} \\
{\cal O}_n (s) & = & {\cal O}_n (0)e^{-4s} + 2\beta e^{-4s}
\int_0^{s} ds' e^{4s'} \sum_{k=1}^{n-1}[{\cal O}_k (s') , {\cal
E}_{n-k} (s')] \label{OSol},
\end{eqnarray}
where the initial conditions are
 \begin{eqnarray}
 {\cal E}_0 (0) = \beta, \; {\cal E}_1(0) = eA_0(\vec{x}), \; {\cal
 E}_n(0)=0 \mbox{ if } n \geq 2 \\
 {\cal O}_1 (0) = \vec{\alpha} \cdot \left(\vec{p} -
 \vec{A}(\vec{x})\right), \; {\cal O}_n(0) = 0 \mbox{ if } n \geq
 2.
 \end{eqnarray}
As expected, we see that ${\cal O}_n (s)$ goes exponentially to
zero as $s \rightarrow \infty$, so that the final Hamiltonian is
indeed block diagonal. We may now proceed to evaluate ${\cal E}_n
(\infty)$. From the recursive solution (\ref{ESol}) we see that
the first two even orders ${\cal E}_0$ and ${\cal E}_1$ are not
affected by the flow. The first odd order term decays
exponentially
\[ {\cal O}_1 (s) = e^{-4s} \vec{\alpha} \cdot \left(\vec{p} -
 \vec{A}(\vec{x})\right).\]
Thus the second order even term is
 \begin{eqnarray}
 {\cal E}_2 (\infty) & = &
 4\beta \int_0^{\infty}ds' e^{-8s'} \left( \vec{\alpha}
 \cdot ( \vec{p} -
 e\vec{A}(\vec{x}))\right)^2 \\
 & = & \frac{1}{2} \beta\left( \vec{\alpha} \cdot ( \vec{p} -
 e\vec{A}(\vec{x}))\right)^2 \\
 & = & \frac{1}{2} \beta \left(( \vec{p} - e\vec{A}(\vec{x}))^2 -
 e\vec{\sigma} \cdot \vec{B}(\vec{x}) \right),
 \end{eqnarray}
where the last line follows from
 \[ (\vec{\sigma} \cdot \vec{a})(\vec{\sigma} \cdot \vec{b}) =
 \vec{a} \cdot \vec{b} + i \vec{\sigma} \cdot (\vec{a} \times
 \vec{b}), \]
and the explicit construction of the $\vec{\alpha}$ matrices. The
same type of index gymnastics gives rise to the third order term
\begin{eqnarray}
{\cal E}_3(\infty) & = & \frac{1}{8} \left[ [{\cal O}_1(0), {\cal
E}_1(0)], {\cal O}_1(0) \right] \\
 & = & -\frac{ie}{8} \vec{\sigma}(\nabla \times \vec{E}(\vec{x}))
 - \frac{e}{4} \vec{\sigma}(\vec{E}(\vec{x}) \times (\vec{p} -
 e\vec{A}(\vec{x})) - \frac{e}{8}\nabla \cdot \vec{E}(\vec{x}).
 \end{eqnarray}
It is clear that the ${\cal E}_n (\infty)$ have term for term
reproduced the standard Foldy-Wouthuysen transformation. One
advantage of the flow equation approach is that all orders in
$1/m$ can be computed in a standard way from the solutions
(\ref{ESol}) and (\ref{OSol}). Another advantage is that whereas
the standard treatment involves an ansatz for the generator
\cite{Gross}
 \[ U = U^{\dagger} = \sqrt{1 - \frac{p^2}{4m^2}} \vec{\alpha} \cdot
 \vec{p} \]
the flow equation approach proceeds in a systematic fashion - the
requirement that the final Hamiltonian commutes with $\beta$
basically fixes the generator. Although the level of computational
complexity involved in each method is ultimately similar, this is
a good example of {\em how} to solve the flow equations exactly,
in a perturbative framework.

\section{The electron-phonon interaction} \label{TEPI}

One of the most intriguing applications of flow equations is the
elimination of the electron-phonon interaction in favour of an
effective electron-electron interaction. In 1957 Bardeen, Cooper
and Schrieffer developed their famous theory of superconductivity
\cite{BCS}, which involved an effective interaction between
electrons of a many-particle system \cite{Cooper}. Fr\"{o}hlich
had showed in 1952 that this effective electron-electron
interaction can have its origin in the interaction between the
lattice phonons and the electrons \cite{Frolich, Kittel}, in the
sense that the effective electron-electron attraction term arises
from eliminating the electron-phonon interaction term in the
original Hamiltonian by a unitary transformation.

Fr\"{o}hlich's approach attempts to find the renormalized
Hamiltonian up to quadratic order in the electron-phonon coupling
coefficients. The unitary transformation employed is highly
singular at certain points in electron momentum space and may
become repulsive or even undefined, due to a vanishing energy
denominator.

The electron-phonon elimination problem was treated in 1996 using
flow equations by the father of flow equations, Franz Wegner, and
a colleague Peter Lenz \cite{LenzWegner}. This paper unleashes the
flow equations on a highly non-trivial problem in a thorough and
comprehensive fashion. The objective is similar to that of
Fr\"{o}hlich : eliminate the electron-phonon interaction to second
order in the electron-phonon coupling. The intriguing outcome is
that the result differs slightly from Fr\"{o}hlich's. The
transformation is less singular and always attractive for
electrons belonging to a Cooper pair.

These statements will become clearer in what follows, where we
shall present an overview of Wegner's treatment. First though, we
review Fr\"{o}hlich's treatment.

\subsection {Fr\"{o}hlich's Transformation} The Hamiltonian we are
concerned with is
 \begin{eqnarray}
 H &=& \sum_q \omega_q a_q^{\dagger}a_q + \sum_k \epsilon_k
 c_k^\dagger c_k + \sum_{k,q}M_q(a_{-q}^\dagger +
 a_q)c_{k+q}^\dagger c_k + E \\
  & = & H_0 + H_{e-p},
 \end{eqnarray}
where the summation index $k=\{\mathbf{k}, \sigma\}$ since the
interaction is not spin-dependent. The $a_q^{(\dagger)}$ are
bosonic annihilation (creation) operators for the phonons and the
$c_k^{(\dagger)}$ are fermionic annihilation (creation) operators
for the electrons. The $M_q(a_{-q}^\dagger + a_q)c_{k+q}^\dagger
c_k$ terms are the electron-phonon interactions we wish to somehow
renormalize into an effective electron-electron interaction. There
is no dependancy on the electron momentum in  the initial coupling
coefficients $M_q$ \cite{Bloch, Nordheim}.  The original Coulomb
interaction has not been included as it plays no significant role
in Fr\"{o}hlich's method. $E$ is a constant energy term which may
be present.

The idea of Fr\"{o}hlich's transformation is a simple brute-force
expansion up to order $M_q^2$ of a unitary transformation via the
BCH formula
 \be \label{BCSS}
 H^F = e^{S}He^{-S} = H + [S, H] + \frac{1}{2}[S, [S, H]] + \cdots
 \ee
where the Fr\"{o}hlich generator is
 \be
 S = \sum_{k,q}M_q\left(\frac{1}{\epsilon_{k+q} - \epsilon_k +
 \omega_q}a^{\dagger}_{-q} + \frac{1}{\epsilon_{k+q} - \epsilon_k -
 \omega_q}a_{q} \right) c^{\dagger}_{k+q}c_k \label{FBCH}.
 \ee
The fact that $S$ contains the coupling coefficient $M_q$ shows
that Eq. (\ref{BCS}) can be arranged in a power series in $M_q$,
and explains the meaning of the phrase `up to order $M_q^2$'. The
expression also contains denominators which may vanish, since the
usual assumption is that $\epsilon_k$ is a quadratic dispersion
while $w_k$ is linear. It is in this sense that the transformation
is said to be highly singular in certain regions of momentum
space. The motivation for using $S$ as a generator is that
 \be \label{genrel}
 H_{e-p} = - [S, H_0],
 \ee
which shows that, at least for the first terms in the BCH
expansion, the electron-phonon interaction term has been
eliminated (This was the requirement that initially determined the
form of $S$). In fact, the relation (\ref{genrel}) shows that only
one commutator needs to be explicitly evaluated. To see this, we
arrange the terms appearing in (\ref{BCSS}) in powers of $M_q$
 \begin{eqnarray}
 \lefteqn{H + [S, H] + \frac{1}{2}[S, [S, H]] =  }\\
 & & \underbrace{H_0}_{\mbox{zeroth order}} + \underbrace{H_{e-p} +
 [S, H_0]}_{\mbox{first order}} +  \underbrace{[S, H_{e-p}] +
 \frac{1}{2}[S, [S, H_0]]}_{\mbox{second order}} \\
 & = & H_0 -\frac{1}{2}[S, H_{e-p}].
 \end{eqnarray}
The evaluation of $[S, H_{e-p}]$ will yield (schematically)
 \be \label{schematics}
 [(a^\dagger + a)c^\dagger c, (a^\dagger + a)c^\dagger c] \rightarrow
 c^\dagger
 c^\dagger c c + (aa + a^\dagger a + a^\dagger a^\dagger)c^\dagger
 c.
 \ee
We will ignore the two-phonon-processes on the right and
concentrate only on the new electron-electron interaction term.
The transformed Hamiltonian reads
 \begin{eqnarray}
 H^F = \sum_q \omega_q^F a_q^{\dagger}a_q + \sum_k \epsilon_k^F c_k^\dagger
 c_k + \sum_{k, k', \delta} V^F_{k, k', \delta}
 c^\dagger_{k+\delta} c_{k'-\delta}^\dagger c_{k'} c_k
 \label{FrolichH},
 \end{eqnarray}
where $V^F_{k, k', \delta}$ is independent of $k'$ and is given by
 \be
 V^F_{k, k', \delta} = -|M_q|^2 \frac{\omega_q}{\omega_q^2-(\epsilon_{k+q} -
 \epsilon_k)^2}.
 \ee
Providing $(\epsilon_{k+q} - \epsilon_k) < |\omega_q|$ we have
thus generated an effective attractive interaction amongst the
electrons. The nature of the denominator shows, however, that for
certain regions of momentum space the interaction may become
repulsive or singular.

\subsection{Flow equations approach}
The starting point is always to choose some kind of
parametrization of the Hamiltonian during the flow. We choose the
simplest possible form
 \begin{eqnarray}
 H(\ell) &=& H_0(\ell) + H_{e-e}(\ell) + H_{e-p}(\ell) \\
 &=& \sum_q \omega_q(\ell) a_q^\dagger a_q + \sum_k \epsilon_k(\ell) c_k^\dagger
 c_k + \sum_{k, k', \delta} V_{k, k', \delta}(\ell) c_{k+\delta}^\dagger c_{k'-\delta}^\dagger c_{k'}
 c_k + \\ & & \qquad \sum_{k,q} \left(M_{k,q}(\ell)a_{-q}^\dagger +
 M^*_{k+q,-q}(\ell)
 a_q\right)c_{k+q}^\dagger c_k,
 \end{eqnarray}
where we have made clear our intention to track only the most
important terms during the flow. Note the special arrangement of
indices for the coupling $M_{k,q}(\ell)$ which is done for later
convenience. The initial conditions on the coefficients are
 \begin{eqnarray}
 M_{k,q}(0) = M^*_{k+q,-q}(0) = M_q, \quad
  \quad \omega_q(0)=\omega_q, \quad \epsilon_k(0)=\epsilon_k,
 \quad V_{k, k', \delta}(0) = 0.
 \end{eqnarray}

The next step is to choose the generator $\eta$. Our intention is
to remove the electron-phonon interaction terms, which is
equivalent to the requirement that the final Hamiltonian
$H(\infty)$ should commute with the total number operators for the
phonon and electron fields. It is more convenient to require the
similar restriction that $H(\infty)$ commutes with $H_0$. Strictly
speaking, the flow equation program instructs us to then adopt as
our generator $\eta(\ell)=[H_0, H(\ell)]$, where the full
$H(\ell)$ should be used on the right hand side. In the name of
simplicity Wegner preferred
 \be \label{genchoice}
 \eta(\ell) = [H_0, H_{e-p}(\ell)]
 \ee
as the generator. Of course, the physics is never violated by
choice of generator since the transformation is still strictly
unitary (up to the given order). It remains to be proved however
that this choice of generator is optimal (in the sense of removing
the electron-phonon interaction) in this case. Evaluating
(\ref{genchoice}) gives
 \be
  \eta(\ell) =\sum_{k,q} \left(M_{k,q}(\ell)\alpha
  _{k,q}a^{\dagger}_{-q}+M^*_{k+q,-q}(\ell)
  \beta_{k,q}a_q\right)c^{\dagger}_{k+q}c_k,
 \ee
where $\alpha_{k,q}$ and $\beta_{k,q}$ are the familiar constants
appearing previously,
 \[ \alpha_{k,q} = \varepsilon_{k+q}-\varepsilon_k +\omega_q , \quad
 \beta_{k,q} = \varepsilon_{k+q}-\varepsilon_k -\omega_q.\]
Evaluating the commutator $[\eta, H]$ gives rise to various new
interactions of the form displayed in Eq. (\ref{schematics}),
which are ignored. Additional terms of the form $a^\dagger
a^\dagger + aa$ also appear. Wegner showed that these can be
transformed away by adding new terms to the flow and choosing
coefficients carefully. After normal ordering, we are finally left
with a system of differential equations for the renormalization of
the Hamiltonian parameters \cite{LenzWegner},
\begin{eqnarray}
  \frac{dM_{k,q}}{dl} & = & -\alpha_{k,q}^2M_{k,q}\nonumber \\
  & & -2
  \cdot \sum_{\delta} V_{k,k+q+\delta ,\delta }M_{k+\delta
    ,q}\alpha_{k+\delta ,q}\cdot \left( n_{k+q+\delta }-n_{k+\delta }\right
  ) \nonumber \\
  & & - 2M_{k,q}\alpha_{k,q}\cdot \sum_{\delta}
  \left(n_{k+\delta }V_{k,k+\delta ,\delta }-n_{k+q+\delta
      }V_{k+q,k+q+\delta ,\delta } \right) \nonumber \\
  & &+ 2 \cdot
    \sum_{k'} V_{k,k'+q,q}M_{k',q}\alpha_{k',q} \cdot \left(n_{k'+q}-n_{k'}
  \right) \nonumber \\
  & & - \frac{M^*_{k+q,-q}}{\omega _q} \sum_{k'}
  M_{k',q}M_{k'+q,-q}\beta_{k',q} \cdot (n_{k'+q}-n_{k'}) \label{fg1}
  \\ \frac{dV_{k,k',q}}{dl} & = &
  M_{k,q}M^*_{k'-q,q}\beta_{k',-q}-M^*_{k+q,-q}M_{k',-q}\alpha_{k',-q}
  \label{fg3} \\ \frac{d\omega_q}{dl} & = & 2 \cdot \sum_{k}
  |M_{k,q}|^2\alpha_{k,q} \cdot (n_{k+q}-n_k) \label{fg4} \\
  \frac{d\varepsilon _k}{dl} & = & -\sum_{q}
  (2n_q|M_{k+q,-q}|^2\beta_{k,q}+2(n_q+1)|M_{k,q}|^2\alpha_{k,q})
  \label{fg5},
\end{eqnarray}
where $n_q$ is a bosonic occupation number whereas $n_k$ and
$n_k+q$ denote the fermionic ones.

The aim is to solve these equations up to order $|M_{k,q}|^2$, in
order to compare the results with Fr\"{o}hlich's treatment. In
this way lines two to five in the flow of $M_{k,q}$ (\ref{fg1})
are irrelevant, and we are left with
\[ \frac{dM_{k,q}}{d \ell} = -\alpha_{k,q}^2 M_{k,q} \]
with solution
 \be \label{MSolution}
 M_{k,q}(\ell) = M_q e^{-\left(\varepsilon_{k+q}(0) -
 \varepsilon_k(0) + \omega_k(0)\right)^2 \ell},
 \ee
which shows that the goal of eliminating $H_{e-p}$ is achieved as
 $\ell \rightarrow \infty$.
This solution is substituted directly into Eqs.
(\ref{fg3})-(\ref{fg5}). Integrating the resulting differential
equations is easy and gives the same values for the renormalized
single-particle energies as in Fr\"{o}hlich's treatment (i.e.
$\omega_k(\infty) = \omega_k^F, \quad
\varepsilon_k(\infty)=\varepsilon_k^F$). The result for the
electron-electron interaction is
\begin{equation}
  V_{k,k',q} (\infty) = |M_q| ^2 \left ( \frac{\beta
    _{k',-q}}{\alpha ^2_{k,q}+\beta ^2_{k',-q}} - \frac{\alpha
    _{k',-q}}{\beta ^2_{k,q}+\alpha ^2_{k',-q}} \right) , \label{sA3}
\end{equation}
which is explicitly a function of $k$, $k'$ and $q$. Since
Fr\"{o}hlich's interaction is independent of $k'$ we must choose a
value for $k'$ for purposes of comparison. The natural choice is
to compare the interaction between the electrons of a Cooper
pair($k'=-k$), where it becomes
\begin{equation}
  V_{k,-k,q}(\infty) =- |M_q|^2\frac{\omega _q}{(\varepsilon
    _{k+q}-\varepsilon_k)^2+\omega _q^2} \label{sA5} .
\end{equation}
The corresponding Fr\"{o}hlich value is
\begin{equation}
  V^{F}_{k,k',q} = V^{F}_{k,-k,q} =-|M_q|^2\frac{\omega
    _q}{\omega_q^2 - (\varepsilon _{k+q}-\varepsilon_k)^2} \quad
  .\label{sA4}
\end{equation}
At this point we realize that a remarkable difference has arisen
between the flow equations approach and the Fr\"{o}hlich
transformation, which is depicted in the relative signs in the
denominators of (\ref{sA5}) and (\ref{sA4}).

\subsection{${\cal L}$ Ordering and the generator expansion}
The flow equation
 \be \label{dH1} \frac{dH}{d\ell} = [\eta(\ell), H(\ell)] \ee
looks formally like the Heisenberg equation of motion with an
explicitly time dependent Hamiltonian. Recall from Section
\ref{WegFE} that the generator $\eta(\ell)$ can be expressed in
terms of the unitary transformation $U(\ell)$ appearing in $
H(\ell) = U(\ell)H(0)U^{\dagger}(\ell)$ via
 \be
 \eta(\ell) = \frac{dU}{d\ell}U^\dagger.
 \ee
Thus the differential equation for $U(\ell)$ is
 \be
 \frac{dU}{d\ell} = \eta(\ell)U(\ell),
 \ee
with the familiar implicit solution
\[ U(\ell) = 1 + \int_0^\ell d\ell ' \eta(\ell ')U(\ell '). \]
This can be written as a formal series
\begin{eqnarray}
 U(\ell) &=& 1 + \int_0^\ell d\ell'\eta(\ell') + \int_0^\ell
 \int_0^{\ell'} d\ell' d\ell'' \eta(\ell') \eta(\ell'') + \cdots
 \\ & = & 1 + \int_0^\ell d\ell'\eta(\ell') + \frac{1}{2!} \int_0^\ell
 \int_0^{\ell} d\ell' d\ell'' {\cal L}\left[\eta(\ell') \eta(\ell'')\right] + \cdots \\
 & = & {\cal L} e^{\int_0^\ell d\ell' \eta(\ell ')},
\end{eqnarray}
where the $\ell$-ordering operator $\cal L$ has been introduced
which orders products of $\ell$-dependent operators in order of
decreasing $\ell$.

In order to compare Fr\"{o}hlich's result with the result from
flow equations, we must first solve the following general problem
: Find an equivalent generator $S(\ell)$ which accounts for the
entire $\ell$-evolution of the generator $\eta(\ell)$ in the sense
that
 \begin{eqnarray}
 e^{S(\ell)} &=& U(\ell) \label{EExp} \\
 & = & 1 + \int_0^\ell d\ell'\eta(\ell') + \frac{1}{2!} \int_0^\ell
 \int_0^{\ell} d\ell' d\ell'' {\cal L}\left[\eta(\ell') \eta(\ell'')\right] +
 \cdots  \label{UExp}
 \end{eqnarray}
In this way one could compare the $l$-independent flow equations
generator $S(\infty)$ and Fr\"{o}hlich's generator $S_F$. To solve
this problem, we first artificially insert a $\theta$ dependence
into the expansion (\ref{UExp}) so that we can group terms of the
same order. In the electron-phonon problem, this process is not
artificial since $M_q$ would serve the role of $\theta$ as it
appears linearly in the generator. The obvious way to insert
$\theta$ is
 \be
 U(\theta, \ell) = 1 + \theta\int_0^\ell d\ell'\eta(\ell') + \frac{\theta^2}{2!} \int_0^\ell
 \int_0^{\ell} d\ell' d\ell'' {\cal L}\left[\eta(\ell') \eta(\ell'')\right] +
 \cdots
 \ee
Now we view the generator $S$ as a function of $\theta$ and $\ell$
, expand it in a power series, and insert this series into the
exponential in (\ref{EExp})
 \begin{eqnarray}
 S(\theta, \ell) &=& \theta S_1(\ell) + \theta^2 S_2
 (\ell) + \cdots \\
 e^{S(\theta, \ell)} &=& 1 + (\theta S_1 + \theta^2 S_2 + \cdots) +
 \frac{1}{2!}(\theta S_1 + \theta^2 S_2 + \cdots)^2 + \cdots
 \end{eqnarray}
After grouping terms order by order, and setting $\theta=1$, we
are faced with simplifying sums of ${\cal L}$-ordered products and
conventional products. For instance, for $S_2(\ell)$ we have
 \begin{eqnarray}
 S_2(\ell) &=& \frac{1}{2} \int_{\ell_1=0}^{\ell_1=\ell}
 \int_{\ell_2=0}^{\ell_2=\ell} d\ell_1 d\ell_2 \bigg( {\cal
 L}[\eta(\ell_1)\eta(\ell_2)] - \eta(\ell_1)\eta(\ell_2) \bigg) \\
 & = & \frac{1}{2} \int_{\ell_1=0}^{\ell_1=\ell}
 \int_{\ell_2=0}^{\ell_2=\ell_1} d\ell_1 d\ell_2
 \left[ \eta(\ell_1), \eta(\ell_2) \right].
 \end{eqnarray}
In this spirit we obtain, to second order
 \be \label{SExpansion}
 S(\ell) = \int_0^\ell d\ell_1 \eta(\ell_1) +
 \frac{1}{2}\int_0^\ell \int_0^{\ell_1} d\ell_1 d\ell_2
 \left[\eta(\ell_1), \eta(\ell_2)\right] + \cdots
 \ee

\subsection{Comparison with Fr\"{o}hlich's Results} \label{LastSec}

 We are now in a position to compare the approaches of
Fr\"{o}hlich and the flow equations. Fr\"{o}hlich expanded
 \be \label{BCS}
 U_F = e^{S_F}
 \ee
up to second order in $M_q$, where $S_F$ was an $\ell$-independent
generator. To compare results we must simply compute $S =
S(\infty)$ up to second order. Since we know $\eta(\ell)$
precisely up to second order from Eq. (\ref{MSolution}), there is
no inconsistency in our approach and we simply substitute
(\ref{MSolution}) into the generator expansion (\ref{SExpansion}).
The first order term returns precisely Fr\"{o}hlich's
transformation:
 \be \label{SSoln} S = S_F + S_2 +
\cdots \ee The second order term involves the commutator
$[\eta(\ell_1), \eta(\ell_2)]$ which is of the schematic form
$[(a^\dagger + a)c^\dagger c, (a^\dagger + a) c^\dagger c]$. This
commutator has been encountered before in Eq. (\ref{schematics}).
The result is a complicated sum of products of the single-particle
energy differences $\alpha_{k,q}$ and $\beta_{k,q}$: \newpage
\begin{eqnarray}
  S_2 & = & - \frac{1}{2}\sum_{k,k',q} |M_q(0)|^2 \nonumber \\ & &
  \left\{ \frac{\alpha _{k,q}}{\beta_{k',-q} (\alpha _{k,q}^2+\beta
      _{k',-q}^2)}-\frac{1}{\alpha_{k,q}\beta_{k',-q}}
    -\frac{\beta_{k,q}}{\alpha_{k',-q}(\beta_{k,q}^2+\alpha_{k',-q}^2)}+
\frac{1}{\beta_{k,q}\alpha_{k',-q}}\right
    \} \nonumber \\ & & \times c^{\dagger}_{k+q} c^{ }_k
    c^{\dagger}_{k'-q}c^{ }_{k'-q}. \nonumber\\ & & + \mbox{ terms of
      the structure }a^{\dagger}a c^{\dagger}c.
\end{eqnarray}
We can now perform a consistency check on our calculations. We do
this by simply exponentiating our result (\ref{SSoln}) up to
second order in $M_q$
 \be
 e^{S_1 + S_2}H_0e^{-(S_1 + S_2)} = H_0 + [S_1 + S_2, H] +
 \frac{1}{2}[[S_1 + S_2, [S_1 + S_2, H]] + \cdots
 \ee
If we have kept track of all orders consistently, we should be
able to account for the difference between Fr\"{o}hlich's
effective electron-electron interaction, and the effective
interaction from the flow equations, by appealing to the extra
term $S_2$ in our generator. Up to order $M_q^2$ the only extra
commutator we need to calculate is $[S_2, \sum_k \varepsilon_k
c_k^\dagger c_k]$. Indeed we find
\begin{equation}
  [S_2, \sum_{k} \varepsilon _k c^{\dagger}_kc_k] = \sum_{k,k',q}
  (V_{k,k',q} (\infty) -V_{k,k',q}^F ) c^{\dagger}_{k+q}
  c^{\dagger}_{k'-q}c_{k'}c_k.
\end{equation}
Thus we have demonstrated the important result that {\em the flow
of $\eta(\ell)$ has dynamically altered the unitary
transformation}. This is because the flow equations are steepest
descent curves which change their direction in operator space as
the flow proceeds. In this way we see that there is no conflict
between Fr\"{o}hlich's transformation and the flow equations :
they are expanding two different unitary transformations. The flow
equations result is more accurate since {\em it proceeds further
up to order $M_q^2$}, by including more terms. This gives a result
which is less singular and always attractive.

Up to now we have always worked only up to order $M_q^2$. Lenz and
Wegner \cite{LenzWegner} go on to show that a study of the full
problem in the asymptotic limit ($\ell \rightarrow \infty$) is
indeed tractable if one makes the ansatz
 \be
 M_{k,q}(\ell) = M_q(0)e^{{-\int_0^\ell d\ell' \left(
 \varepsilon_{k+q} - \varepsilon_k + \omega_q(\ell') \right) }^2},
 \ee
which by including $\ell$-dependency in $\omega_q(\ell)$ considers
terms of all orders in $M_q(0)$. In this way it is possible to
show that the electron-phonon coupling is always eliminated, even
in the case of degeneracies.

We will return to the electron-phonon problem in Chapter
\ref{RenormChapter}.

%% file: Chapter3.tex
\chapter{The Lipkin model} \label{Chap3}

\section{Introduction}

Originally introduced in nuclear physics in 1965 by Lipkin,
Meshkov and Glick \cite{Lipkin1, Lipkin2, Lipkin3}, the Lipkin
model is a toy model that describes in its simplest version two
shells for the nucleons and an interaction between nucleons in
different shells. It has proved to be a traditional testing ground
for new approximation techniques \cite{Hatsuda, Walet} since it is
numerically solvable. There has been renewed interest in it
recently in the context of finite temperatures and as a test of
self-consistent RPA-type approximations \cite{TZeng, DShuck}. In
this chapter we will give a short introduction to the Lipkin model
and present some exact numerical results for the flow equations.
This will be followed by three recent approaches to the model via
flow equations \cite{PF, Stein, MielkeA}. Finally we shall present
our own work on the Lipkin model, which attempts to find an
effective Hamiltonian valid for the entire coupling range by
tracking the ground state during the flow. This is achieved
firstly by employing a variational calculation as an auxiliary
step. A more sophisticated method then dispenses with this
requirement by utilizing a self-consistent calculation, which
delivers good results.

\subsection{The model}

In the Lipkin model $N$ fermions distribute themselves on two
$N$-fold degenerate levels which are separated by an energy $\xi
_0$. The interaction $V_0$ introduces scattering of pairs between
the two shells.

 \begin{figure}
 \begin{center}
 \includegraphics[width=0.5\textwidth]{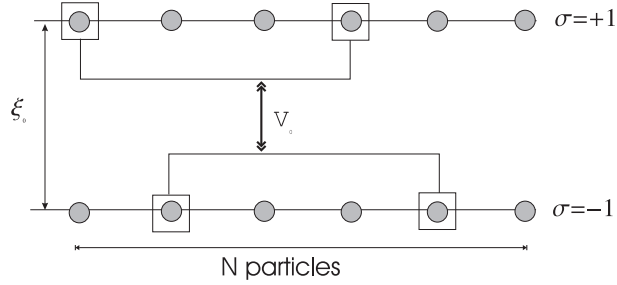}
 \end{center}
 \caption[The Lipkin Model]{\label{ModelFig} The Lipkin model. Note how the interaction only couples
 states in the lower level to states in the upper level.}
 \end{figure}
The Hamiltonian is
 \be \label{InitHamilt}
  {\cal H}(\xi_0, V_0)=\frac{1}{2}\xi_0\sum_{\sigma,p}\sigma a^\dagger_{p\sigma} a_{p\sigma}
 +\frac{1}{ 2} V_0\sum_{pp',\sigma} a^\dagger_{p\sigma}
 a^\dagger_{p'\sigma} a_{p'- \sigma} a_{p-\sigma},
 \ee where
$\sigma= \pm 1$ labels the levels (see Fig. \ref{ModelFig}).

A spin representation for $\cal H$ may be found  by setting
 \be
 \label{spinreps} J_z=\frac{1}{2}\sum_{p,\sigma}\sigma
 a^\dagger_{p\sigma} a_{p\sigma},\ J_+=\sum_p a^\dagger_{p,+1} a_{p,-1},\
 J_-=\sum_p a^\dagger_{p-1} a_{p+1},
 \ee
which satisfy the SU(2) algebra
 \be [J_z,J_\pm]=\pm J_\pm \;,\; [J_+,J_-]=2J_z.
 \ee
The resulting Hamiltonian,
 \be \label{resHam}
 {\cal H}(\xi_0, V_0) = \xi_0J_z + \frac{1}{2}V_0(J_+^2 + J_-^2),
 \ee
commutes with $J^2$ and its irreducible representation breaks up
into blocks of size $2j+1$, where $j$ is the total angular
momentum quantum number. The low-lyings states occur in the
multiplet $j=N/2$ \cite{Lipkin1}, which is a matrix of dimension
$N+1$. This is the reason that the model is numerically solvable.
Without a quasi-spin representation the dimension of the bare
Hamiltonian (\ref{InitHamilt}) scales exponentially with the
number of particles since there are $2^N$ basis states, which have
the form
 \be
 \prod\nolimits^N_{i=1} a_{i\pm}^+ \left|0\right>.
 \ee
The Hamiltonian (\ref{resHam}) depends linearly on two parameters.
To remove a trivial scaling factor, from now on we divide by
$\xi_0$ and for convenience drop the prime on the rescaled
Hamiltonian (i.e. all our results will be expressed in units of
$\xi_0$). Defining $\beta_0 = 2jV_0/\xi_0$ we obtain
 \be \label{Hoo1}
 {\cal H}(\beta _0) = J_z + \frac{\beta_0}{4j}(J_+^2 + J_-^2).
 \ee

With no interaction the ground state is the product state of all
particles in the lower level
$\left|\psi_0\right>=\prod\nolimits^N_{i=1} a_{i-}^+
\left|0\right>$ which is written in the spin basis as
$\left|\psi_0\right>=\left|-j\right>$, and $<\psi_0| {\cal H}_0
|\psi_0> = -j$.

\subsection{Phase transition}

The model shows a phase change in the $N\rightarrow \infty$ limit
above $\beta_0=1$ where the ground state becomes a condensate of
pairs, where each pair consists of one particle from the lower
level ($\sigma=-1$) and one particle from the upper level
($\sigma=+1$). To see this, we use the Holstein-Primakoff
representation of SU(2) to cast the problem in bosonic language,
 \be \label{100}
 (J_z)_B = -j+b^\dagger b, \quad (J_+)_B = (J_-)^\dagger_B
 = b^\dagger\sqrt{2j-b^\dagger b}.
 \ee
Substituting these into the Hamiltonian (\ref{Hoo1}) gives the
bosonized version \be \label{101} {\cal H}_0(\beta_0) = b^\dagger
b + \beta_0(bb + b^\dagger b^\dagger) + O(1/N) ,\ee where we have
also dropped constant terms. Now we perform a standard Bogolubov
transformation to rewrite the Hamiltonian (\ref{101}) in terms of
new boson operators $B$ and $B^\dagger$,
 \be
 b=\cosh\!\phi \,B + \sinh\!\phi \,B^{\dagger},\quad  b^{\dagger} =
 \sinh\!\phi\, B + \cosh\!\phi\, B^{\dagger},
 \ee
where $B$ and $B^{\dagger}$ satisfy $[B, B^+] = 1$. Choosing
$\phi$ so that the $BB$ and $B^+B^+$ terms vanish gives \be {\cal
H}_0(\beta_0) = \sqrt{1-\beta_0^2}\;B^{\dagger}B + O(1/N) \ee from
which we conclude that in the limit $N \rightarrow \infty$, the
energy gap $\Delta$ between the ground and first excited states is
given by
 \be \label{ExactD}
 \Delta(\beta_0)=\left\{
 \begin{array}{r@{\quad \quad}l} \sqrt{1-\beta_0^2} & \beta_0 \leq
 1 \\ 0 & \beta_0 > 1
 \end{array}\right. .
 \ee
clearly showing a non-analytic phase transition at $\beta_0 = 1$.

\section{Some preliminary numerics} \label{SPN}
\subsection{Brute force diagonalization}
\begin{figure}[h]
 \begin{center}
 \includegraphics[width=0.5\textwidth]{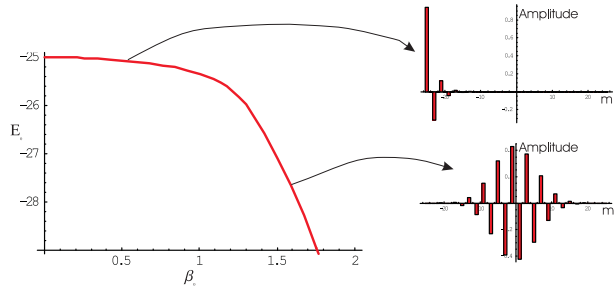}
 \end{center}
 \caption {\label{GroundFig} Exact ground state energy and its associated state, as a function of
 $\beta_0$, for $N=50$ particles.}
 \end{figure}

\begin{figure}[t]
 \begin{center}
 \includegraphics[width=0.4\textwidth]{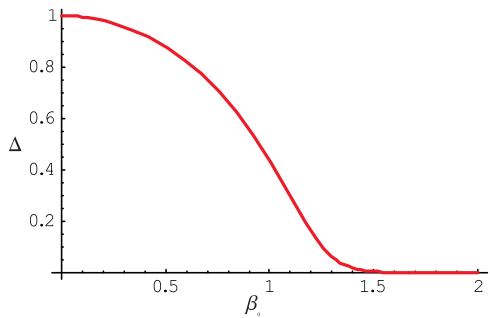}
 \end{center}
 \caption {\label{GapFig} Band gap $E_1-E_0$ as a function of $\beta_0$, for $N=50$ particles.}
 \end{figure}

In order to make completely clear the dynamics behind the model,
and eliminate any possible confusion that may be lingering in the
reader's mind at this point, we present numerical results for the
ground state energy $E_0$ and the band gap $\Delta$, as functions
of $\beta_0$ in Fig. \ref{GroundFig} and Fig. \ref{GapFig}. These
were obtained by numerical diagonalization routines in the case of
$N=50$ particles, so that the $j=N/2$ multiplet is a matrix of
dimension $51$. It is clear that there is a sharp change in the
nature of the ground state, and in the value of the band gap,
around the phase transition line $\beta_0=1$. This is made
explicit by diagrams of the ground state for two representative
$\beta_0$ values. These diagrams are to be understood as the
components of the ground state $\left<m | \Psi \right>$. In the
first phase the ground state has almost all particles on the lower
level (i.e. $m=-j$ amplitude is dominant), with only a few
excitations. In the second (deformed) phase the ground state is a
condensate of pairs (i.e. values of $m$ around zero are dominant).
It is important to note that there are no components of the ground
state for $m=-j+i$, where $i$ is odd, since the dynamics of the
model only promotes/demotes particles in pairs.

\subsection{Exact numerical solution of flow equations}
The exact solution for the flow equations in certain important
cases will now be obtained numerically. This exercise is highly
instructive as it shows which operators are important during the
flow, as a function of the interaction $\beta_0$.

\subsubsection{Choice of generator} \label{ChoiceofGen}

\begin{figure}
 \begin{center}
 \includegraphics[width=0.5\textwidth]{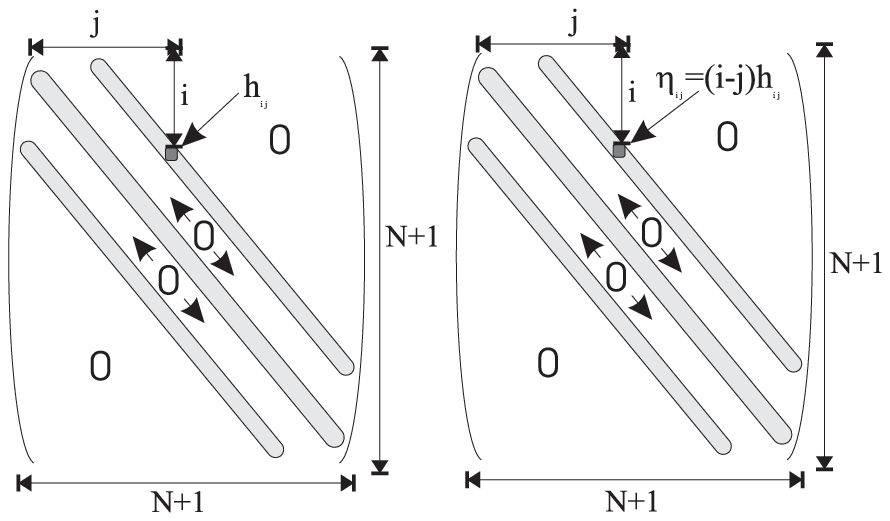}
 \caption{\label{banddiag} The matrix elements $\left<i\right| H_0 \left|j\right>$ are zero except for
 $i=j$, $i=j\pm2$.}
 \end{center}
 \end{figure}

The first step is to choose a parametrisation of ${\cal H}(\ell)$
that is closed under the flow. The answer to this question depends
largely on the choice of generator $\eta(\ell)$. In this model we
are interested in the energy spectrum and thus we want to
completely diagonalize the Hamiltonian, and not just reduce it
into block-diagonal form. Now the operator $X$ in
 \be
 \eta(\ell) = [X, {\cal H}(\ell)]
 \ee
represents the destination of the steepest descent flow (see
Chapter \ref{MathChapter}). This presents us with two natural
choices:
 \begin{enumerate}
 \item $X =\mbox{Diag}\left( {\cal H}(\ell)\right)$ - the original Wegner
 prescription, or
 \item $X = J_z$,
 \end{enumerate}
both of which will ensure flow to diagonality (the former slightly
faster). To make our decision we appeal to the form of the
original Hamiltonian (\ref{Hoo1}), which is band diagonal (see
Fig. \ref{banddiag}). By ``band diagonal'' we mean that there are
only three bands of nonzero entries in the matrix:
 \be
 {\cal H}_{ij} = 0  \mbox{ unless } i=j \mbox{ or } i=j\pm 2.
 \ee

We shall employ the second option and choose our generator as
 \be \label{genc}
 \eta(\ell) = [J_z, {\cal H}(\ell)] \quad \eta_{ij} = (i-j){\cal
 H}_{ij}
 \ee
since, contrary to $\eta=[\mbox{Diag}({\cal H}), {\cal H}]$ , this
choice preserves the band-diagonal structure of the Hamiltonian.
This can be proved directly from multiplying out the types of
matrices involved \cite{Drie2}. A more illuminating procedure is
to show that the following parametrization of ${\cal H}(\ell)$
 \be \label{Ba5}
 {\cal H}(\ell) = \sum_{\mbox{\scriptsize odd } i} \lambda_i(\ell) J_z^i +
 \sum_i \chi_i(\ell)(J_z^i J_+^2 + J_-^2 J_z^i)
 \ee
remains closed after utilizing (\ref{genc}) as the generator. This
is the most general matrix of band-diagonal form, as defined
above, except that no even powers of $J_z$ need be included. The
reason for this is as follows. The initial Hamiltonian
(\ref{Hoo1}) can be rewritten in terms of the $J_x$ and $J_y$
angular momentum operators
 \be
 J_x = \frac{1}{2}(J_+ + J_-), \, J_y = \frac{1}{2i}(J_+ - J_-)
 \ee
in the following way,
 \be \label{JxHam}
 {\cal H}(\beta _0) = J_z + \frac{\beta_0}{2j}(J_x^2 - J_y^2).
 \ee
The rotational symmetry operation
 \be
 J_x \rightarrow J_y, \, J_y \rightarrow J_x, \, J_z \rightarrow
 -J_z
 \ee
transforms ${\cal H}_0$ into $-{\cal H}_0$, so if $E$ is an
eigenvalue of $H_0$, then so is $-E$. Consequently the eigenvalues
must occur in positive/negative pairs symmetrically situated
around the zero point energy. In this way we see that no even
powers of $J_z$ need be included in (\ref{Ba5}), as such terms
would shift the center of the eigenspectrum positively, away from
zero, violating the initial symmetry shown in Eq. (\ref{JxHam}).

To show that the parametrisation (\ref{Ba5}) remains closed we
consider substituting it into the flow equation $\frac{d{\cal
H}}{d\ell}=[[J_z, {\cal H}], {\cal H}]$. Since the commutator of
$J_z$ with a power of $J_{\pm}$ takes the schematic form
 \be
 [J_z, J_{\pm}^i] = f_i^{\pm}(J_z)J_{\pm}^i,
 \ee
where $f_i^{\pm}$ is some computable function of $J_z$, we see
that no new terms will be generated during the flow. In this way
the problem has been simplified as the flow has been restricted
only to the band-diagonal matrix elements.

\subsubsection{Numerical results} \label{NumR}
Let us first tidy up things by normalizing the diagonal
coefficients (powers of $J_z$) to the scale of $J_z$, and the
off-diagonal coefficients to the scale of $J_+^2 + J_-^2$:
 \be \label{NormB}
 {\cal H}(\ell) = \sum_{\mbox{\scriptsize odd }i} \lambda_i(\ell) \frac{||J_z||}{||J_z^i||}  J_z^i +
 \sum_i
 \chi_i(\ell)\frac{||J_+^2 + J_-^2 ||}{||J_z^i J_+^2 + J_-^2 J_z^i||} (J_z^i J_+^2 + J_-^2
 J_z^i).
 \ee
The idea is to get a feeling for the path of the Hamiltonian
through operator space. The normalization employed above attempts
to eliminate artificial effects due to some of the operators in
the basis having larger matrix elements than the others.

\thispagestyle{headings}

\begin{figure}[p]
 \begin{minipage}[t]{0.9\textwidth}
 \begin{center}
 \includegraphics[width=0.9\textwidth]{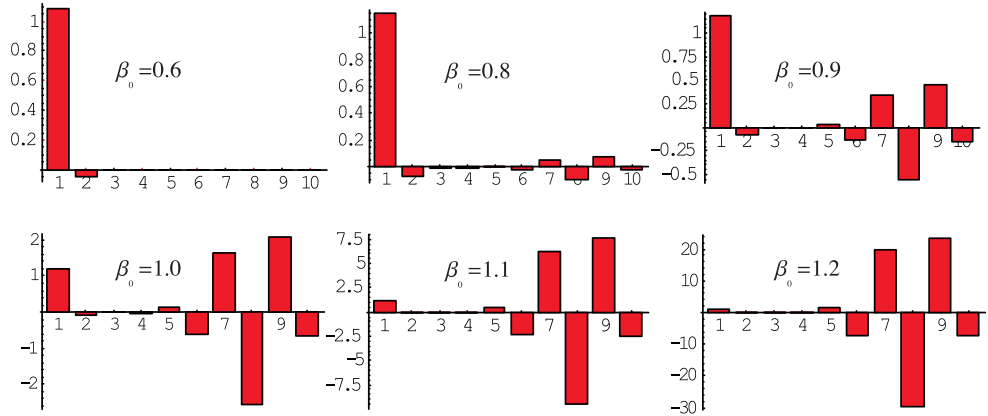}
 \\(a) Powers of $J_z$ in ${\cal H}(\infty)$\\ \mbox{}
 \end{center}
 \end{minipage}
 \begin{minipage}[t]{0.9\textwidth}
 \begin{center}
 \includegraphics[width=0.9\textwidth]{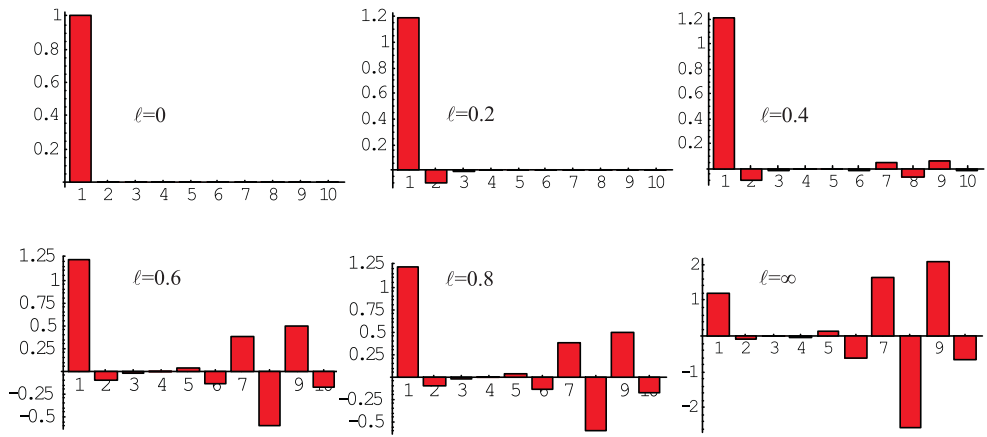}
 \\(b) Flow of different powers of $J_z$ for $\beta_0 = 1.0$ \\ \mbox{}
 \end{center}
 \end{minipage}
 \begin{minipage}[t]{0.9\textwidth}
 \begin{center}
 \includegraphics[width=0.9\textwidth]{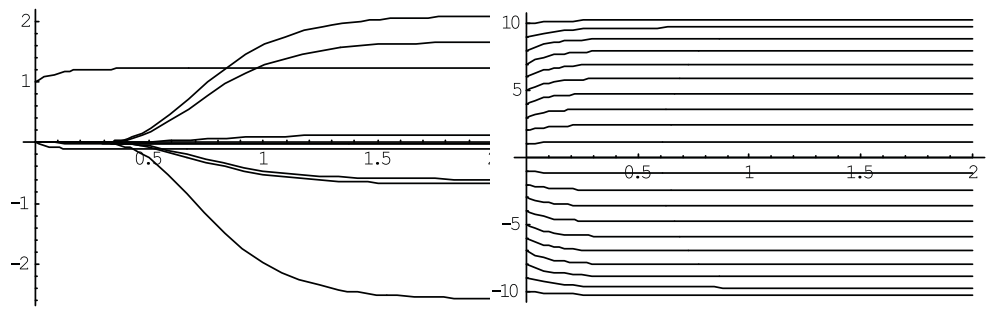}
 \\(c) Flow of powers of $J_z$ (left) vs flow of diagonal matrix elements (right) for $\beta_0=1.0$ \\ \mbox{}
 \end{center}
 \end{minipage}
 \caption[Diagonal Flow in powers of $J_z$ basis]{\label{barchartsdiag} The powers of $J_z$ present for various coupling strengths $\beta_0$ and stages $\ell$
 during the flow, for $N=20$ particles. The numbers $1\ldots10$ label the powers of $J_z$.
 Only odd powers need be considered.}
 \end{figure}

\begin{figure}[p]
 \begin{center}
 \includegraphics[width=0.9\textwidth]{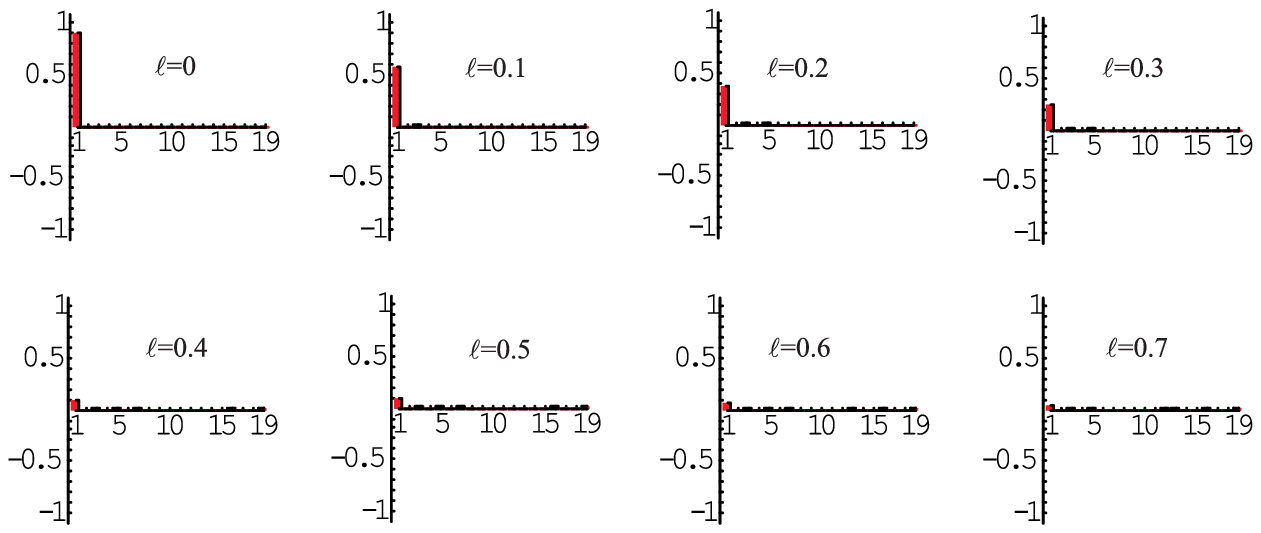}
 \\(a) $\beta_0=0.9$\\ \thispagestyle{empty} \mbox{}
 \includegraphics[width=0.9\textwidth]{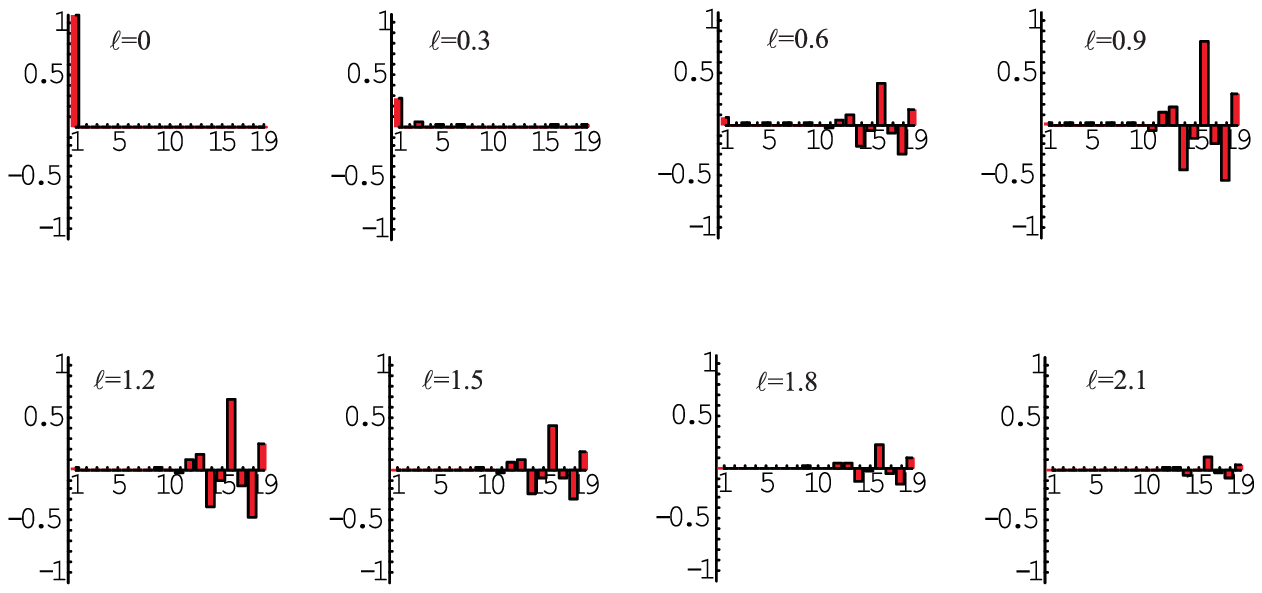}
 \\(b) $\beta_0=1.1$ \\ \mbox{}
 \includegraphics[width=0.7\textwidth]{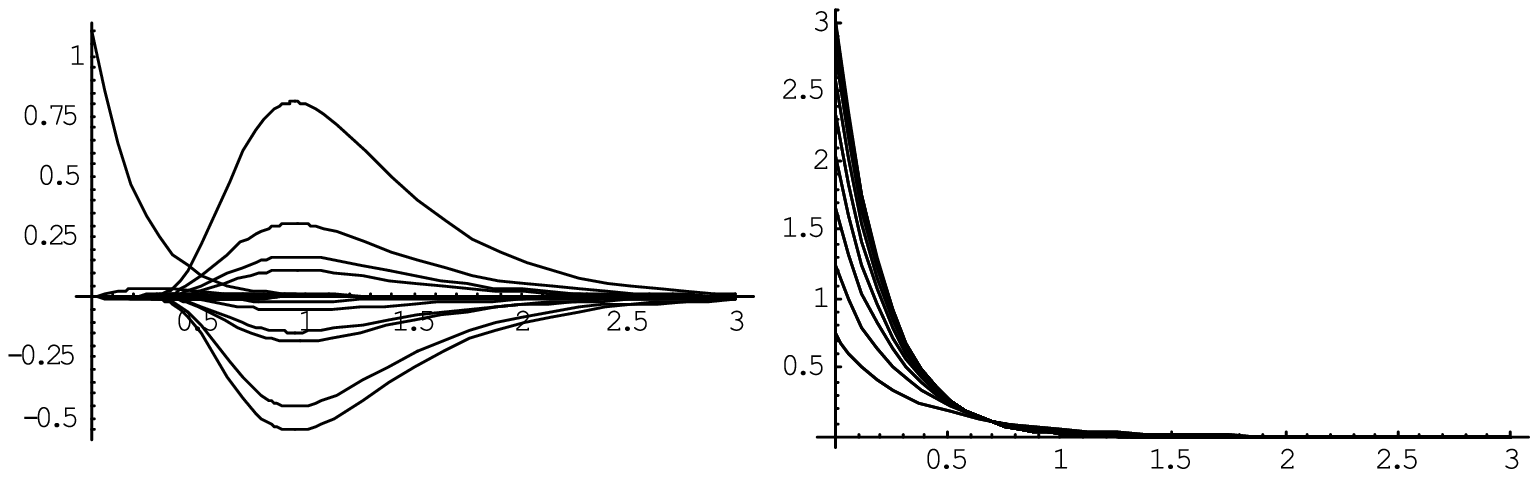}
 \\(c) Flow of ${\cal H}_{\mbox{\scriptsize off-diagonal}}$ in $J_z$ basis (left) and
 in terms of actual matrix elements (right) for $\beta_0=1.1$ \\ \mbox{}
 \caption[Off-diagonal flow in powers of $J_z$ basis]{\label{barchartsoffdiag} Flow of ${\cal H}_{\mbox{\scriptsize off-diagonal}}$ for (a) $\beta_0=0.9$ and (b)
 $\beta_0=1.1$. The numbers $1\ldots20$ label the powers of $J_z$
 in $(J_z^i J_+^2 + J_-^2 J_z^i)$. Figure (c) contrasts flow in
 two different bases.}
 \end{center}
 \end{figure}

The next step is to numerically integrate the flow equation
 \be
 \frac{d{\cal H}}{d\ell} = [[J_z, {\cal H}], {\cal H} ], \quad {\cal
 H}(0) = {\cal H}_0(\beta_0) = J_z + \frac{\beta_0}{4j} (J_+^2 +
 J_-^2).
 \ee
Since the matrices are $N+1$ dimensional, this may be viewed as a
set of $N+1+N-1=2N$ nonlinear coupled ordinary differential
equations in the diagonal matrix elements $\varepsilon_i$ and the
two-off-the-diagonal matrix elements $b_i$, of the form
 \begin{eqnarray}
  \dot{\varepsilon}_i &=& 4(b_{i-2}^2 - b_{i}^2) \label{ElEqs1}\\
  \dot{b}_i &=& 2b_i(\varepsilon_i - \varepsilon_{i+2}).
 \end{eqnarray}
After integrating these equations numerically, we perform a change
of variables to $\lambda_i$ and $\chi_i$, to make it clear which
operators are excited during the flow. The results are graphed in
Fig. \ref{barchartsdiag} and Fig. \ref{barchartsoffdiag}. In this
plot we chose $N=20$ particles, a number large enough to show the
general trend but small enough to discriminate meaningfully
between the operators.

Fig. \ref{barchartsdiag}(a) informs us that the powers of $J_z$
which constitute the final Hamiltonian $\cal H(\infty)$ are very
different in the two phases. In the first phase ($\beta_0<1$) the
final Hamiltonian is completely dominated by $J_z$, while in the
second phase ($\beta_0>1$) it is dominated by the high powers of
$J_z$, which occur with large coefficients. Fig.
\ref{barchartsdiag}(b) graphs the path of the Hamiltonian through
operator space for $\beta_0=1$, the phase change line. The initial
part of the flow is characterized by a gradual increase in $J_z$,
and it is only later that the flow direction changes, and the
higher powers of $J_z$ are excited. This is a good illustration of
the $\ell$-dependency of the generator, and the non-linear nature
of the flow equations.

Fig. \ref{barchartsdiag}(c) contrasts the flow in the powers of
$J_z$ basis ($\lambda_i$) with the flow of the actual matrix
elements ($\varepsilon_i$). The former shows the different regimes
of the flow referred to above. It is clear that the transformation
between the two is ``highly geared'' since a small change in the
matrix elements basis may manifest itself as a large change in the
powers of $J_z$ basis. Notice again how only the odd powers of
$J_z$ participate in the flow.

The same analysis is done for the flow of the off-diagonal
elements in Fig. \ref{barchartsoffdiag}. Since the final
Hamiltonian is always diagonal, we have concentrated here on the
flow of the off-diagonal elements for two values of the
interaction parameter on either side of the phase change boundary,
$\beta_0=1.0 \pm 0.1$. For $\beta_0=0.9$ the $J_+^2 + J_-^2$ term
is by far the dominant term in the flow and decreases to zero very
quickly, with the higher order terms $J_z^iJ_+^2 + J_-^2J_z^i$
hardly participating. For $\beta_0=1.1$ the first section of the
flow is again dominated by the decrease of $J_+^2 + J_-^2$, but
later on in the flow the higher order terms are highly excited,
until they eventually flow to zero too.

This process is made explicit in Fig. \ref{barchartsoffdiag}(c),
which, similarly to Fig. \ref{barchartsdiag}(c), contrasts the
off-diagonal flow in the $J_z^iJ_+^2 + J_-^2J_z^i$ basis
($\chi_i$) with the flow of the actual off-diagonal matrix
elements ($b_i$), for $\beta_0=1.1$. It is important to note the
distinction that, while the actual matrix elements must decrease
monotonically from the analysis presented in Section \ref{WegFE},
various operator terms may be excited during the flow.

After this preliminary numerical exercise, there should hopefully
be no confusion left in the readers minds as to the flow equations
program. It has also become clear that the problem will be far
more difficult to handle in the second phase, where the flow
displays its non-linear behaviour and enlists a range of higher
order operators as it evolves. With this in the back of our minds,
let us now review some recent treatments of the Lipkin model using
flow equations.

\section{Pirner and Friman's treatment} \label{PirnerT}

Pirner and Friman were the first to apply flow equations to the
Lipkin model \cite{PF}. Their method was to deal with unwanted
newly generated operators by linearizing them around their ground
state expectation value. There are two schemes, the second more
sophisticated than the first in that it includes a new operator
into the flow. The idea, as always, is not to try to solve the
Lipkin model (this can be done numerically) or even to try to
solve the flow equations in the Lipkin model exactly (this was
done in Section \ref{SPN}). Rather, one is more interested in
finding an effective Hamiltonian for the lower lying states, in
such a way that shows promise for application to other systems.
\thispagestyle{plain}

\subsection{First scheme}
The first step is to choose a parametrisation of the flow. For the
first scheme we will employ
 \begin{eqnarray} \label{6}
 {\cal H}(\ell) &=& \alpha(\ell)J_z + \beta(\ell)
 \frac{1}{4j}(J_+^2 + J_-^2) + \delta(\ell)j  \\
  \alpha(0)&=&1\\
  \beta(0)&=&\beta_0 \\
  \delta(0)& =&0,
 \end{eqnarray}
which simply makes the couplings in front of the original
Hamiltonian (\ref{Hoo1}) $\ell$-dependent. In addition a $\delta$
term proportional to the identity, normalized to the scale of
$\alpha$, has been included. In the exact case such a term is
never present in the flow since it would shift the centre of the
eigenspectrum away from zero. However, an approximation that will
be made later will generate such a term(see Eq. (\ref{8})), and it
is necessary if we want to compute the ground state energy. It was
not included in the original Pirner and Friman treatment where
they were only interested in the gap $\Delta$ between the ground
state and the first excited state. We have inserted it here for
completeness.

\subsubsection{Linearizing newly generated operators}
We now employ the generator choice (\ref{genc}) in the flow
equations. That is, we attempt to solve the differential equation
 \be \label{2}
 \frac{d{\cal H}(\ell)}{d\ell}=[[J_z,{\cal H}(\ell)],{\cal H}(\ell)], \;
 {\cal H}(0) = {\cal H}_0(\beta_0).
 \ee
The first commutator $\eta(\ell) = [J_z, {\cal H}(\ell)]$ gives
 \be \label{etaeval}
 \eta(\ell) = \frac{\beta(\ell)}{2} (J_+^2 - J_-^2).
 \ee
Inserting the Hamiltonian (\ref{6}) into the flow equation
(\ref{2}) yields
 \be \label{7}
 [[Jz,{\cal H}(\ell)], {\cal H}(\ell)] = \beta^2\frac{2j(j+1)-1}{4j^2}J_z -
 \beta^2\frac{1}{2j^2}J_z^3 - 4\alpha\beta\frac{1}{4j}(J_+^2 +
 J_-^2),
 \ee
in which a term $\sim J_z^3$ has been generated. This generation
of ever new operators during the flow is a generic feature of the
flow equations; indeed to fully capture the flow we must use many
more parameters in the Hamiltonian as in Eq. (\ref{Ba5}). Pirner
and Friman dealt with such operators by linearizing them around
the ground state expectation value, and neglecting higher-order
fluctuations:
 \be \label{8}
 J_z^3 \mapsto 3\langle J_z \rangle^2 J_z -2\langle J_z \rangle^3.
 \ee
In this way we aim to provide an effective theory for the
low-lying states, where the approximation (\ref{8}) is most valid.
If we are interested {\em only} in the ground state energy, then
there is a more accurate linearisation scheme (see Appendix
\ref{LinApp}). In the present case we are interested also in
properties like the band gap $\Delta$ between the first excited
state and the ground state, which makes the linearization
(\ref{8}) the best choice. Notice that the term $-2\langle J_z
\rangle^3$ on the RHS of (\ref{8}) has generated a term
proportional to the identity, as promised earlier. The presence of
this term attests to the fact that we are finding an effective
Hamiltonian for the lower lying states, and hence our window must
be `displaced' from zero in order to center on the ground state.

It is clear that some further approximation must be employed to
evaluate the expectation value in the linearization (\ref{8}),
since the ground state is unknown. In Pirner and Friman's
treatment, the expectation value of $J_z$ was evaluated with
respect to the zero interaction ground state, that is
$\left|\Psi\right> = \left|-j\right>$ which gives
 \be \label{Jzqj}
 \left<J_z\right>=-j.
 \ee
 Substituting the approximation (\ref{8}) into the double commutator (\ref{7})
 yields
 \begin{eqnarray}
 \dot{\alpha} & = & -\beta^2(\frac{6 \left< J_z \right>^2 -2j(j+1) + 1}{j^2})  \label{Before3}\\
 \dot{\beta}  & = & -4\alpha \beta \label{Before4} \\
 \dot{\delta} & = & 4\beta^2 \frac{ \left< J_z \right>^3}{j^3}.
 \label{Before5}
 \end{eqnarray}
The second equation implies that the magnitude of the off-diagonal
matrix element $\beta$ decreases in the course of the evolution,
providing $\alpha$ remains positive. These equations may easily be
combined to give two invariants of the evolution:
 \be \label {invars1}
 \delta = (\alpha - 1)\frac{4j^2}{4j^2-2j+1} \, , \qquad \alpha^2 - \beta^2
 \frac{4j^2-2j+1}{4j^2} = 1 - \beta_0^2 \frac{4j^2-2j+1}{4j^2}.
 \ee
The $j$ dependent terms approach unity for large $j$ (large $N$).
 \begin{figure}
 \begin{center}
 \includegraphics[width=0.6\textwidth]{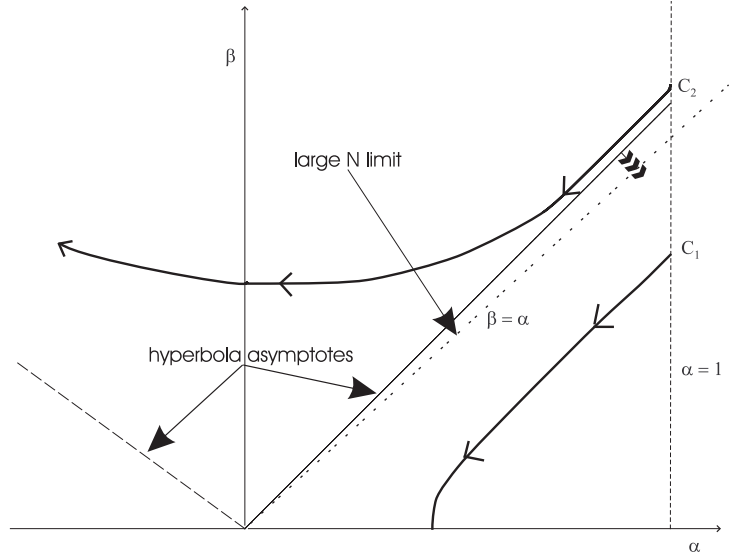}
 \end{center}
 \caption {\label{FFirstFlow}The flow of the Hamiltonian parameters in the $\alpha$-$\beta$ plane
 for the first scheme. }
 \end{figure}
Since $\delta$ is related to $\alpha$ in a simple way, it is
instructive to consider the projection of the flow on the
$\alpha$-$\beta$ plane. This is illustrated in Fig.
\ref{FFirstFlow}. The Hamiltonian begins at the point $(1,
\beta_0)$. Providing $\beta_0<\sqrt{\frac{4j^2}{4j^2-2j+1}} \sim
1$, as in $C_1$, it flows down the hyperbola (\ref{invars1}),
thereby reducing the off-diagonal $\beta$ term to zero and
asymptotically intercepting the $\alpha$ axis at $(\alpha(\infty),
0)$. If $\beta_0>\sqrt{\frac{4j^2}{4j^2-2j+1}} \sim 1$, as in
$C_2$, flow occurs along the other branch of the hyperbola and
diverges to infinity. This failure of the linearization
approximation (\ref{8}) and the expectation value approximation
(\ref{Jzqj}) will be discussed later. For the first phase the
final Hamiltonian is of the form
 \be
 {\cal H}(\infty) = \alpha(\infty) J_z + j\delta(\infty),
 \ee
which gives the expressions for the ground state energy $E_0$ and
the gap $\Delta$, for large $N$, as
 \begin{eqnarray}
 E_0 &=& \left<-j\right| {\cal H}(\infty) \left|-j\right> \\
     &=& \frac{-4j^3 + j(2j-1)\sqrt{1-\beta_0^2(1 - \frac{2j-1}{4j^2})}}{2j(2j-1)+1}
     \label{e1}\\
  \Delta &=& \left<-j+1\right| {\cal H}(\infty) \left|-j+1\right> -
  \left<-j\right| {\cal H}(\infty) \left|-j\right>  \\
  &=& \sqrt{1 - \beta_0^2 \frac{4j^2-2j+1}{4j^2}} \label{Delt1}\\
     &\approx& \sqrt{1-\beta_0^2}  \qquad \mbox{(large $j$)} ,\label{DeltaGap}
  \end{eqnarray}
where the gap $\Delta$ agrees with the exact result (\ref{ExactD})
in the large $N$ limit.

\subsection{Second scheme}
In the second scheme we attempt more accuracy by including a
$J_z^3$ term in our parametrization of the Hamiltonian:
 \begin{eqnarray} \label{SecSchemeH}
 {\cal H}(\ell) = \alpha(\ell)J_z + \gamma(\ell)\frac{1}{j^2}J_z^3 +
 \beta(\ell)\frac{1}{4j}(J_+^2 + J_-^2)  \\
 \alpha(0)=1, \quad \gamma(0)=0, \quad \beta(0)=\beta_0.
 \end{eqnarray}
The $\delta$ identity term has been dropped as our linearization
scheme will not generate such a term. Substitution into the double
bracket commutators of the flow equations will yield the same
generator (\ref{etaeval}), since $J_z^3$ commutes with $J_z$:
 \be \label{etaeval}
 \eta(\ell) = \frac{\beta(\ell)}{2} (J_+^2 - J_-^2).
 \ee
For the second commutator we will be faced with evaluating
 \be
 [J_+^2 - J_-^2, J_z^3].
 \ee
However, applying our rule $J_z^3 \rightarrow 3\left< J_z
\right>^2 J_z -2\left<J_z\right>^3$ does not commute with applying
the commutator:
\begin{eqnarray}
[\cdot,\cdot]\circ \mbox{linearize} &\rightarrow& -6\left<J_z \right>^2 (J_+^2 + J_-^2) \label{Lin1} \\
 {}\mbox{linearize} \circ [\cdot,\cdot] &\rightarrow& \left( 12(1-\left<J_z
\right>)J_z + 6 \left<J_z \right>^2 - 8 \right) J_+^2 + conj.
\label{Lin2}
\end{eqnarray}
The operators are understood to be applied from right to left, as
is conventional. We choose to linearize first, as Pirner and
Friman did. Linearizing second generates additional off-diagonal
terms. This point will be discussed later in Section
\ref{ImpSection}. As promised, in either case no term proportional
to the identity is generated. An interesting difference with the
first scheme is that in the second scheme the approximation $J_z^3
\rightarrow 3\left< J_z \right>^2 J_z -2\left<J_z\right>^3$ gets
tagged with the {\em off-diagonal elements} while in the first
scheme it gets tagged with the {\em diagonal} elements. The flow
equations are
\begin{eqnarray}
\dot{\alpha}&=&\beta^2\left(\frac{2j(j+1)-1}{j^2}\right) \label{Second1} \\
\dot{\beta} &=& -4 \left( \alpha \beta + \frac{3\beta\gamma \left<
J_z \right>^2}{j^2} \right)
 \label{Second2} \\
\dot{\gamma} &=& -2\beta^2 .\label{Second3}
\end{eqnarray}
Combining the first and third equations gives
 \be \label{gamminv}
 \gamma = \frac{2j^2}{2j(j+1)-1} (1-\alpha).
 \ee
Eliminating $\gamma$ gives rise to another more complicated
invariant
 \be \label{compinv}
 (4k-24)\alpha^2 + 48\alpha + k^2\beta^2 = 4k + 24 + k^2\beta_0^2,
 \ee
where
 \be
 k=\frac{2j(j+1)-1}{j^2} \rightarrow 2 \mbox{ (large j)}.
 \ee
The invariant (\ref{compinv}) may be interpreted as a shifted
hyperbola by setting $\bar{\alpha}=\alpha-\frac{24}{24-4k}$
 \be
 (24-4k)^2\bar{\alpha}^2 - (24-4k)k^2\beta^2 = 4k^3\beta_0^2
 -24k^2\beta_0^2+16k^2.
 \ee
The asymptotes of this hyperbola are given by the lines
 \be \label{asymptote2}
 \beta_{\mbox{\scriptsize asymptotes}}^{(2)} = \pm\left(
 -2\frac{6-k}{k^2}\alpha + \frac{12}{\sqrt{6-k}\,k^2}
 \right) \rightarrow \pm \left(-2\alpha + 3\right)\mbox{ (large
 j)}.
 \ee

\begin{figure}
 \begin{center}
 \includegraphics[width=0.5\textwidth]{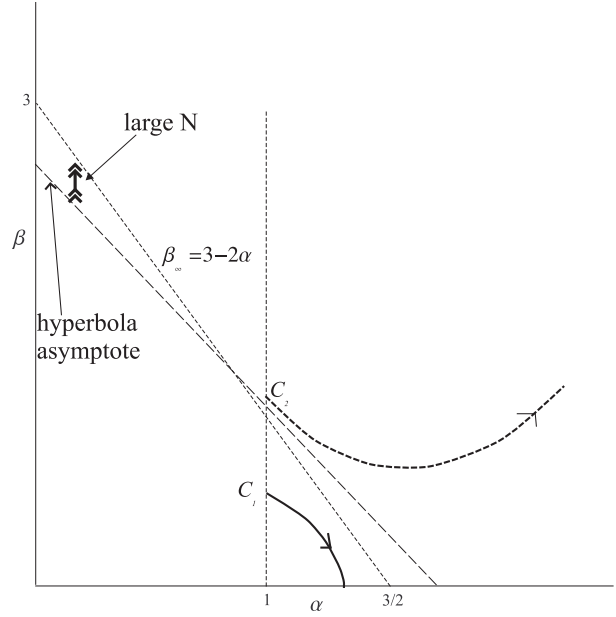}
 \end{center}
 \caption {\label {NewFlow2}The flow of the Hamiltonian parameters in the $\alpha$-$\beta$ plane
 for the second scheme. }
 \end{figure}

The analysis runs similarly as before, and the flow in the
$\alpha$-$\beta$ plane is illustrated in Fig. \ref{NewFlow2}. It
begins at $(1, \beta_0)$, but this time $\alpha$ is set to {\em
increase}. Providing
 \be
 \beta_0< \beta_{\mbox{\scriptsize asymptotes}}^{(2)}(1) = \frac{2j}{\sqrt{4j^2-2j+1}} \rightarrow 1 \mbox{ (large
 $j$)},
 \ee
as in ${\cal C}_1$, the Hamiltonian flows down the lower branch of
the hyperbola,
decreasing the off-diagonal term $\beta$, and intercepts the
$\alpha$ axis at
 \be \label{ainf2}
 \alpha(\infty) = \frac{6}{6-k}-\frac{k\sqrt{4 - (6 -
 k)\beta_0^2}}{2(6-k)}.
 \ee
If $\beta_0 > \frac{2j}{\sqrt{4j^2-2j+1}}$ as in ${\cal C}_2$, the
Hamiltonian flows along the upper hyperbola and diverges to
infinity.

In the first phase, the final Hamiltonian is of the form
 \be
 {\cal H}(\infty) = \alpha(\infty) J_z +
 \gamma(\infty)\frac{1}{j^2}J_z^3,
 \ee
which together with the $\alpha$-intercept (\ref{ainf2}) and the
$\gamma$ invariant (\ref{gamminv}) gives
 \begin{eqnarray}
 E_0 &=& \frac{-4j^3 + j(2j-1)\sqrt{1-\beta_0^2(1 - \frac{2j-1}{4j^2})}}{2j(2j-1)+1}
 \label{E02} \\
 \Delta &=& \left((-j+1)\alpha(\infty) +
  {\frac{(-j+1)^3}{j^2}}\gamma(\infty)\right) - \left(-j\alpha(\infty) +
  {\frac{-j^3}{j^2}}\gamma(\infty)\right) \\
  &=& \frac{6j-2+(4j(j-2)+3)\sqrt{1-\beta_0^2\left(1 -
  \frac{2j-1}{4j^2}\right)}}{2j(2j-1) + 1} \label{Delt2}\\
  &\approx&  \sqrt{1-\beta_0^2} \mbox{ (large $j$)}.
 \end{eqnarray}
The result for the ground state is the same as in the first scheme
(\ref{e1}). Due to the $J_z^3$ term, the new gap (\ref{Delt2})
differs from the previous result (\ref{Delt1}).

\subsection{Discussion}
\begin{figure}
 \begin{minipage}[t]{0.4\textwidth}
 \begin{center}
 \includegraphics[width=\textwidth]{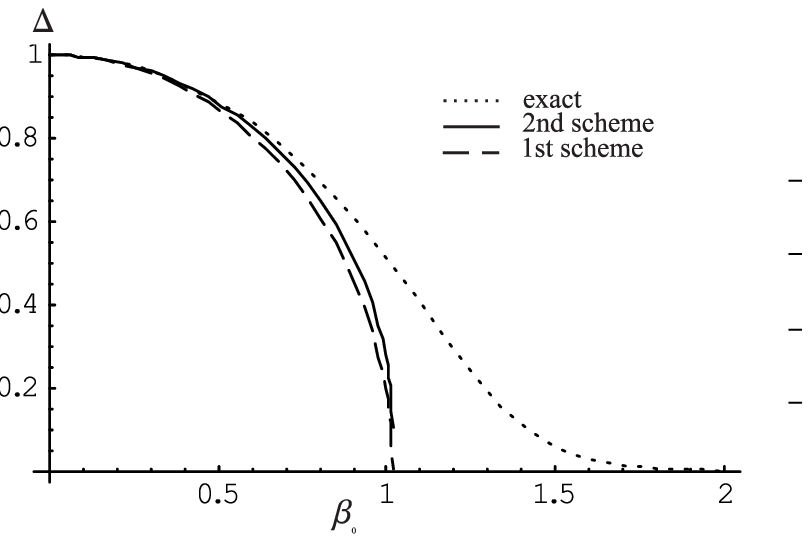}
 (a) Gap $\Delta$ vs $\beta_0$ for $N=30$ particles.
 \end{center}
 \end{minipage}
 \hfill
 \begin{minipage}[t]{0.4\textwidth}
 \begin{center}
 \includegraphics[width=\textwidth]{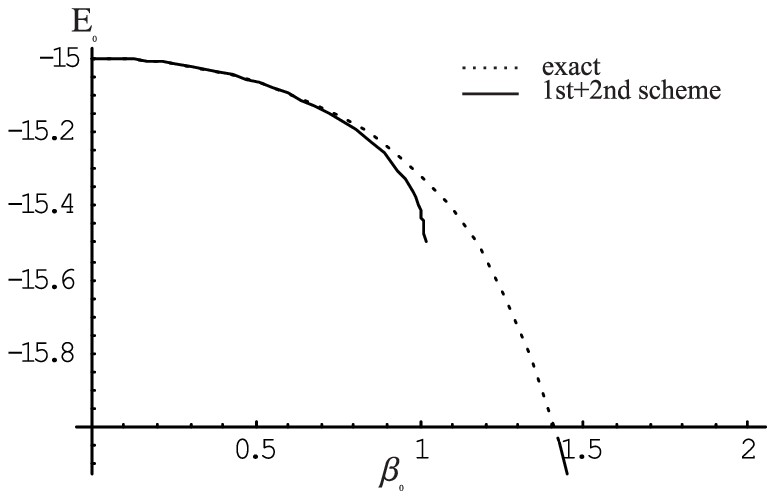}
 (b) Ground state $E_0$ vs $\beta_0$ for $N=30$ particles.
 \end{center}
 \end{minipage}
 \begin{minipage}[t]{0.4\textwidth}
 \begin{center}
 \includegraphics[width=\textwidth]{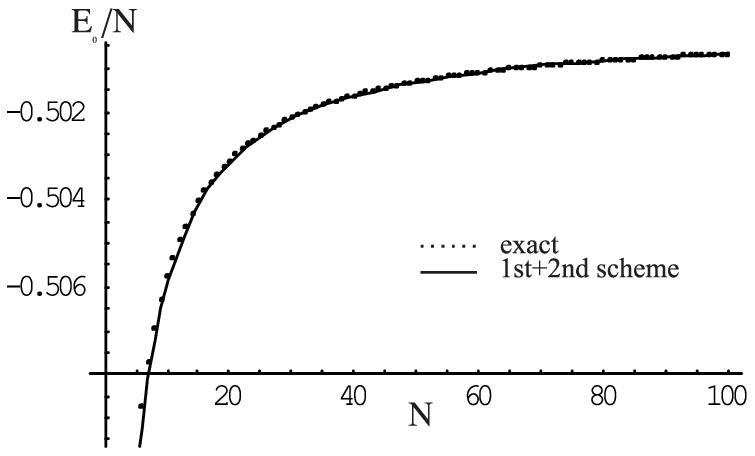}
 (c) Ground state energy per particle vs $N$. $\beta_0=0.5$.
 \end{center}
 \end{minipage}
 \hfill
 \begin{minipage}[t]{0.4\textwidth}
 \begin{center}
 \includegraphics[width=\textwidth]{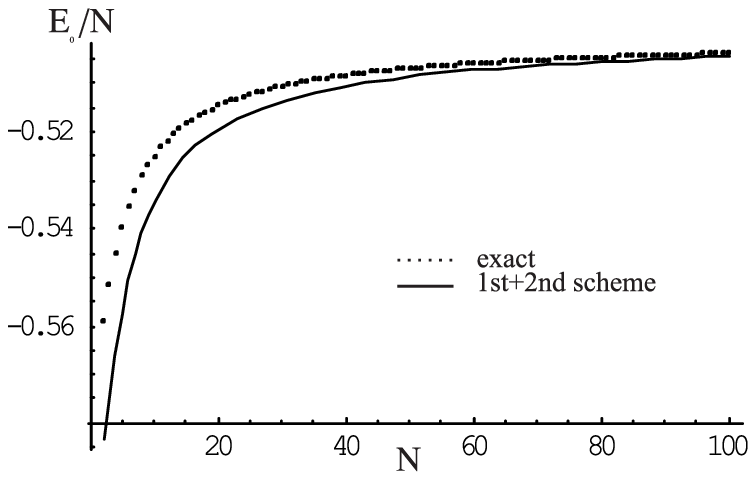}
 (d) Ground state energy per particle vs $N$. $\beta_0=1.0$.
 \end{center}
 \end{minipage}
 \begin{minipage}[t]{0.4\textwidth}
 \begin{center}
 \includegraphics[width=\textwidth]{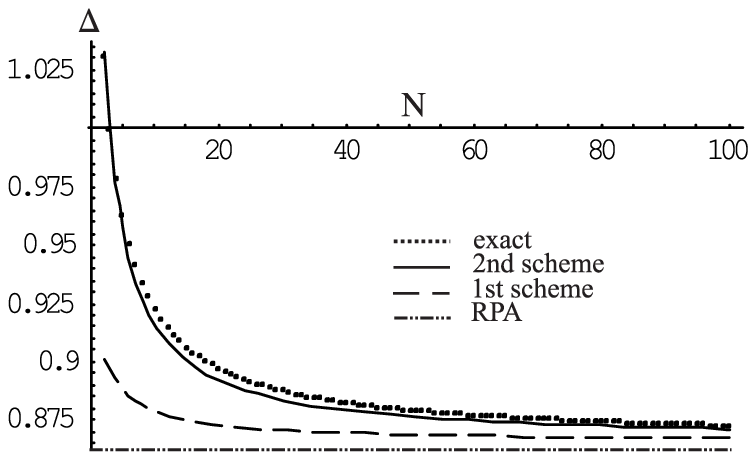}
 (e) Gap $\Delta$ vs $N$. $\beta_0=0.5$
 \end{center}
 \end{minipage}
 \hfill
 \begin{minipage}[t]{0.4\textwidth}
 \begin{center}
 \includegraphics[width=\textwidth]{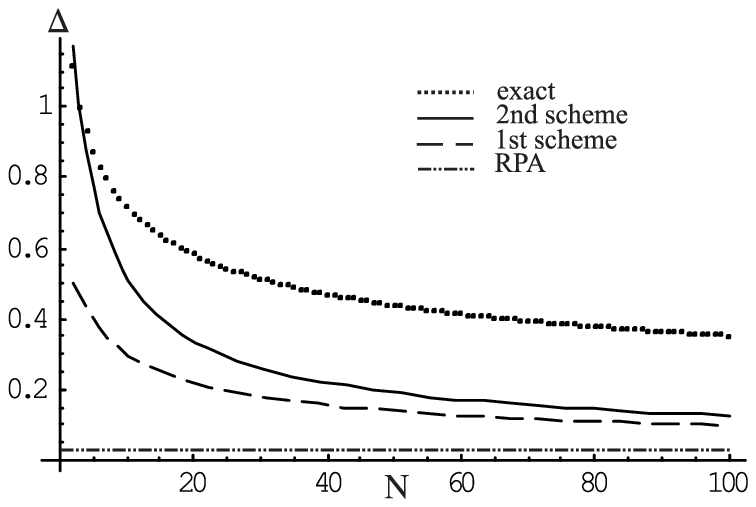}
 (f) Gap $\Delta$ vs $N$. $\beta_0=1.0$
 \end{center}
 \end{minipage}
 \begin{minipage}[t]{0.4\textwidth}
 \begin{center}
 \includegraphics[width=\textwidth]{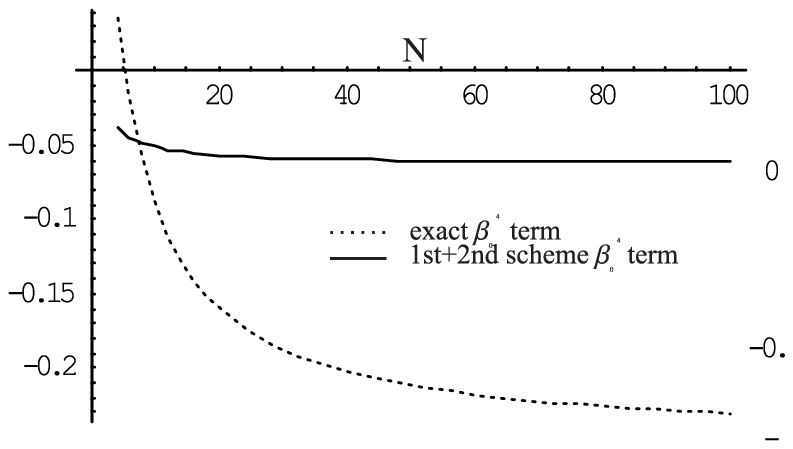}
 (g) Coefficient of $\beta_0^4$ for ground state $E_0$.
 \end{center}
 \end{minipage}
 \hfill
 \begin{minipage}[t]{0.4\textwidth}
 \begin{center}
 \includegraphics[width=\textwidth]{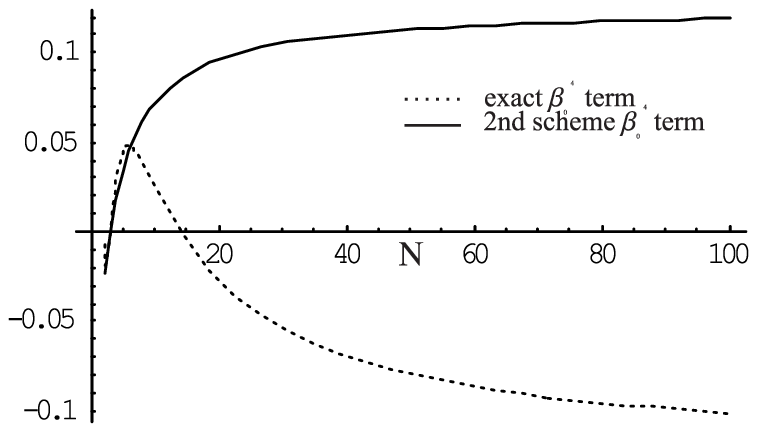}
 (h) Coefficient of $\beta_0^4$ for gap $\Delta$.
 \end{center}
 \end{minipage}
 \caption[Comparison of exact results and the two flow equations
 schemes.]{\label{ExacFS} Comparison of exact results and the two flow equations
 schemes. Both schemes give divergent results in the second phase
 $\beta_0 > 1$.}
 \end{figure}

At this stage it is instructive for purposes of comparison to
consider an exact power series expansion of the ground state
energy in the coupling $\beta_0$. Only even powers of $\beta_0$
will appear due to the symmetry of the Hamiltonian (as explained
underneath Eq. (\ref{Ba5}) . To fourth order, the result from
perturbation theory is \cite{Lipkin1}
 \be
 E_0^{\mbox{ex}}(\beta_0, j) = -j + \frac{1}{8}(\frac{1}{j} - 2)\beta_0^2 -
 \-\frac{(2j-1)(4j^2-14j+9)}{32j^3}\beta_0^4 + \cdots
 \ee
On the other hand, a series expansion of our result (both schemes
gave the same value for the ground state energy) gives:
 \be
 E_0(\beta_0, j) = -j + \frac{1}{8}(\frac{1}{j} - 2)\beta_0^2 +
 \frac{(2j-1)(4j^2-2j+1)}{128j^2}\beta_0^4 + \cdots \label{Expans}
 \ee
The flow equations result is correct up to order $\beta_0^2$. This
fact is not entirely trivial since it is not clear precisely how
many orders of $\beta_0$ are accounted for by the linearization
procedure (\ref{8}) and the expectation value approximation
(\ref{Jzqj}). The fourth order term as a function of $N=2j$ in the
exact case and in the flow equations result, for both the ground
state energy and the gap energy, are plotted in Fig.
\ref{ExacFS}(g) and Fig. \ref{ExacFS}(h). The flow equations
result is close to the exact value, but becomes less accurate with
increasing particle number.

The results for the ground state energy (\ref{E02}), (\ref{e1})
and the energy gap (\ref{Delt1}), (\ref{Delt2}) are plotted in
Fig. \ref{ExacFS}, where the exact results are also shown. Figs.
\ref{ExacFS}(a) and (b) plot the ground state energy, and the gap
against $\beta_0$, for $N=30$ particles. For small $\beta_0$ both
schemes deliver accurate results. As $\beta_0$ increases to unity,
the ground state energy starts to diverge. The gap energy from the
second scheme is more accurate than the result in the first
scheme.

This is made more apparent in Figs. \ref{ExacFS}(c)-(f) which plot
the behavior of the ground state energy per particle, and the gap,
as a function of $N$, for fixed $\beta_0$. For comparison the RPA
result \cite{Lipkin2} is also shown. We see that including a
$J_z^3$ term during the flow has considerably improved the
accuracy of the method. A perturbative treatment here would
require an expansion of the ground state energy as a function of
$1/N$, for fixed $\beta_0$. This is not possible using ordinary
perturbation theory since $1/N$ enters into the Hamiltonian
(\ref{Hoo1}) in both the coupling {\em and} implicitly in the
dimensions of the matrices involved. It is here that flow
equations have an advantage since they successfully interpolate
between perturbation theory for small $N$ and RPA for large $N$
\cite{PF}. Indeed, instead of expanding the flow equations result
in $\beta_0$ as in Eq. (\ref{Expans}), one may expand it in $1/j$.
This exercise gives
 \be \label{jExpansion}
 E_0(\beta_0, j) = -j -\frac{1}{2}(1 - \sqrt{1-\beta_0^2}) +
 \frac{\beta_0^2}{8\sqrt{1-\beta_0^2}}\frac{1}{j} + \cdots
 \ee
The failure of the linearisation scheme when $\beta_0 > 1$ is due
to the fluctuations in $J_z$ becoming stronger, so the expansion
of the new operators in powers of fluctuating operators is not so
good anymore. This point becomes clearer from examining the powers
of $J_z$ involved in the flow in the second phase, from Section
\ref{NumR}. Nevertheless, we have developed a simple way to modify
the linearisation scheme to be able to deal with this phase, which
will be presented in Section \ref{ImpSection} . For now we
continue our review of treatments of the Lipkin model using flow
equations.

\section{Mielke's matrix element treatment} \label{MielkeSec}
Dissatisfied with the fact that the original Wegner generator
$\eta = [\mbox{Diag}({\cal H}),{\cal H}]$ did not conserve the
band structure of a matrix, Mielke proposed in 1998 a new choice
of generator \cite{Mielke1},
 \be \label{MielkeGen}
 \eta_{ij} = \mbox{sign}(i-j){\cal H}_{ij},
 \ee
constructed to ensure that band diagonality is preserved. This
choice of generator cannot be written as a commutator of some
matrix with ${\cal H}$. The proof that it conserves band
diagonality follows from
 \be \label{BigSum}
 \frac{d{\cal H}_{ij}}{d \ell} = [\eta, {\cal H}]_{ij} =
 -\mbox{sign}(i-j)({\cal H}_{ii}-{\cal H}_{jj}){\cal H}_{ij} +
 \sum_{k \neq (n,m)} \left( \mbox{sign}(i-k) + \mbox{sign}(j-k)
 \right) {\cal H}_{ik} {\cal H}_{kj}.
 \ee
If the Hamiltonian is band diagonal (${\cal H}_{ij} = 0$ if
$|i-j|>M$) then the sum of the sign functions in the second term
will vanish for the matrix elements outside the band. Mielke was
obviously not aware that precisely this problem had been studied
before in the mathematical literature \cite{Drie2}, where it was
proved that writing the generator as the commutator of ${\cal H}$
with a diagonal matrix with a constant difference between its
diagonal elements would conserve band diagonality. In the Lipkin
model, $J_z$ is an operator with such a property. Indeed,
 \be
 \eta_{ij} = [J_z, {\cal H}]_{ij} = (i-j){\cal H}_{ij}
 \ee
clearly resembles Mielke's generator (\ref{MielkeGen}) and also
causes the second term in Eq. (\ref{BigSum}) to vanish. For a
matrix with a single band (i.e. only one band other than the
diagonal is non-zero), the two choices of generator will differ
only by a constant factor. Therefore, rather than discussing
Mielke's generator, we shall focus on his direct method of solving
the flow equations for the Lipkin model.

\begin{figure}
 \begin{minipage}[t]{0.3\textwidth}
 \begin{center}
 \includegraphics[width=\textwidth]{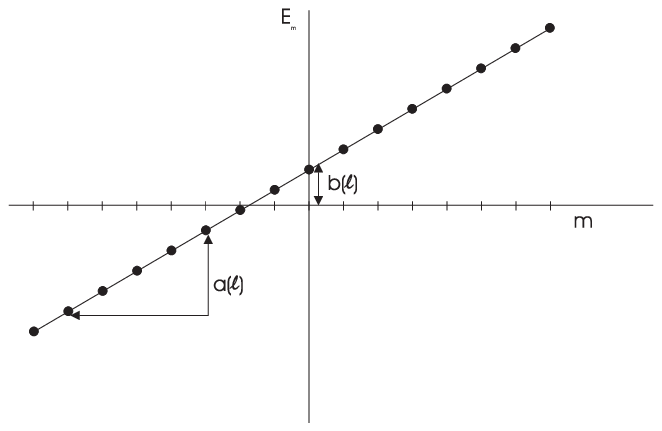}
 (a)
 \end{center}
 \end{minipage}
 \hfill
 \begin{minipage}[t]{0.3\textwidth}
 \begin{center}
 \includegraphics[width=\textwidth]{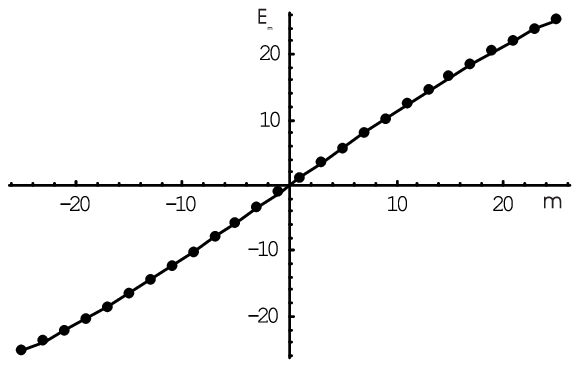}
 (b)
 \end{center}
 \end{minipage}
 \hfill
 \begin{minipage}[t]{0.3\textwidth}
 \begin{center}
 \includegraphics[width=\textwidth]{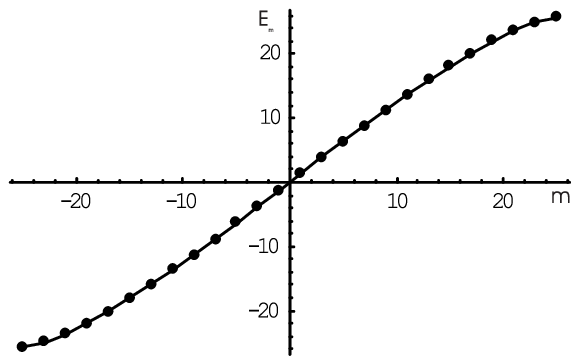}
 (c)
 \end{center}
 \end{minipage}
 \caption[Schematic illustration of ansatz (\ref{linans})]{\label{Lin}(a) Schematic illustration of ansatz (\ref{linans}). (b) Exact spectrum for $\beta_0 = 0.8$.
 (c) Exact spectrum for $\beta_0 = 1.2$. $N=50$ particles. Note how boundary effects enter for larger $\beta_0$.}
 \end{figure}

We want to consider the flow equations directly from the matrix
elements. The Hamiltonian is
 \be
 {\cal H}(\beta _0) = J_z + \frac{\beta_0}{4j}(J_+^2 + J_-^2).
 \ee
Since the interaction only connects $\left|m\right>$ with $\left|m
\pm 2 \right>$, we now rearrange the basis $\{ \left|-j \right>,
\left|-j+1\right>, \cdots, \left|+j\right>\}$ into odd and even
$m$. For example, for even $N$ we form
 \be
 \{ \left|-j \right>,
 \left|-j+2 \right>, \cdots,
 \left|+j\right>\}\cup\{\left|-j+1\right>, \left|-j+3\right>,
 \cdots, \left|j-1\right>\}.
 \ee
In this way $\cal H$ splits into two tridiagonal (${\cal H}_{ij} =
0$ if $|i-j|>1$) submatrices, the dimension of which depends on
$N$. If $N$ is even, one of the matrices has dimension $N/2+1$,
and the other $N/2$. If $N$ is odd, both matrices have dimension
$\frac{N+1}{2}$. We write the matrix elements as
 \be
 \varepsilon_n = {\cal H}_{nn}, \quad \theta_n = {\cal H}_{nn+1},
 \quad n=0\ldots \mbox{Dim}({\cal H}).
 \ee
The initial values are easily computed to be
\begin{eqnarray}
\varepsilon_n(0) &=& -j+2n \\
\theta_n(0) &=&
\frac{\beta_0}{4j}\sqrt{j(j+1)-(j-2n)(j-2n-1)}\sqrt{j(j+1)-(j-2n-1)(j-2n-2)}
\label{offs1},
\end{eqnarray}
or
\begin{eqnarray}
\varepsilon_n(0) &=& -j+2n+1 \\
\theta_n(0) &=&
\frac{\beta_0}{4j}\sqrt{j(j+1)-(j-2n-1)(j-2n-2)}\sqrt{j(j+1)-(j-2n-2)(j-2n-3)}
\label{offs2}.
\end{eqnarray}
Using our generator as $\eta=[J_z, {\cal H}]$, where ${\cal H}$ is
expressed in the rearranged basis, the flow equations are in both
cases
 \begin{eqnarray}
 \frac{d \varepsilon_n}{d\ell} &=& -2(\theta_n^2 - \theta_{n-1}^2)
 \label{MQ1}
 \\
 \frac{d \theta_n}{d\ell} &=&
 -\theta_n(\varepsilon_{n+1}-\varepsilon_n) \label{MQ2},
 \end{eqnarray}
which are a simple modification of the direct matrix element
equations for the original combined matrix (\ref{ElEqs1}). The
problem is now to solve these equations. As Mielke pointed out, a
first possibility is to solve them iteratively, by starting with
the ansatz $\varepsilon_n^{(0)}(\ell) = \varepsilon(0)$ and
$\theta_n^{(0)}(\ell) = \theta_n(0)e^{-2\ell}$. These expressions
could then be inserted onto the right hand side of Eqs.
(\ref{MQ1}) and (\ref{MQ2}), which would yield a first iterative
solution, which could again be inserted onto the right hand side
and so on. This procedure rapidly becomes more complex. It follows
the philosophy of perturbation theory, and works well for small
$\beta_0$ and small $N$. A non-perturbative solution can be
obtained in the limit of large $N$, which is the regime we are
interested in. The first step is noticing  that the off-diagonal
matrix elements initially satisfy
 \be
 \theta_n(0)^2 - \theta_{n-1}(0)^2 = 2\beta_0^2(n+\frac{1}{4})(1 +
 \mbox{O}(1/j))
 \ee
in the first case \ref{offs1}, or
 \be
 \theta_n(0)^2 - \theta_{n-1}(0)^2 = 2\beta_0^2(n+\frac{3}{4})(1 +
 \mbox{O}(1/j))
 \ee
in the second case \ref{offs2}. The reason this is important is
because this expression is used in the right hand side of the flow
equations (\ref{MQ1}). The idea is to use the large $N$ limit to
reduce our set of dynamical variables:
 \be \label{reduction}
 \underbrace{\{ \varepsilon_n(\ell), \theta_n(\ell)\}}_{\mbox{\scriptsize $N/2
 +1+ N/2-1 = N$ variables}} \rightarrow \qquad \underbrace{\{a(\ell), b(\ell),
 f(\ell)\}}_{\mbox{\scriptsize $3$ variables}}.
 \ee
This is achieved by making the ansatz:
 \begin{eqnarray}
 \varepsilon_n(\ell) &=& na(\ell) + b_{1,2}(\ell) \label{linans} \\
 \theta_n(\ell) &=& f(\ell)\theta_n(0),
 \end{eqnarray}
where the subscript on $b_{1,2}$ refers to the submatrix being
diagonalized. This ansatz attempts to track the flow of the
diagonal part of $\cal H$ in a linear fashion. This is illustrated
in Fig. \ref{Lin}, which also shows for comparison the exact
numerical calculation of the form of the eigenvalues, for the odd
$m$ submatrix, with $N=50$ particles and two values of the
coupling on either side of the phase boundary, $\beta_0=0.8$ and
$\beta_0=1.2$. This figure shows that the linear ansatz
(\ref{linans}) should work well in the first phase, but poorly in
the second phase where there are boundary effects\footnote{Indeed,
this is precisely the reason why higher powers of $J_z$ are
necessary in the second phase: The Taylor series expansion of Fig.
\ref{Lin} involves higher order terms which are not necessary in
the first phase.}.

The ansatz (\ref{linans}), in the large $N$ limit, leads to the
following differential equations for $a$, $b_{1,2}$ and $f$:
 \begin{eqnarray}
 \frac{df}{d\ell} &=& -af \label{Before1}\\
 \frac{da}{d\ell} &=& -4\beta_0^2 f^2 \label{Before2} \\
 \frac{db_1}{d\ell} &=& \frac{1}{4}\frac{da}{d\ell} \mbox{(using
 \ref{offs1}),} \qquad  \frac{db_2}{d\ell} = \frac{3}{4}\frac{da}{d\ell}
 \mbox{(using \ref{offs2})}  \label{bs} \\
 f(0) &=& 1, \quad a(0)=2,\quad b_1(0)=-j, \quad b_2(0)=-j+1
 \label{initcond}.
 \end{eqnarray}
Since the first two equations leave $a^2 - 4\beta_0^2$ invariant,
and since $f\rightarrow 0$ as $\ell \rightarrow \infty$, this
yields directly
 \be
 a(\infty) = 2\sqrt{1-\beta_0^2},
 \ee
which, for one thing, shows that the reduction of variables for
large $N$,  using (\ref{reduction}), is only valid in the first
phase, $\beta_0 < 1$. This result also yields immediately, from
the initial conditions (\ref{initcond}) and the equations
(\ref{bs})
 \begin{eqnarray}
 b_{1}(\infty) &=& -j - \frac{1}{2} + \frac{1}{2}\sqrt{
 1-\beta_0^2} \\
 b_{2}(\infty) &=& -j - \frac{1}{2} + \frac{3}{2}\sqrt{
 1-\beta_0^2}.
 \end{eqnarray}
In this way the spectrum for large $N$ is given as
 \be
 \varepsilon_{n 1,2} = 2\sqrt{1-\beta_0^2}(n+\frac{1}{2} \pm
 \frac{1}{4})- j - \frac{1}{2},
 \ee
which yields the ground state as
 \be \label{Ground11}
 E_0 = \varepsilon_{0,2} = -j - \frac{1}{2}(1 -
 \sqrt{1-\beta_0^2}),
 \ee
and the gap
 \be \label{Gap11}
 \Delta = \sqrt{1-\beta_0^2}.
 \ee
This expression for the gap corresponds to Eq. (\ref{ExactD}),
obtained from a Bogolubov transformation. The expression for the
ground state verifies the power series obtained from the flow
equations previously in Eq. (\ref{jExpansion}). Mielke's direct
approach to the flow equations via the matrix elements themselves
has readily yielded the correct results for large $N$ in the first
phase. Extending the approach to the next order in $1/N$ would
require a polynomial ansatz for $\varepsilon_n$ as opposed to the
linear one in Eq. (\ref{linans}), and extending the expression for
$\theta_n^2(0) - \theta_{n-1}^2(0)$ to the corresponding power of
$1/N$. The complexity of this approach rapidly increases and,
although systematic, it cannot be viewed as a miracle method for
the $1/N$ expansion of the properties of the model.

\section{Stein's bosonization method}
Stein \cite{Stein} has recently employed the flow equations in the
Lipkin model by considering the flow in the Holstein-Primakoff
boson representation of the angular momentum operators. In this
way it is possible to systematically solve the flow to any order
in $1/N$. In this section we review his procedure, presenting it
in a unified and improved manner.

Since we will be studying properties such as the ground state in
the large $N$ limit, we firstly rescale our Hamiltonian to ensure
that the ground state energy is of the order of unity : ${\cal H}'
= \frac{1}{j}{\cal H}$. This means that
 \be
 \frac{d{\cal H}'}{d\ell} = \frac{1}{j} \frac{d{\cal H}}{d\ell} =
 \frac{1}{j} [[J_z, {\cal H}], {\cal H}] = [[jJ_z, {\cal H}'],
 {\cal H}'],
 \ee
from which it is clear that if we wish to maintain the traditional
double bracket form of the flow equations, we must rescale $J_z$
by a factor of $j$. Dropping primes in our subsequent work, our by
now familiar Hamiltonian reads
 \be
 {\cal H}_0 = \frac{1}{j}J_z + \frac{\beta_0}{4j^2}(J_+^2 +
 J_-^2).
 \ee
We now employ the Holstein-Primakoff realization of angular
momentum(as in Eq. (\ref{100}))
 \be \label{New100}
 (J_z)_B = -j+a^\dagger a, \quad (J_+)_B = (J_-)^\dagger_B
 = \sqrt{2j}a^\dagger\sqrt{1-\frac{1}{2j}a^\dagger a},
 \ee
 where $b^{(\dagger)}$ are bosonic annihilation (creation)
 operators. Expanding the square root to order $\frac{1}{j^2}$ gives
  \be \label{InitHamm}
  {\cal H}_0 = -1 + \frac{1}{j}\left(a^{\dagger}a +
  \frac{\beta_0}{2}(1-\frac{1}{4j})(a^{\dagger}a^{\dagger} + aa) \right) -
  \frac{1}{j^2} \frac{\beta_0}{4}\left((a^{\dagger}a^{\dagger}a^{\dagger}a +
  a^{\dagger} aaa )\right) - \cdots
  \ee
where we have announced our intention to normal order all terms,
and also to list each operator in the series only once - at that
order of $1/j$ where it first appeared. Subsequent generations of
the same operator from normal ordering or the flow will be grouped
together with the initial one. As the Hamiltonian evolves, we will
track it in the following form
 \be
 {\cal H}(\ell) = -1 + \sum_{k=1}^{\infty} \frac{:{\cal
 H}^D_k(\ell) + {\cal H}^{\pm2}_k(\ell):}{j^k}.
 \ee
Since we choose $\eta=[jJ_z, {\cal H}]$ as the generator, the
Hamiltonian will remain tridiagonal, or in bosonic language, only
contain functions of the number operator (${\cal H}_k^D$) or
operators that change the occupation number by two (${\cal
H}_k^{\pm2}$). This point has been confused in Ref. \cite{Stein}
where extra terms were added to the generator for each order of
$1/j$ in order to conserve tridiagonality. The choice $\eta=[jJ_z,
{\cal H}]$ produces these extra terms automatically. To be
explicit, the flow of ${\cal H}$ up to order $\frac{1}{j^2}$ will
be parameterized as
 \be \label{Parameter}
 {\cal H}(\ell) = -1 + \frac{1}{j}\left(E(\ell) + f(\ell) a^{\dagger}a
 + \frac{\beta_0}{2}g(\ell)(a^{\dagger}a^{\dagger} + aa)  \right)
 + \frac{\beta_0}{j^2}\left( h(\ell) a^{\dagger}a^{\dagger}aa -
 \frac{1}{4}\zeta(\ell)(a^{\dagger}a^{\dagger}a^{\dagger}a +
 a^{\dagger}aaa) \right).
 \ee
Due to our policy of listing each operator only once, the flow
coefficients must also be viewed as being $j$ dependent, and hence
can be arranged as a power series in $1/j$.

We first compute everything up to order $1/j$. In this case the
initial conditions are, from (\ref{InitHamm}),
 \be
 f(0) = 1, \quad g(0) = 1, \quad E(0) = 0.
 \ee
The generator is
 \be
 \eta = [ja^{\dagger}a, {\cal H}] =
 \beta_0g(a^{\dagger}a^{\dagger}+ aa).
 \ee
The computation of the flow equations commutator yields, to order
$1/j$,
 \be \label{Computation}
 [\eta, {\cal H}] = \frac{1}{j}\left(-2\beta_0^2g^2 -
 4\beta_0^2g^2a^{\dagger}a -
 \frac{\beta_0}{2}4\beta_0fg(a^{\dagger}a^{\dagger}+ aa) \right).
 \ee
The identity term arises from the normal ordering of
$[a^{\dagger}a^{\dagger}-aa, a^{\dagger}a^{\dagger}+aa]$, and is
not generated in higher orders. Comparison with the Hamiltonian
(\ref{Parameter}) gives
 \begin{eqnarray}
 \frac{df}{d\ell} &=& -4\beta_0^2g^2 \\
 \frac{dg}{d\ell} &=& -4fg \\
 \frac{dE}{d\ell} &=& -2\beta_0^2g^2.
 \end{eqnarray}
We have seen similar equations before (Eqs. (\ref{Before3}),
(\ref{Before4}), (\ref{Before1}) and (\ref{Before2})). The
familiar asymptotic behavior is, providing $\beta_0 < 1$,
 \be
 f(\infty) = \sqrt{1-\beta_0^2}, \quad g(\infty) = 0, \quad
 E(\infty) = -\frac{1}{2j}(1 - \sqrt{1-\beta_0^2}).
 \ee
The final Hamiltonian takes the form
 \be
 {\cal H}(\infty) = -1 + \frac{1}{j}(E(\infty) +
 f(\infty)a^{\dagger}a),
 \ee
and since the lowest two states are $\left|0\right>$ and
$\left|1\right>$, we recover our previous expressions(Eqs.
(\ref{Ground11}) and (\ref{Gap11})) for the ground state energy
$E_0$ and the gap $\Delta$ to order $1/j$.

We now work to order $1/j^2$. In this case the relevant initial
conditions are
 \be
 f(0) = 1, \quad g(0) = 1-\frac{1}{4j}, \quad E(0) = 0, \quad
 h(0)=0, \quad \zeta(0)=1.
 \ee
We only need to evaluate the generator up to order $1/j$ since the
second commutator introduces another $1/j$:
 \be
 \eta = \beta_0g(a^{\dagger}a^{\dagger}-aa) -
 \frac{\beta_0\zeta}{2j}(a^{\dagger}a^{\dagger}a^{\dagger}a -
 a^{\dagger}aaa).
 \ee
Evaluation of the second commutator yields
 \begin{eqnarray} \label{BigComm}
 [\eta_, {\cal H}] = \frac{1}{j}\left(-2\beta_0^2g^2 + (-
 4\beta_0^2g^2 + \frac{6\beta_0^2g\zeta}{j})a^{\dagger}a
 -\frac{\beta_0}{2}(4fg +
 \frac{4\beta_0gh}{j})(a^{\dagger}a^{\dagger}+ aa) \right) \\-
 \frac{\beta_0}{j^2}\left(6\beta_0g\zeta(a^{\dagger}a^{\dagger}aa)
 - (4f\zeta - 16\beta_-gh)\frac{1}{f}(a^{\dagger}a^{\dagger}a^{\dagger}a +
 a^{\dagger}aaa)\right).
 \end{eqnarray}
In evaluating this commutator, interactions of the form
$a^{\dagger}a^{\dagger}a^{\dagger}a^{\dagger} + aaaa$ were
generated. These terms were cancelled by similar terms arising
from extending the generator to order $1/j$. This important
observation will be elaborated on later. For now we content
ourselves with writing down the differential equations, using
(\ref{BigComm}) and the expression for the Hamiltonian
(\ref{Parameter}),
 \begin{eqnarray}
 \frac{df}{d\ell} = -4\beta_0^2g^2 + \frac{1}{j}6\beta_0^2g\zeta
 &\quad& \frac{dg}{d\ell} = -4fg -\frac{1}{j}4\beta_0gh \label{EEqn2}    \\
 \frac{dE}{d\ell} = -2\beta_0^2g^2 &\quad& \frac{dh}{d\ell} =
 6\beta_0g\zeta \label{EEqn}\\
 \frac{d\zeta}{d\ell} = -4f\zeta + 16\beta_0gh.
 \end{eqnarray}
These equations may be solved explicitly for all $\ell$ by finding
an integral basis and subsequent lengthy algebra \cite{Stein}. We
are only interested in the asymptotic values. As expected, the
off-diagonal terms go to zero while the diagonal terms obey
 \begin{eqnarray}
 f(\infty) &=& \sqrt{1-\beta_0^2} +
 \frac{1}{j}\frac{\beta_0^2(3\sqrt{1-\beta_0^2})}{2(1-\beta_0^2)}
 \label{finf}
 \\
 h(\infty) &=& \frac{3\beta_0}{4(1-\beta_0^2)} \label{hinf}.
 \end{eqnarray}
From glancing at the differential equations (\ref{EEqn}) and
(\ref{EEqn2}) one sees that $-2E+f-\frac{1}{j}\beta_0h=1$ is an
invariant of the flow. Using the asymptotic forms of $f(\ell)$
(\ref{finf}) and $h(\ell)$ (\ref{hinf}), and remembering that the
final Hamiltonian takes the form
 \be
 {\cal H}(\infty) = -1 + \frac{1}{j}\left(E(\infty) +
 f(\infty)a^{\dagger}a\right) +
 \frac{\beta_0}{j^2}h(\infty)a^{\dagger}a^{\dagger}aa,
 \ee
gives the ground state energy and gap, up to order $1/j$ (not
$1/j^2$ since we express our final results in the original
unprimed Hamiltonian ${\cal H} = j {\cal H}'$), as
\begin{eqnarray}
E_0 &=& -j -\frac{1}{2}(1-\sqrt{1-\beta_0^2}) +
\frac{1}{8j}\frac{\beta_0^2(3-2\sqrt{1-\beta_0^2})}{1-\beta_0^2}
\\
\Delta &=& \sqrt{1-\beta_0^2} +
\frac{1}{2j}\frac{\beta_0^2(3-\sqrt{1-\beta_0^2})}{1-\beta_0^2}.
\end{eqnarray}
Expressing the Hamiltonian in the Holstein-Primakoff
representation has allowed for a systematic solution of the flow
equations in $1/N$. One may ask if the same result could have been
obtained with the Dyson mapping
 \be
 (J_z)_B = -j+a^\dagger a, \quad (J_+)_B =
 a^{\dagger}(2j-a^{\dagger}a), \quad (J_-)_B = a,
 \ee
which would, for example, deliver as initial Hamiltonian
 \be
 {\cal H}(0) = -1 + \beta_0(1-\frac{1}{2j})a^{\dagger}a^{\dagger}
 + \frac{1}{j}\left(a^{\dagger}a +
 \beta_0(\frac{1}{2j}-1)a^{\dagger}a^{\dagger}a^{\dagger}a\right)
 + \frac{1}{j^2}\frac{\beta_0}{4}(aa +
 a^{\dagger}a^{\dagger}a^{\dagger}a^{\dagger}aa).
 \ee
 Of course, any method when carefully and correctly executed will give
the same results. The question here is ease of computability.
There are two problems with the Dyson mapping. The first is that,
although the initial Hamiltonian is given by a finite expression,
the orders in $1/j$ are misleading. In other words, when
calculating $E_0$ to a desired order in $1/j$, one will have to
take into account higher order terms. This is due to the second
problem, which is that the flow equations are not closed order for
order in $1/j$. The remarkable property of the Holstein-Primakoff
mapping, as commented on later(see (\ref{BigComm})), is that one
in fact obtains a closed set of equations for each order in $1/j$
since newly generated terms cancel out.

The same problem appears if one attempts to use the Schwinger
mapping, which uses two boson types $a$ and $b$:
 \be
 (J_z)_B = \frac{1}{2}(a^\dagger a-b^\dagger b), \quad (J_+)_B =
 a^{\dagger}b, \quad (J_-)_B = b^\dagger a.
 \ee
The initial Hamiltonian would be
 \be
 {\cal H}(0) = \frac{1}{j}\left( \frac{1}{2}(a^\dagger a -
 b^\dagger b) + \frac{1}{4j}(a^\dagger a^\dagger bb - b^\dagger
 b^\dagger aa)\right).
 \ee
After substituting this into the flow equations one generates
boson interaction terms of the form $a^\dagger a^\dagger aa +
b^\dagger b^\dagger b b$. While in the Holstein-Primakoff picture
these terms come with their $1/j$ dependence ``built-in'', it is
far from clear in the Schwinger mapping, since the equations are
not closed.

But why can't one apply the order for order method with the
Hamiltonian expressed in the customary angular momentum form? Here
it is even more difficult, because as in the Dyson mapping, the
equations are not closed order for order in $1/j$. The other
problem is that the commutators themselves can produce new orders
of $j$, eg.
 \be
 [J_+^2, J_-^2] = \left(8j(j+1)-4\right)J_z - 8J_z^3,
 \ee
which result in rational $j$ dependent fractions as in the flow
equations (\ref{Before3}). This makes it difficult, but of course
not impossible, to systematically solve the flow equations order
for order. Having reviewed three recent approaches to the Lipkin
model via flow equations, we now present some extensions of our
own.

\section{Tracking the ground state} \label{ImpSection}
The original method of Pirner and Friman was dealt with in Section
\ref{PirnerT}. The basic idea was to close the flow equations by
linearizing newly generated operators around the ground state
expectation value. Specifically, the operator $J_z^3$ was replaced
by
 \be \label{JzCubedMaps}
 J_z^3 \mapsto 3\langle J_z \rangle^2 Jz -2\langle J_z \rangle^3.
 \ee
The expectation value was taken with respect to the
zero-interaction ground state, $\left| \Psi \right> =
\left|-j\right>$, which more importantly is the ground state of
${\cal H}(\infty)$. Since the Hamiltonian is undergoing a
continuous unitary transformation, the ground state also undergoes
a continuous unitary transformation. The flow equations must be
viewed as transforming the Hamiltonian while keeping the basis
vectors constant. In this way the true ground state of the system
evolves from $\left| \Psi_0 \right>$ at $\ell=0$ to
$\left|-j\right>$ at $\ell=\infty$. To be strictly correct, the
expectation value of $J_z$ should be taken with respect to the
{\em dynamical ground state}.

This observation leads one to consider a simple approximate scheme
for tracking the ground state during the flow, and adjusting the
expectation value accordingly. We shall employ a variational
calculation $\left|\psi_v(\alpha(l), \beta(l)) \right>$ to
continuously modify the ground state used in Eq.
(\ref{JzCubedMaps}) as the flow proceeds.

\subsection{First scheme}

The variational ansatz we will use is the familiar coherent state
\cite{BlaizotAndRipka},
 \be \label{9}
 \left|\psi_v(z) \right> = e^{zJ_+} \left|-j \right>,
 \ee
which is simple to evaluate and fairly reliable (see Appendix
\ref{AVarCalc}). The value of the complex parameter $z$ which
minimizes the energy is
 \be \label{varstate}
 z_v(\alpha, \beta) = \left\{
 \begin{array}{c@{\quad \quad}l} 0 & \beta \leq \frac{1}{1-1/N}
 \alpha \\ \pm i \sqrt{\frac{1-\alpha/\beta - 1/N}{1 + \alpha/\beta
 - 1/N}} & \beta > \frac{1}{1-1/N} \alpha \end{array}\right. .
  \ee
There is an $N$ dependent phase separation line
$\beta_{\mbox{\scriptsize variational}}^{(1)} =
\frac{1}{1-1/N}\alpha$ across which there is a continuous but
non-analytic jump in the behaviour of $z_v$. In the deformed
phase, the two variational states are degenerate in energy and we
artificially break the symmetry by choosing the positive imaginary
value. This plays no role in what follows since the relevant
quantity $\left<z_v\right|J_z\left|z_v\right>$ is invariant under
conjugation $z_v\rightarrow z_v^*$. The result for $\left<J_z
\right>$ is
 \be \label{Jz}
 \left<z_v(\alpha, \beta)\right| J_z
 \left| z_v(\alpha, \beta)  \right> = \left\{
 \begin{array}{c@{\quad \quad}l} -j & \beta \leq \beta_{\mbox{\scriptsize
 variational}}^{(1)}
 \\ -j (\frac{\alpha}{\beta})(\frac{1}{1-1/N}) & \beta > \beta_{\mbox{\scriptsize
 variational}}^{(1)}
 \end{array} \right. .
 \ee
In the first phase $\left<J_z \right> = -j$, the exact unperturbed
ground state value used by Pirner and Friman. In the deformed
phase $\left<J_z \right>$ increases and approaches zero for large
$\beta$. This is due to the pairing interaction which promotes
particles from the lower to the higher level until both levels are
equally filled and the resulting expectation value of $J_z$ is
zero due to the structure of its original definition
(\ref{spinreps}). Substituting the variational result (\ref{Jz})
into the flow equations (\ref{Before3})-(\ref{Before5}), results
in the following differential equations
 \be \label{phase1eqns}
 \begin{array}{rcl@{\quad \quad}rcl} \multicolumn{3}{c}{ \beta \leq \beta_{\mbox{\scriptsize variational}}^{(1)}} & \multicolumn{3}{c}{ \beta > \beta_{\mbox{\scriptsize variational}}^{(1)}}  \\ \hline
 \dot{\alpha} & = & -\beta^2(\frac{4j^2-2j+1}{4j^2}) &
 \dot{\alpha} & = & 2(\beta^2 \frac{j^2+j-1/2}{j^2} -3 \alpha^2 \frac{4j^2}{4j^2-4j+1}) \\
 \dot{\beta}  & = & -4\alpha \beta &
 \dot{\beta}  & = & -4\alpha \beta \\
 \dot{\delta} & = & -4\beta^2  & \dot{\delta} & = & -4\frac{(\alpha
 ^3 / \beta)}{(1-1/2j)^3}
 \end{array}.
 \ee

The first phase equations are identical to those obtained
previously, so that the variational scheme does not deliver any
new results in this region.

\begin{figure}
 \begin{center}
 \includegraphics[width=0.5\textwidth]{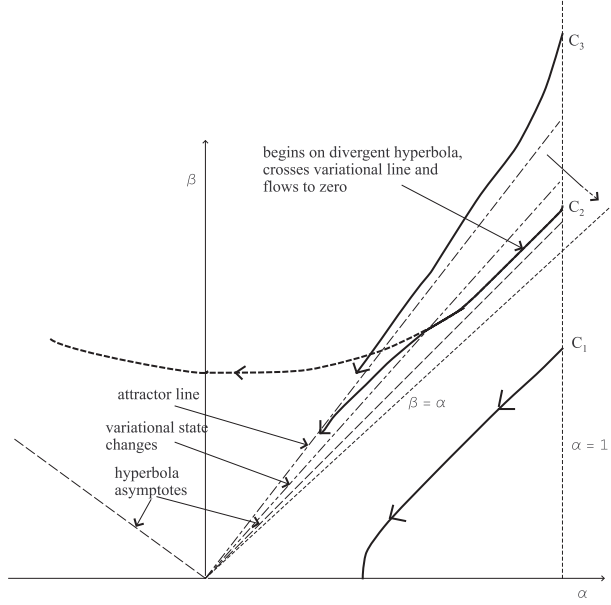}
 \end{center}
 \caption {\label{Extension1}Phase space diagram for finite $N$ in the first scheme,
 with tracking of the ground state expectation value.}
 \end{figure}

We refer the reader to Fig \ref{Extension1}, where the flow is
plotted for three different values of $\beta_0$. For finite $N$,
there are three lines of interest, which listed in increasing
order are
 \begin{eqnarray}
 \beta_{\mbox{\scriptsize asymptotes}}^{(1)} &=& \pm \sqrt{\frac{4j^2}{4j^2-2j+1}}\alpha \\
 \beta_{\mbox{\scriptsize variational}}^{(1)} &=& \frac{1}{1-1/{2j}}\alpha \\
 \beta_{\mbox{\scriptsize attractor}}^{(1)} &=&
 \frac{4j^2(2j(2+j)-1)}{2j(4j^2-j-1)+1}\alpha.
 \end{eqnarray}
The asymptotes of the hyperbola followed in the first phase are
familiar from the original Pirner and Friman scheme. The
variational line is the point at which the variational state
changes from the constant $\left|\Psi \right> = \left|-j\right>$
everywhere in the first phase into a continuously varying paired
state in the second phase, as in the variational result
(\ref{varstate}). Above this line the new equations on the right
hand side of Eqs. (\ref{phase1eqns}) come into play. Although
there is no simple first integral for these equations, as there
are for the first phase, it is not difficult to show that the
attractor line serves as an attractor for the flow, along which
the flow will remain. Thus we have removed the original divergent
results of Pirner and Friman, since now in the second phase all
curves must eventually end at the origin. The only complication is
that, since $\beta_{\mbox{\scriptsize asymptotes}} <
\beta_{\mbox{\scriptsize variational}}$ for finite $N$, there is a
small region of $\beta_0$ close to $1$ in which the flow initially
starts out on the divergent branch of the hyperbola, only to flow
to zero when the variational state begins to change (this shall be
termed the initial hyperbola effect). All three lines tend toward
$\beta=\alpha$ for large $N$.

\begin{figure}
 \begin{minipage}[t]{0.5\textwidth}
 \begin{center}
 \includegraphics[width=\textwidth]{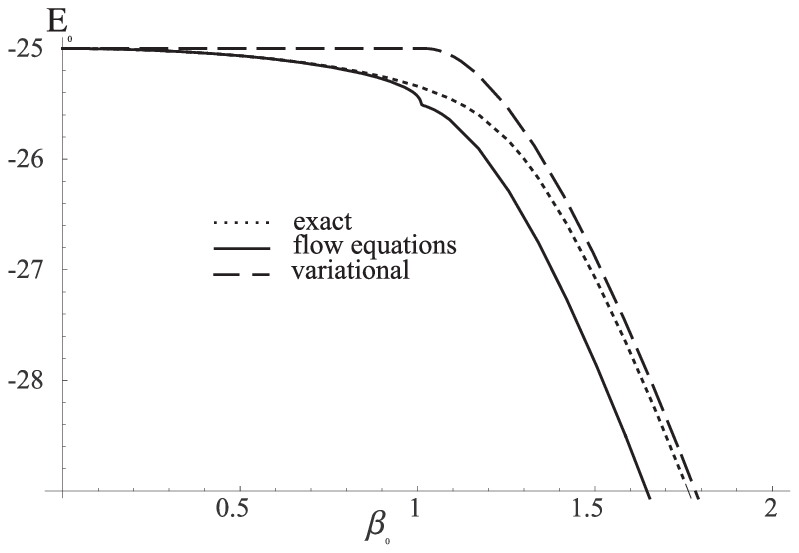}
 (a)
 \end{center}
 \end{minipage}
 \hfill
 \begin{minipage}[t]{0.5\textwidth}
 \begin{center}
 \includegraphics[width=\textwidth]{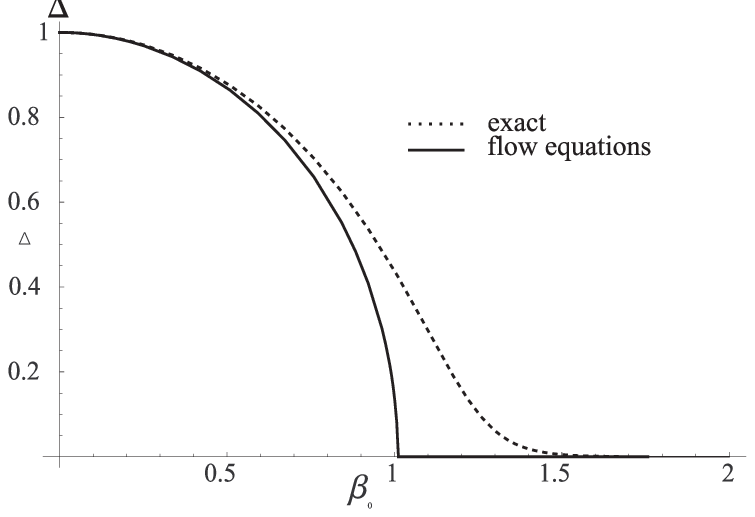}
 (b)
 \end{center}
 \end{minipage}
 \caption[(a) Ground state energy $E_0$ and (b) gap $\Delta$ for $N=50$ particles]{\label{ImprovedFirstScheme}(a) Ground state energy $E_0$ and (b) gap $\Delta$ for $N=50$ particles. The squiggle
 which occurs for finite $N$ is due to initial hyperbola effect. }
 \end{figure}

In this way tracking the evolution of the ground state during the
flow has led to sensible results for all $\beta_0$. The ground
state energy $E^1_N(\beta_0)$ and the gap $\Delta^1_N(\beta_0)$
reproduce Pirner and Friman's results (\ref{e1}) and (\ref{Delt1})
in the first phase, as the variational state is not sensitive to
changes there. $E_0$ must be calculated numerically in the second
phase, while $\Delta$ is identically zero there (see Fig.
\ref{ImprovedFirstScheme}).

\subsection{Second scheme}
The same game can now be played for the second scheme where we
include a $J_z^3$ term in the flow with coefficient $\gamma$, as
we did before in Eq. (\ref{SecSchemeH}). The variational state
must be recalculated since the form of the Hamiltonian has
changed. The result from Appendix \ref{AVarCalc} shows that there
is again a line
 \be
 \beta_{\mbox{\scriptsize variational}}^{(2)} = g(j)\alpha + f(j)
 \ee
across which the variational state begins to change,
 \be \label{JzHoo3}
 \left<z_v(\alpha, \beta)\right| J_z
 \left| z_v(\alpha, \beta)  \right> = \left\{
 \begin{array}{c@{\quad \quad}l} -j & \beta \leq \beta_{\mbox{\scriptsize variational}}^{(2)} \\
  -j\left(1 - 2\frac{h(\alpha, \beta, j)^2}{1+h(\alpha, \beta, j)^2}\right) &
 \beta > \beta_{\mbox{\scriptsize variational}}^{(2)}
 \end{array} \right. ,
 \ee
with $f,g,h$ having a rather complicated $j$-dependency:
 \begin{eqnarray}
 f(j) &=& \frac{2j(6j^2-6j+2)}{4j^3+2j^2-4j+1} = 3 + {\cal
 O}(1/j) \\
 g(j) &=& \frac{-2j(4j^2-8j+3)}{4j^3+2j^2-4j+1} = -2 + {\cal
 O}(1/j) \\
 h(\alpha, \beta, j) &=& \sqrt{\frac{3-4\alpha +
 \sqrt{12(\alpha-1)\alpha + \beta^2}}{3-2\alpha+\beta}} + {\cal
 O}(1/j).
 \end{eqnarray}
Before we use \ref{JzHoo3}, we shall first eliminate the $J_z^3$
coefficient $\gamma$ from the flow equations using Eq.
(\ref{gamminv}) so as to reduce our variables only to $\alpha$ and
$\beta$. The flow equations (\ref{Second1}), (\ref{Second2}) and
(\ref{Second3}) become
 \begin{eqnarray}
 \dot{\alpha}&=&\beta^2\left(\frac{2j(j+1)-1}{j^2}\right) \\
 \dot{\beta} &=& -4\beta \left( \alpha - \frac{6(\alpha-1)\left< J_z \right>^2}{2j(j+1)-1} \right)
 .\end{eqnarray}
Now we use the variational result (\ref{JzHoo3}) to obtain the
\begin{figure}
 \begin{center}
 \includegraphics[width=0.6\textwidth]{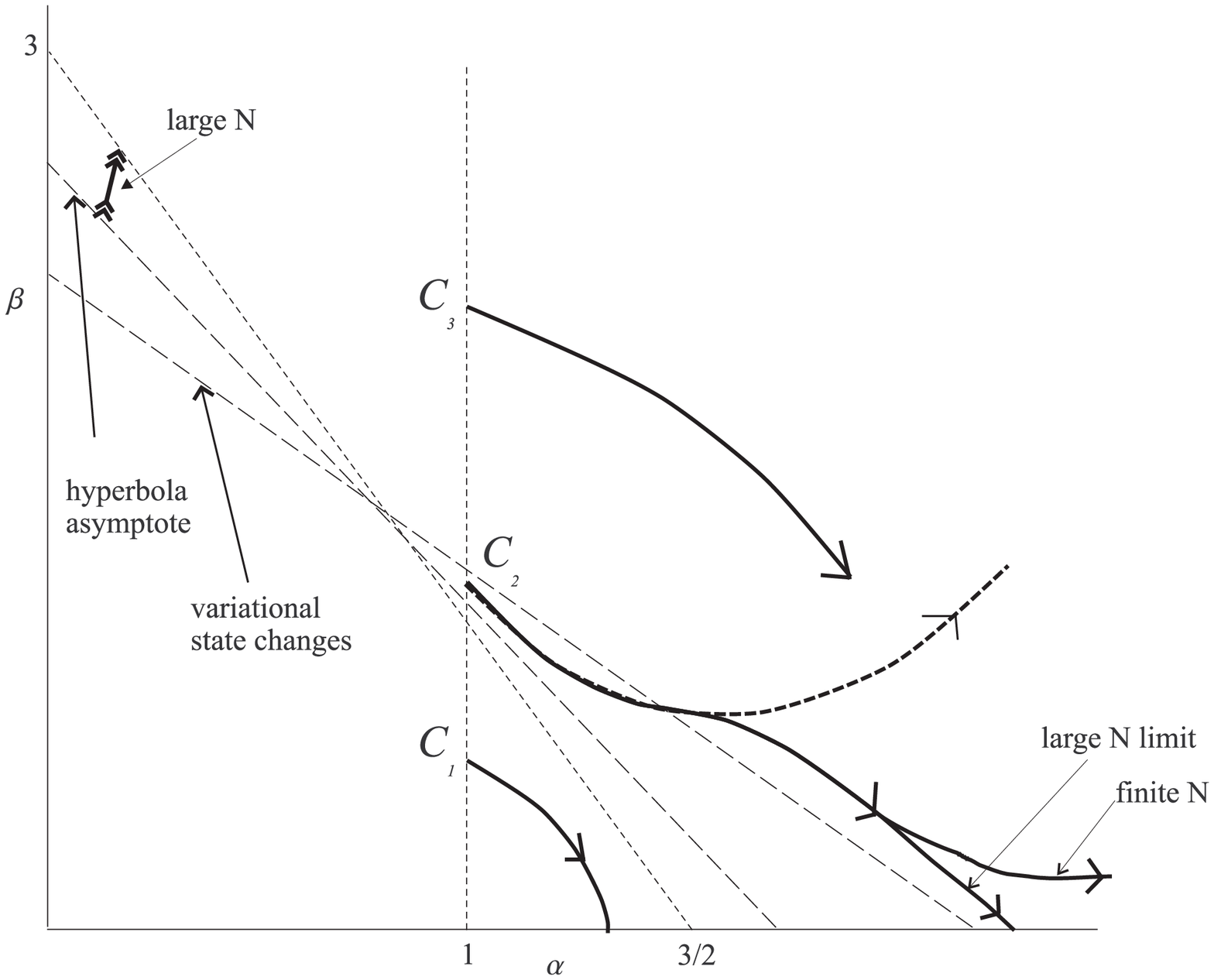}
 \end{center}
 \caption{\label{Improved2}Phase space diagram for finite $N$ in the second scheme,
 with tracking of the ground state expectation value.}
 \end{figure}
flow equations on both sides of the $\beta_{\mbox{\scriptsize
variational}}$ line. The equations shall only be displayed in the
large $N$ limit for simplicity; the reader is welcome to inspect
the full $j$-dependent expression for $h(\alpha, \beta, j)$
(\ref{Bigh}) if desired! \be
 \begin{array}{rcl@{\quad \quad}rcl} \multicolumn{3}{c}{ \beta \leq \beta_{\mbox{\scriptsize variational}}} & \multicolumn{3}{c}{ \beta > \beta_{\mbox{\scriptsize variational}}} \label{BET} \\ \hline
 \dot{\alpha} & = & 2\beta^2 &
 \dot{\alpha} & = & 2\beta^2  \\
 \dot{\beta}  & = & -4\left( \alpha\beta + 3\beta(1-\alpha) \right)
 & \dot{\beta}  & = & -4\left( \alpha\beta + 3\beta(1-\alpha)
 \left(\frac{\beta+2\alpha - \sqrt{12\alpha(\alpha-1) +
 \beta^2}}{\beta - 6\alpha + 6 + \sqrt{12\alpha(\alpha-1) +
 \beta^2}}\right)^2 \right)
 \end{array}.
 \ee
The structure of the phase space is shown in Fig. \ref{Improved2}.
The first phase equations (the left hand side of (\ref{BET})) are
identical to those obtained previously in Eqs.
(\ref{Second1})-(\ref{Second3}). For $\beta_0 <
\beta_{\mbox{\scriptsize asymptotes}}^{(2)}$ as in ${\cal C}_1$
the flow begins at $(1, \beta_0)$ and flows to the right along the
lower branch of the hyperbola with asymptote
 \be
 \beta_{\mbox{\scriptsize asymptotes}}^{(2)}=
 \frac{12j^4}{(2j(j+1)-1)^2}.
 \ee
The off-diagonal $\beta$ term decreases and $\cal H$ is
diagonalized as $\ell \rightarrow \infty$. For
$\beta_{\mbox{\scriptsize asymptotes}}^{(2)}< \beta_0 <
\beta_{\mbox{\scriptsize variational}}^{(2)}$, the flow begins to
flow to infinity along the upper divergent hyperbola. Soon it
crosses the line $\beta_{\mbox{\scriptsize variational}}^{(2)}$
and starts to obey the flow equations on the RHS of (\ref{BET}).
At this point the flow changes direction, and begins to flow
toward diagonal form (i.e. towards the $\alpha$ axis).

At this point it is crucial to investigate the behaviour of the
RHS of the $\dot{\beta}$ equation in (\ref{BET}), so as to
determine if $\beta$ will ultimately decrease to zero (thus
diagonalizing ${\cal H}$), or if $\beta$ will somehow diverge away
as before. The answer to this question lies in Fig.
\ref{FunctionSigns}, which plots the sign of $\dot{\beta}$ in the
phase space. For finite $N$ there is a small sliver just above the
$\alpha$ axis in which $\dot{\beta}$ becomes positive. Thus, for
finite $N$, the matrix approaches diagonal form very closely, only
  \begin{figure} \label{FunctionSigns}
 \begin{center}
 \includegraphics[width=0.5\textwidth]{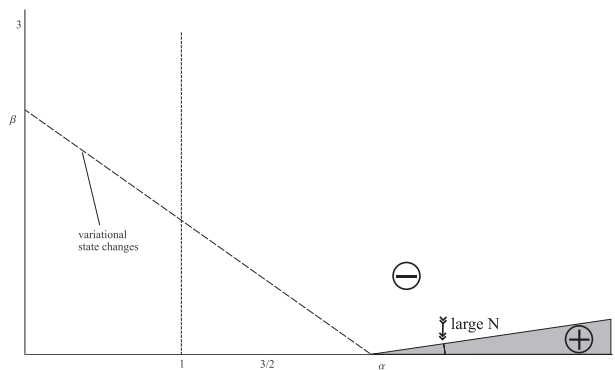}
 \end{center}
 \caption{\label{FunctionSigns}The sign of the full $j$ dependent $\dot{\beta}$.}
 \end{figure}
to diverge away at the last moment, as shown in ${\cal C}_2$. The
positive region becomes ever thinner as $N$ increases, so that in
the large $N$ limit ${\cal H}$ does indeed flow to diagonal form,
as in ${\cal C}_3$. In this case we produce the same results for
$E_0$ and $\Delta$ as in Fig. \ref{ImprovedFirstScheme}.

The task remains to explain why our method, of tracking the ground
state during the flow and adjusting the linearization accordingly,
has come close but failed to diagonalize $\cal H$ in the second
phase. The answer probably lies in the linearization step
(\ref{Lin1}). Recall that the order in which linearization or
commutation is applied makes a difference:
\begin{eqnarray}
[\cdot,\cdot]\circ \mbox{linearize} &\rightarrow& -6\left<J_z \right>^2 (J_+^2 + J_-^2) \label{LLin1} \\
 {}\mbox{linearize} \circ [\cdot,\cdot] &\rightarrow& \left( 12(1-\left<J_z
\right>)J_z + 6 \left<J_z \right>^2 - 8 \right) J_+^2 + conj.
\label{LLin2}
\end{eqnarray}
To obtain a measure of which order is more accurate, we define
 \begin{eqnarray}
 A_{\mbox{\scriptsize first}}(\alpha) &=& || \, [J_+^2 - J_-^2, J_z^3]
 \Psi - \left(-6\left<J_z \right>^2 (J_+^2 + J_-^2) \right) \Psi
 || \\
 A_{\mbox{\scriptsize second}}(\alpha) &=& || \,[J_+^2 - J_-^2, J_z^3]
 \Psi - \left(\left( 12(1-\left<J_z
\right>)J_z + 6 \left<J_z \right>^2 - 8 \right) J_+^2 + conj.
\right) \Psi ||,
 \end{eqnarray}
 \begin{figure}[t]
 \begin{center}
 \includegraphics[width=0.35\textwidth]{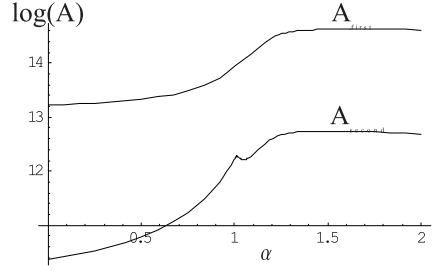}
 \end{center}
 \caption{\label{Distances}The error involved in linearizing first ($A_{\mbox{\scriptsize
 first}}$) or linearizing second ($A_{\mbox{\scriptsize second}}$).
 $N=50$ particles.}
 \end{figure}
where the subscripts refer to whether linearization is applied
first or second with respect to commutation. $\Psi$ is the exact
ground state of the Hamiltonian. The $\alpha$ dependency enters
because we intend to evaluate these functions along the line
$(\beta = 1, \alpha=0 \ldots 2)$, which passes through both
regions of phase space. Along this line $\Psi$ will change, and so
too will  $\langle J_z \rangle$ since it is a function of $\alpha$
and $\beta$ from Eq. (\ref{JzHoo3}). The idea is simply to measure
which method comes closest to approximating the action of $[J_+^2
- J_-^2, J_z^3]$ on the ground state. The results are plotted on a
logarithmic scale in Fig. \ref{Distances}. It is clear that
linearizing second is orders of magnitude more accurate than
linearizing first. This was to be expected, since it is
theoretically the sounder scheme. Unfortunately it is not easy to
implement, as it necessarily involves adjoining new operators to
the flow ($J_zJ_+^2 + J_-^2J_z$), which themselves produce new
operators, and so on. Fig. \ref{Distances} also clearly shows how
the linearization procedure is poorer in the second phase
($\beta_0 > 1$), where the structure of the ground state is more
complex.

In conclusion we see that the algebraic structure of the
Hamiltonian has made it difficult to consistently follow a
linearization procedure, when a $J_z^3$ term is added to the flow.
One is forced to linearize in an ad hoc fashion, which introduces
finite $N$ errors that ultimately prevent the Hamiltonian from
being diagonalized in the second phase.

\section{Self-consistent linearization}
The drawback of the previous scheme is that a variational
calculation was necessary as an outside addition to the process.
It would be desirable to eliminate this step by calculating
$\langle J_z \rangle (\ell)$ in a self-consistent manner. This is
indeed possible, in a scheme that has recently been outlined by
one of the supervisors of this work \cite{Scholtz}. To do this we
note that it cannot have an explicit dependence on $\ell$, but
rather can only depend on $\ell$ implicitly though its dependence
on the $\ell$-dependent parameters $\alpha(\ell)$ and
$\beta(\ell)$. This in turn is because the dynamical ground state
$\left|G(\alpha(\ell) \right>$ cannot have an explicit dependence
on $\ell$ since the Hamiltonian $H(\alpha(\ell),\beta(\ell),
\delta(\ell))$ does not have such a dependence. Furthermore, we
note that $\langle J_z \rangle (\ell)$ can only be a function of
the dimensionless coupling constant $x(\ell) \equiv
\beta(\ell)/\alpha(\ell)$, since the constant $\delta$ and
rescaling by $1/\alpha$ do not affect expectation values. This
implies that we can write
 \be
 \langle J_z \rangle (\ell) \equiv \langle G, \ell | J_z | G, \ell
 \rangle \equiv f( \beta(\ell)/\alpha(\ell)).
 \ee
Now we differentiate $f(\beta(\ell)/\alpha(\ell))$ with respect to
$\ell$, and use the flow equations (\ref{Before3})-(\ref{Before5})
to obtain
 \begin{eqnarray}
 \left[ \frac{\beta^3}{j^2\alpha^2} \left(6f^2(\beta/\alpha) -
 2j(j+1) + 1\right) - 4\beta \right] f'(\beta/\alpha) &=& -
 \langle G, \ell| [\eta, J_z] |G, \ell \rangle \\
 &=& 4(E_g - \alpha f(\beta/\alpha) - j\delta)
 \end{eqnarray}
where the second line follows from substituting the form of the
generator $\eta$ from \ref{etaeval} and the parameterization of
the original Hamiltonian \ref{Hoo1}. Now we differentiate this
result again with respect to $\ell$ and use the flow equations
(\ref{Before3})-(\ref{Before5}) again. The remarkable result is
that we now obtain a closed differential equation for $f(x)$,
where $x=\beta/\alpha$ is the dimensionless coupling constant,
\begin{figure}
 \begin{center}
 \includegraphics[width=0.5\textwidth]{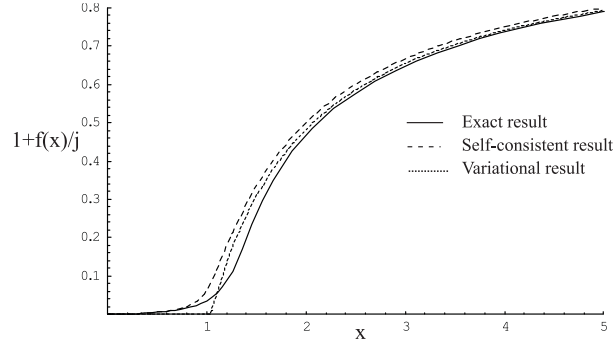}
 \end{center}
 \caption[Order parameter $1+f(x)/j$ as a function of the dimensionless coupling constant
 $x$, for $N=30$ particles.]{\label{SCFig1}Order parameter $1+f(x)/j$ as a function of the dimensionless coupling constant
 $x$, for $N=30$ particles. The legend distinguishes the exact
 result, variational result and self-consistent result.}
 \end{figure}
 \begin{eqnarray}
 \left[\frac{x^2}{j^2}\left(6 f^2(x) - 2j(j+1) + 1\right) -
 4\right]^2 f''(x) + (\frac{2x^3}{j^4} - \frac{8x}{j^2})\left(6 f^2(x) - 2j(j+1) +
 1\right)^2 f'(x) \nonumber \\
 +\frac{12x^4}{j^4}\left(6 f^2(x) - 2j(j+1) + 1\right)f'^2(x)f(x)
 - \frac{48x^2}{j^2}f'^2(x)f(x) \hspace{30mm} \nonumber\\
 -\frac{4}{j^2}\left(6 f^2(x) - 2j(j+1) +1\right)f(x) + \frac{16}{j^2}f^3(x) = 0.
 \hspace{40mm} \label{SC5}
 \end{eqnarray}
This is a non-linear differential equation that uniquely
determines the function $f(x)$ once boundary conditions have been
specified. It is useful to define $1 + f(x)/j$ as an order
parameter, since in the thermodynamic limit this quantity vanishes
everywhere for $x<1$ and is equal to unity for $x>1$. This last
statement follows from the fact that the expectation value
$\langle J_z \rangle$ scales linearly with $j$ in the weakly
coupled system($x<1$) but does not scale with $j$ in the strongly
coupled system($x>1$). In principle Eq. (\ref{SC5}) contains all
information on the phase structure of the system. All that remains
is to determine the boundary conditions on $f(x)$. These are
easily derived by noting that at $x=0$ the Hamiltonian is ${\cal
H}=J_z$ so that $f(0) = -j$. Furthermore, since $f(-x)=f(x)$, one
easily sees that $f(x)$ attains its minimum value at $x=0$. Thus
we have the boundary conditions
 \be \label{SCBound}
 f(0)=-j, \quad f'(0)=0
 \ee
In Fig. \ref{SCFig1} the order parameter $1+f(x)/j$ is plotted as
a function of $x$ for $j=15$. The exact result(from numerical
diagonalization) and the variational result from (\ref{Jz}) is
also shown for comparison. We note excellent agreement, even in
the transitional regions where the fluctuations are large. In the
two different phases where the fluctuations are expected to be
small, the self-consistent result indeed converges to the exact
result. The variational result is a marginally better
approximation in the second phase, but is noticeably poorer in the
transitional region.
\begin{figure}
 \begin{minipage}{0.5\textwidth}
 \begin{center}
 \includegraphics[width=\textwidth]{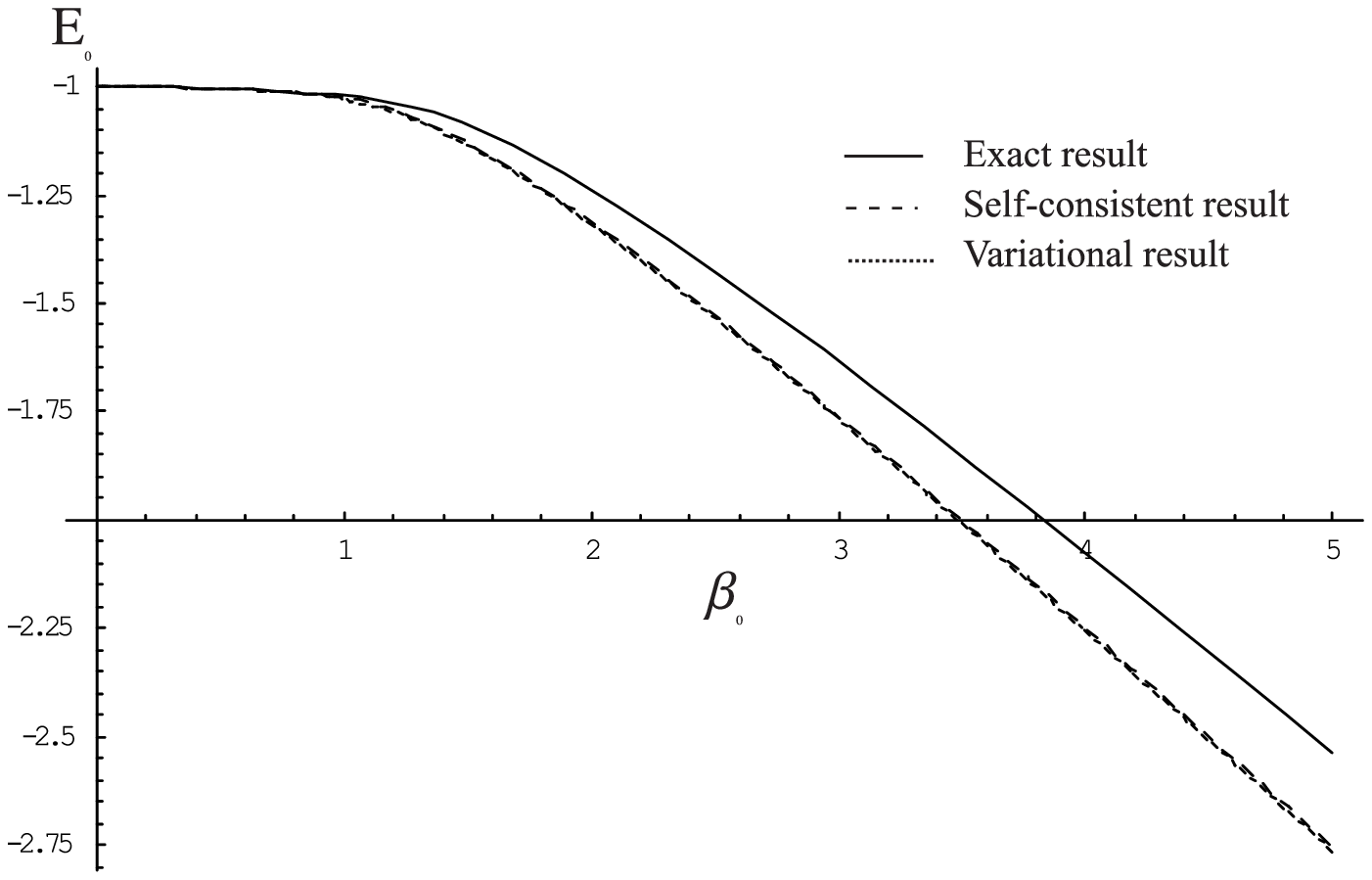}
 (a)
 \end{center}
 \end{minipage}
 \hfill
 \begin{minipage}{0.5\textwidth}
 \begin{center}
 \includegraphics[width=\textwidth]{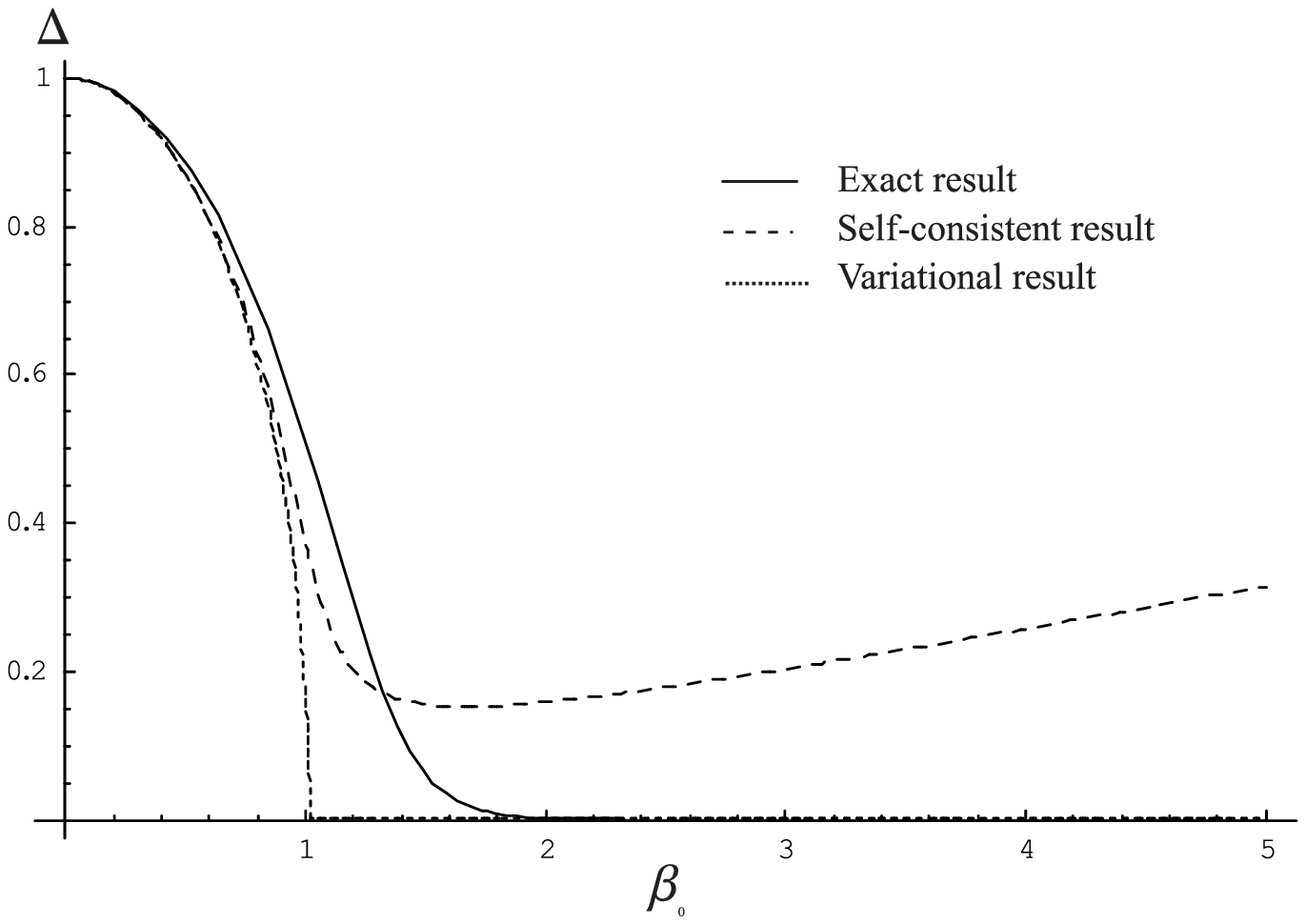}
 (b)
 \end{center}
 \end{minipage}
 \caption[a) Ground state energy $E_0$ and (b) gap $\Delta$ for $N=30$ particles] {\label{SCFig2}(a) Ground state energy $E_0$ and (b) gap $\Delta$ for $N=30$ particles.
 The legend distinguishes the exact result, the flow equations result using the variational calculation,
 and the flow equations result using the self-consistent calculation.}
 \end{figure}

Once $f(x)=\langle J_z \rangle$ has been numerically computed from
(\ref{SC5}), the flow equations (\ref{Before3})-(\ref{Before5})
can be integrated, as was done previously in Section
\ref{ImpSection} using the variational calculation of $\langle J_z
\rangle$. This allows us to compare the results for the ground
state energy and the gap, which are presented in Fig.
\ref{SCFig2}(a) and Fig. \ref{SCFig2} (b). The result for the
ground state is virtually indistinguishable from that obtained
previously when a variational calculation was employed. This is a
remarkable success of the method, since it used no external
information and had to generate ``its own'' $\langle J_z \rangle$.
However the result from the gap, while faring better than that
obtained from a variational calculation in the first phase, fares
markedly poorer in the second phase. This result shows how the
small difference in $\langle J_z \rangle$ present between the two
approaches in Fig. \ref{SCFig1} can create a notable difference
after the flow equations have been integrated. Both the
self-consistent and variational calculations approach the exact
result in the limit of large $N$.

Clearly the success of the method can be generalized to many other
systems. The central idea is that the flow equations are closed by
linearizing around ground state expectation values. These
expectation values are always only implicitly dependent on $\ell$,
and can only be an explicit function of one less than the number
of parameters in the Hamiltonian, since they are invariant under
multiplication of the Hamiltonian by a constant factor. This means
that differentiation of these functions together with the use of
the flow equations should yield a closed differential equation for
the order parameter(in this case the expectation value), in a
self-consistent manner. Integration of this equation then provides
one with accurate knowledge of the order parameter and ground
state energy, {\em without ever diagonalizing the Hamiltonian or
performing an auxiliary calculation}.

\section{Other flow equations approaches to the Lipkin model}
Besides the method of tracking the ground state during the flow,
we have also investigated other approaches. Although they did not
lead to concrete results, they are listed here since all of them
are new methods of dealing with the flow equations that haven't
been attempted before.

\subsection{Operator differential equations}
\begin{figure}
 \begin{minipage}[t]{0.4\textwidth}
 \begin{center}
 \includegraphics[width=\textwidth]{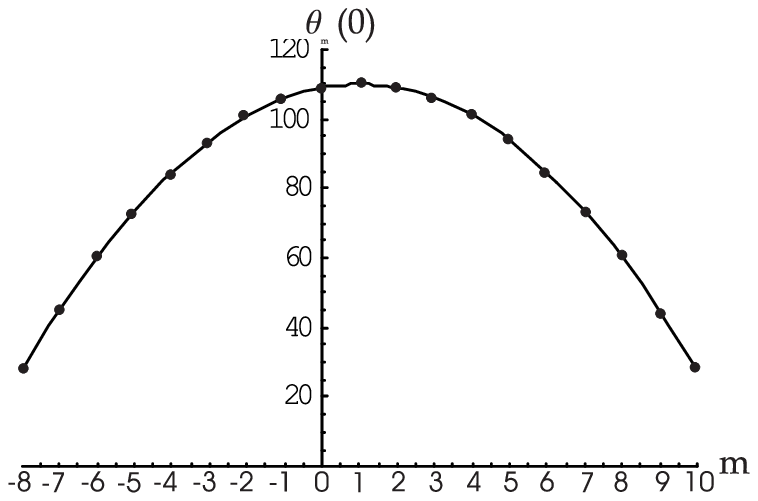}
 (a)
 \end{center}
 \end{minipage}
 \hfill
 \begin{minipage}[t]{0.4\textwidth}
 \begin{center}
 \includegraphics[width=\textwidth]{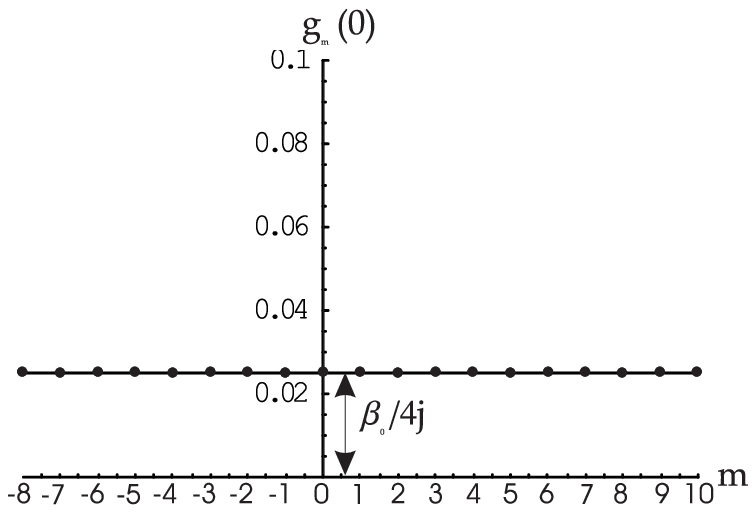}
 (b)
 \end{center}
 \end{minipage}
 \caption{\label{OpFlowFig} Comparison between initial conditions, contrasting the curve (a) $\theta_m(0)
 $ with (b) the horizontal line $g_m(0)$.  }
 \end{figure}

Instead of parameterizing the Hamiltonian during the flow with
scalar coefficients, we might try to track the flow with {\em
operator valued} coefficients. To be precise, we will write the
Hamiltonian during the flow as
 \be \label{opflow1}
 {\cal H}(\ell) = f(J_z, \ell) J_z + g(J_z, \ell)J_+^2 +
 J_-^2g(J_z, \ell).
 \ee
The functions $f(J_z, \ell)$ and $g(J_z, \ell)$ are operator
valued and have initial conditions
 \be
 f(J_z, 0) = J_z, \quad g(J_z, 0) = \beta_0/4j.
 \ee
Evaluating the double bracket yields
 \begin{eqnarray}
 [[J_z, {\cal H}], {\cal H}] = 4\left[ g^2(J_z)B(J_z) +
 A(J_z)A(J_z+1)(g^2(J_z) - g^2(J_z+2))\right] \\
 -\left[2g(J_z)(f(J_z)-f(J_z-2))\right]J_+^2 -
 J_-^2\left[2g(J_z)(f(J_z)-f(J_z-2))\right],
 \end{eqnarray}
where $B(J_z) = -8J_z^3 + (8j(j+1) - 4)J_z$ and $A(J_z) = -J_z^2 -
J_z + j(j+1)$. We have thus obtained a closed set of differential
equations for $f$ and $g$:
 \begin{eqnarray}
 \frac{\partial f(J_z, \ell)}{\partial \ell} &=& 4\left[ g^2(J_z)B(J_z) +
 A(J_z)A(J_z+1)(g^2(J_z) - g^2(J_z+2))\right] \label{of2} \\
 \frac{\partial f(J_z, \ell)}{\partial \ell} &=&
 -2g(J_z)(f(J_z)-f(J_z-2))\label{of3}.
 \end{eqnarray}
The fact that the equations are closed was to be expected since
Eq. (\ref{opflow1}) is the most general band-diagonal matrix, and
we have already shown that the Hamiltonian remains band-diagonal
during the flow. It is important to compare Eqs. (\ref{of2}) and
(\ref{of3}) with the direct equations for the diagonal elements
$\varepsilon_n$ and off-diagonal elements $\theta_n$, as in
Mielke's treatment\footnote{In Mielke's treatment the basis was
rearranged to produce a tridiagonal matrix. We show here the
equations in the original basis, which is why they differ slightly
from Eqs. (\ref{MQ1}) and (\ref{MQ2}).}(Eqs. (\ref{MQ1}) and
(\ref{MQ2})),
 \begin{eqnarray}
 \frac{d \varepsilon_n}{d\ell} &=& 4(\theta_{n+2}^2 - \theta_n^2)
 \label{MMQ1}
 \\
 \frac{d \theta_n}{d\ell} &=&
 -2\theta_n(\varepsilon_{n}-\varepsilon_{n-2}) \label{MMQ2}.
 \end{eqnarray}
If we write $f_m = \langle m|f(J_z)|m \rangle$ and $g_m = \langle
m |g(J_z) | m \rangle$, then the two are related by a change of
variables
 \be
 \varepsilon_m = mf_m, \quad
 \theta_m = \sqrt{(j-m+2)(j-m+1)(j+m)(j+m-1)}g_m.
 \ee
What we have gained, though, in writing the flow equations in the
variables $f_m$ and $g_m$, is that the initial conditions are much
simpler. This is illustrated in Fig. \ref{OpFlowFig}, for the case
of the off-diagonal entries.

Nevertheless, the new system of equations (\ref{of2})-(\ref{of3}),
was still unable to cope with the second phase($\beta_0>1$) in a
powerful way, after using the same type of nonperturbative
approximation that Mielke used in Section \ref{MielkeSec}. The
reason is that in the second phase, though $g_m$ begins as a
straight horizontal line at $\ell=0$, the exact solution shows
that it rapidly develops two sharp bumps on each end, which
invalidate the approximation made. This can be traced back to the
boundary effects present in Fig. \ref{ExacFS}.

In a more general context, by writing the Hamiltonian using $\ell$
dependent operators instead of scalars it may be possible to find
a closed operator form of the Hamiltonian during the flow. This
immediately eliminates redundant parameters and may give a deeper
insight into the problem.

\subsection{Moment preservation}
Under the exact unitary flow of the Hamiltonian, all moments will
remain invariant:
 \be
 \mbox{Tr}({\cal H}^i(\ell)) = \mbox{Tr}({\cal H}_0).
 \ee
Indeed, this can be regarded as a {\em definition} of unitary
equivalence. When approximations are made in the flow equations,
the invariance of the moments is lost and the traces begin to
change. One may attempt to choose the parameters in the
Hamiltonian in such a way as to strictly maintain one of the
moments during the flow. Out of the myriads of possible schemes
that present themselves in this way, we will focus here on the
simplest choice. We use the first scheme parametrization of the
Hamiltonian,
 \be \label{mypam}
 {\cal H(\ell)} = \alpha J_z + \frac{\beta}{4j}(J_+^2 + J_-^2) +
 \delta.
 \ee
$\alpha(\ell)$ and $\beta(\ell)$ are taken from the first scheme
solution to the flow equations when the ground state is tracked
during the flow, as in Eq. (\ref{phase1eqns}). However,
$\delta(\ell)$ is chosen so that the second order moment
 \be
 \mbox{Tr}({\cal H}^i) = \alpha^2 \frac{(2j+1)(j+1)j}{3} +
 \beta^2 \frac{(2j+3)(2j+1)(2j-1)(j+1)}{60j} + \delta^2 (2j+1)(j)
 \ee
is preserved. The second order moment is chosen due to its
relative simplicity; other orders are obviously also possible.
Explicitly, we set
 \be \label{opset}
 \delta = -\frac{\sqrt{20j^2(2j+1)(j+1)(1 - \alpha^2) +
 (4j(1+j)-3)(\beta_0^2 + \beta^2)}}{2j\sqrt{15(2j+1)}}.
 \ee
 \begin{figure}
 \begin{minipage}[t]{0.4\textwidth}
 \begin{center}
 \includegraphics[width=\textwidth]{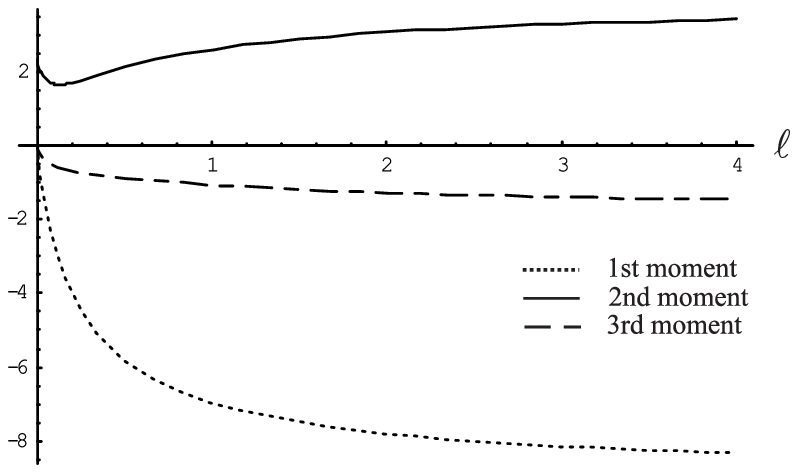}
 (a)
 \end{center}
 \end{minipage}
 \hfill
 \begin{minipage}[t]{0.4\textwidth}
 \begin{center}
 \includegraphics[width=\textwidth]{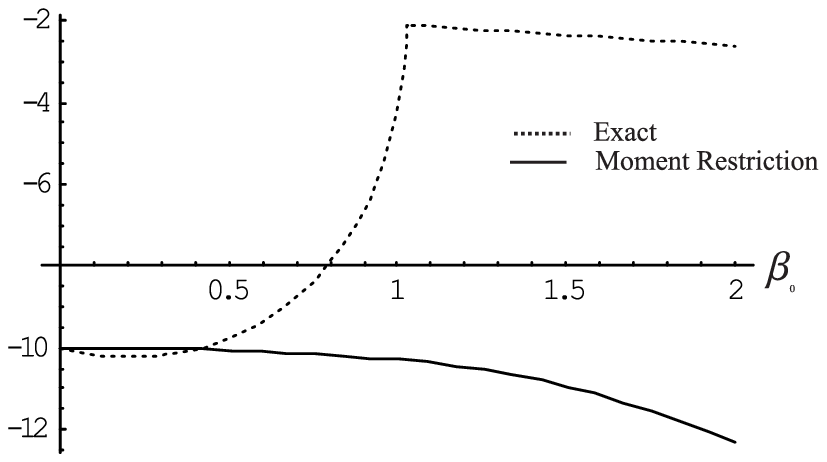}
 (b)
 \end{center}
 \end{minipage}
 \caption[Effect of preserving the moments during the flow]{\label{OpFlowFig2} (a) Flow of first three moments, for $\beta_0=1$. Moments are normalized by $1/N$, $1/N^2$ and
 $1/N^3$.
 (b) Effect of utilizing the moment preservation equation (\ref{opset}). }
 \end{figure}
The results are shown in Fig. \ref{OpFlowFig2}. For the reader to
get an idea of how the various moments change with $\ell$, Fig.
\ref{OpFlowFig2}(a) plots the flow of the first three moments,
normalized appropriately. Fig. \ref{OpFlowFig2}(b) plots the
result, using the moment preservation equation (\ref{opset}) and
the known first scheme solution for $\alpha(\infty)$ from Section
\ref{ImpSection}, for the ground state $E_0$. It is not surprising
to see that preserving the second moment in a parametrization like
Eq. (\ref{mypam}) is a very poor approximation if one is
interested in the ground state properties. The reason is that
preserving the moments is attempting to find the closest fit to
the whole exact Hamiltonian during the flow, whereas we are only
interested in the ground state. More sophisticated methods,
involving different parametrizations, and different definitions of
the trace operation so as to zero in on the lowest lying states,
produce slightly better results, although the complexity involved
is also significantly increased.

\subsection{Generator flow}
\begin{figure}
 \begin{minipage}[t]{0.25\textwidth}
 \begin{center}
 \includegraphics[width=\textwidth]{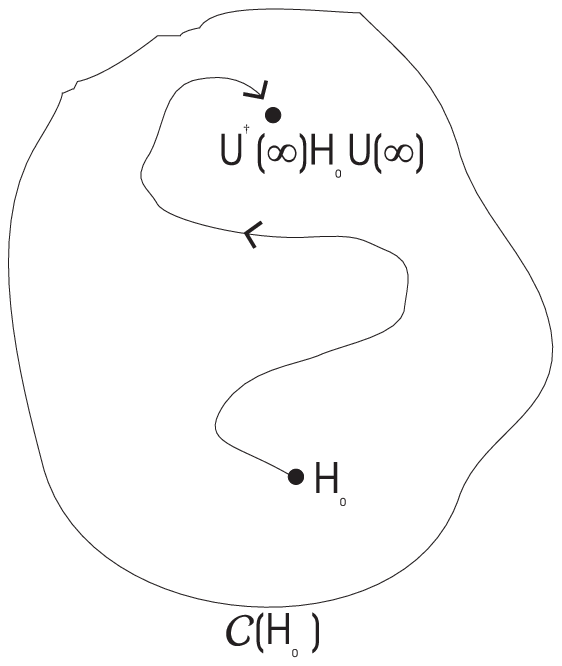}
 (a)
 \end{center}
 \end{minipage}
 \hfill
 \begin{minipage}[t]{0.25\textwidth}
 \begin{center}
 \includegraphics[width=\textwidth]{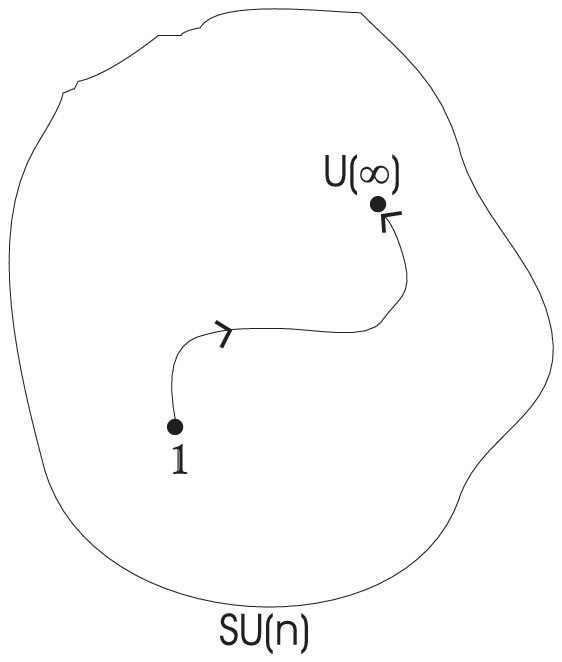}
 (b)
 \end{center}
 \end{minipage}
 \hfill
 \begin{minipage}[t]{0.35\textwidth}
 \begin{center}
 \includegraphics[width=\textwidth]{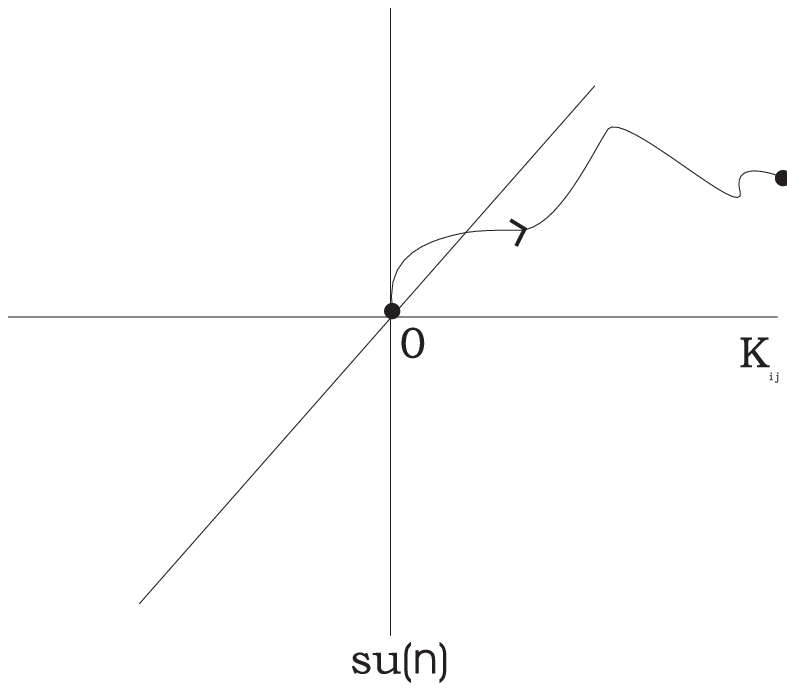}
 (c)
 \end{center}
 \end{minipage}
 \caption{\label{FlowSets} The flow in (a) $\C({\cal H}_0)$, (b) $SU(N)$ and (c) $su(N)$  }
 \end{figure}

So far we have viewed the flow as taking place either in $
\C({\cal H}_0) = \{H \; : \; H \mbox{ is unitarily equivalent to }
{\cal H}_0 \}$, or in $SU(N)$. We may go a step further and {\em
consider the flow as taking place in the Lie Algebra of $SU(N)$},
namely {\em su}$(N)$(See Fig. \ref{FlowSets}).

In other words, if we write the unitary transformations in the
exponential form
 \be
 U(\theta_i) = e^{\sum_i \theta_i K_i},
 \ee
then we consider the flow as taking place in the generator space
$K_i$. The hope is to remove that part of the non-linearity of the
flow equations which is due to the exponential. For instance,
consider a $2 \times 2$ matrix, with real coefficients:
 \be
 H_0 = \left(
 \begin{array}{ccc}
 1 & t \\
 t & 2 \\
 \end{array}
 \right).
 \ee
This may be viewed as a reduced Lipkin model with
$J_z=\mbox{Diag}(1,2)$. Let $Q$ be the real orthogonal matrix that
transforms the Hamiltonian. We know from Section \ref{sdf} that
$\dot{Q} = -Q\eta$, where $\eta=[J_z, Q^TH_0Q]$. Thus if we write
 \begin{equation}
 Q(\theta) = e^{\theta K} \leftrightarrow
 \left(
 \begin{array}{cc}
 \cos \theta & \sin\theta \\
 -\sin\theta & \cos\theta
 \end{array}
 \right)
  = \mbox{Exp} \left(
 \begin{array}{cc}
 0 & 1 \\
 -1 & 0
 \end{array}
 \right) ,
 \end{equation}
then the differential equation for $\dot{\theta}$ is
 \begin{eqnarray}
 \dot{\theta} &=& t\cos(2\theta) - \frac{1}{2}\sin(2\theta) \label{gen50} \\
 \theta(0) &=& 0  .
 \end{eqnarray}
If we were hoping for a simpler differential equation, then we
have failed dismally, as this equation is more complex than the
direct Hamiltonian matrix elements version
 \begin{eqnarray}
 \dot{\varepsilon_1} = -2v^2, \quad \dot{\varepsilon_2} &=& 2v^2,
 \quad \dot{v} = -v(\varepsilon_2 - \varepsilon_1) \label{gen51} \\
 \varepsilon_1(0) = 1, \quad \varepsilon_2(0) &=& 2, \quad v(0) =
 t,
 \end{eqnarray}
where $\varepsilon_i = H_{ii}$ and $v=H_{12}$. But we have gained
in two areas. Firstly, the flow has been reduced from three
variables to one. For a general real Hamiltonian of size $N$,
there would be $\frac{1}{2}N(N+1)$ equations in the Hamiltonian
matrix elements, and $\frac{1}{2}N(N-1)$ equations in $so(N)$.
Secondly, the boundary condition has been shifted into the
equation for $\dot{\theta}$. This may be useful if one is
interested in the onset of a phase change, as investigation of the
instability of the equations would give direct information on the
parameter values involved at the transition point.

In addition, it might be possible by a clever choice of the
generators $K_{i}$ to effectively decouple the equations for the
$\theta_i$, and hence ``straighten the flow''. We know this is
always mathematically possible from differential geometry (eg. see
Ref. \cite{Frankel}). In the Lipkin model, one might use
 \be
 K_{ij} = \frac{J_z^{j-1}J_+^i -
 J_-^iJ_z^{j-1}}{\left(N/2\right)^{i+j-1}} \qquad i=1\ldots N, \,
 j=1\ldots N+1-i
 \ee
as a basis for the generator space. Unfortunately, again, this
choice works well for small $\beta_0$ where the unitary
transformation is approximately of the form
 \be
 U = e^{\theta_{11} K_{11}} = e^{\theta_{11}( J_+^2 - J_-^2)}.
 \ee
Indeed, this is familiar from the first scheme in Section
\ref{ImpSection}, where $\eta = [J_z, {\cal H}] \propto J_+^2 -
J_-^2$. However for larger $\beta_0$ near the phase change, other
$\theta_{ij}$ become important and the approximation is not so
good anymore.

\section{Discussion}
We have presented various different methods of dealing with the
Lipkin model by flow equations. The first replaced newly generated
operators by linearizing them around the zero interaction ground
state. The second considered the matrix elements themselves and
provided a non-perturbative solution correct for large $N$. The
third used the Holstein-Primakoff mapping to systematically solve
the flow equations order for order in $1/N$. The fourth method
improved on the first by employing a variational calculation to
track the ground state, and hence the expectation value relevant
to the linearization, during the flow. The specific choice of a
variational state to approximate the ground state is not crucial
to the method, and any approximate input as to the nature of the
ground state during the flow wold have produced similar results.
The fifth method dispensed with the outside input necessary in the
previous technique by utilizing a self-consistent calculation of
the expectation value. This resulted in very similar results to
the previous method for the ground state energy, as well as
providing a simple and exciting new way to calculate an order
parameter such as the expectation value.

Only the last two mentioned methods were able to deal with the
second phase in a meaningful way, since the flow equations were
dynamically altered there due to a changing expectation value. The
reason the second phase has proved more difficult to analyze is
clear from Figs. \ref{barchartsdiag}-\ref{barchartsoffdiag}. The
flow is highly nonlinear in this region and the direction of the
flow, as specified by the generator $\eta$, changes appreciably as
it evolves.

We have also outlined three other possible methods of dealing with
the operator flow. Some ideas here might be valuable in a
different context.

%% file: Chapter4.tex
\chapter{Flow equations and renormalization} \label{RenormChapter}

In the early twenties and thirties a gloomy problem hovered over
relativistic quantum field theory. Calculation of properties of
even the simplest systems such as the energy of an electron in an
atom yielded divergent results for anything but the lowest orders
in perturbation theory. Hans Bethe's seminal paper in 1947, which
calculated the Lamb shift between the $2s$ and $2p$ levels in a
hydrogen atom, was the first to lead to a finite, accurate result.
Renormalization - in its modern perturbative sense - was born.
Initially some of the procedures seemed ad hoc and specific to the
model under consideration, but modern developments have unified
the theory with the language of condensed matter and many body
physics. In this chapter we discuss renormalization in a
Hamiltonian framework. By the term `renormalization' we mean that
the parameters present in the initial bare Hamiltonian will be
altered so as to provide a new, effective theory which is
equivalent to the old one, but more accessible to study. Two of
the pioneers of the modern theory of renormalization, Stan Glazek
and Kenneth Wilson, have recently made significant progress in
this field by using flow equations to dynamically renormalize the
Hamiltonian \cite{GlazekWilson1, GlazekWilson2}. In this chapter
their method will be introduced and compared with Wegner's flow
equation. This will be followed by two examples of applications of
the formalism.

\section{Glazek and Wilson's similarity renormalization}
 \label{Sect1}
Recall that, from a mathematical viewpoint, the primary motivation
for Wegner's flow equation
 \be \label{Weg1}
 \frac{d {\cal H}}{d\ell} = [[\mbox{Diag}{\cal H}, {\cal H}],
 {\cal H}]
 \ee
was that it defines the {\em steepest path to diagonality}. To
first order in the coupling, the off-diagonal terms flow as
 \be \label{1storderV}
 v_{ij}^{(1)}(\ell) = V_{ij}e^{-(E_i-E_j)^2\ell},
 \ee
where $E_i$ and $V_{ij}$ are the initial diagonal and off-diagonal
elements respectively. In this way we see that interaction terms
coming from states with large energy differences are decoupled
first. However, to second order the matter is not so clear-cut,
and the renormalization properties of Wegner's generator
(\ref{Weg1}) are not entirely clear.  Although Glazek and Wilson
developed their similarity renormalization scheme at the same time
as Wegner invented his flow equation, in hindsight it is clear
that their scheme follows from {\em imposing stricter
renormalization requirements than Wegner's scheme}.

The idea is to construct a continuum of unitarily equivalent
Hamiltonians $H_\lambda$, which interpolate between the initial
Hamiltonian ($\lambda = \Lambda$) and the effective Hamiltonian
($\lambda = \lambda_0$). The parameter $\lambda$ has the
dimensions of energy and is a measure of the size of the largest
energy differences that play a role in $H_\lambda$. These
statements will be made more precise in what follows.

As usual, we separate $H_\lambda$ into its diagonal and
interacting (off-diagonal) parts
 \be
 H_\lambda = H_{0 \lambda} + H_{I \lambda}.
 \ee
The diagonal entries of $H_{0 \lambda}$ are denoted by
$E_\lambda$:
 \be
 \langle i | H_{0 \lambda} | i \rangle = E_{i \lambda}.
 \ee
As discussed before in Section \ref{WegFE}, an arbitrary unitary
flow can be described as
 \be \label{arbU}
 \frac{dH_\lambda}{d \lambda} = [T_\lambda, H_\lambda],
 \ee
where $T_\lambda$ is anti-Hermitian. The problem is to choose
$T_\lambda$ so as to satisfy our requirements. The aim is to
eliminate matrix elements $\langle i | H_\lambda | j \rangle
\equiv H_{\lambda i j}$ whenever these matrix elements would cause
large jumps in energy beyond the scale set by $\lambda$, i.e. when
$E_{i \lambda}$ is sufficiently separated from $E_{j \lambda}$ and
the larger of these is well above $\lambda$ itself. In this way,
as $\lambda$ is reduced, the far off-diagonal part of $H_{I
\lambda}$ is reduced systematically, in that as $\lambda$
decreases it is mainly terms which jump from much lower energies
to energies of order $\lambda$ that are being eliminated - terms
with energies of order higher than $\lambda$ have already been
eliminated, while terms with energies of order lower than
$\lambda$ will be eliminated later.

In order to set up the machinery which will produce the desired
effect, we first define an auxiliary function
 \be \label{auxfunction}
 x_{\lambda i j} = \frac{E_{i \lambda} - E_{j \lambda}}{E_{i
 \lambda} + E_{j \lambda} + \lambda}.
 \ee
This function encodes the ideas discussed in the previous
paragraph. Namely, its modulus is close to 1 when one of the
 \begin{figure}
 \begin{center}
 \includegraphics[width=0.35\textwidth]{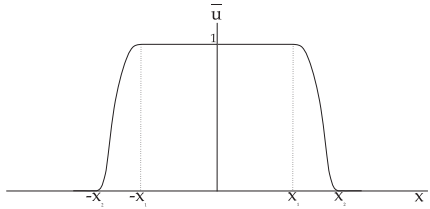}
 \end{center}
 \caption{\label{UFig}The zone function $\bar{u}(x)$}
 \end{figure}
energies is much larger than the other and large in comparison to
the cutoff $\lambda$. On the other hand, $x_{\lambda i j}$ is
close to 0 when the energies are similar or small in comparison to
the cutoff. With this function defined, we may now define {\em
zones} of the Hamiltonian. Let $\bar{u}(x)$ be a suitable ${\cal
C}^\infty$ function which satisfies:
 \be \label{uproperties}
 \bar{u}(x)= \bar{u}(|x|) =  \left\{
 \begin{array}{c@{\quad \quad}l}
 1 & |x| \leq x_1 \\
 0 & |x| \geq x_2
 \end{array}\right. ,
 \ee
and which drops smoothly from 1 to 0 between $|x|=x_1$ and
$|x|=x_2$ (see Fig. \ref{UFig}). $x_1$ and $x_2$ can be arbitrary,
unless they lie outside the energy range of the problem, in which
case no flow will take place.

Now we define the zone matrices
 \begin{eqnarray}
 u_{\lambda i j} & \equiv & \langle i | u_{\lambda} | j \rangle
 \equiv
 \bar{u}(x_{\lambda i j}) \\
 r_{\lambda i j} &\equiv& \langle i | r_{\lambda} | j \rangle
 \equiv
 1 - \bar{u}(x_{\lambda i j}).
 \end{eqnarray}
We shall use these zone matrices to identify zones of an operator
$u_\lambda[\hat{O}]$ in the following way:
 \be
 \left(u_\lambda [\hat{O}]\right)_{ij} = \langle i | u_\lambda | j
 \rangle \langle i | \hat{O} | j \rangle = u_{\lambda i j}
 \hat{O}_{ij},
 \ee
which is simply a notation for multiplying a matrix by a two index
function, and should not be confused with matrix multiplication,
although we shall write $u_\lambda \hat{O}$ as shorthand for
$u_\lambda[\hat{O}]$. The regions of $u_\lambda \hat{O}$ where
$u_\lambda = 1$, $0 < u_\lambda < 1$ and $u_\lambda = 0$ will be
named zone 1, zone 2 and zone 3 respectively.

\begin{figure}
 \begin{minipage}[t]{0.6\textwidth}
 \begin{center}
 \includegraphics[width=\textwidth]{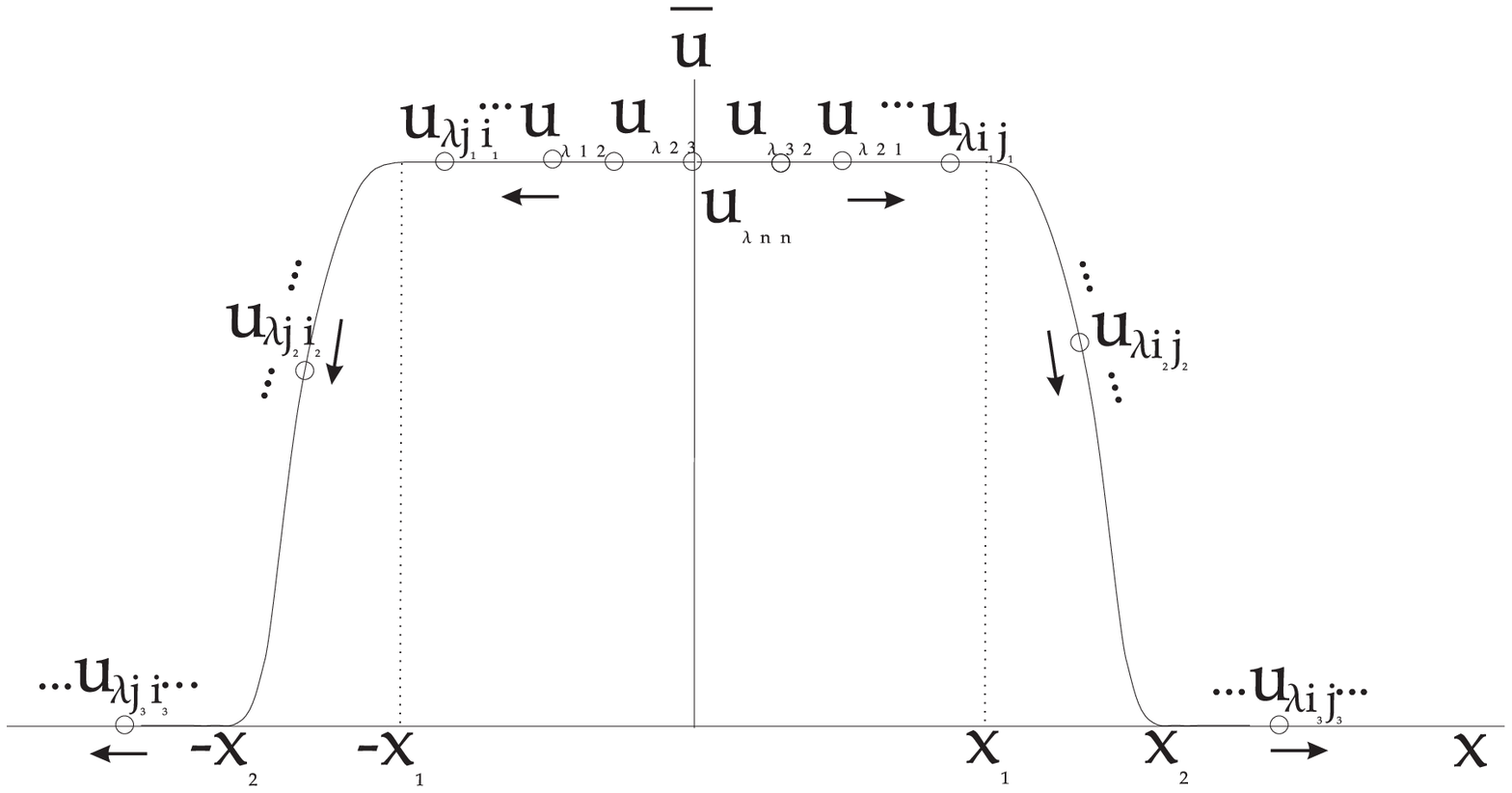}
 (a)
 \end{center}
 \end{minipage}
 \hfill
 \begin{minipage}[t]{0.28\textwidth}
 \begin{center}
 \includegraphics[width=\textwidth]{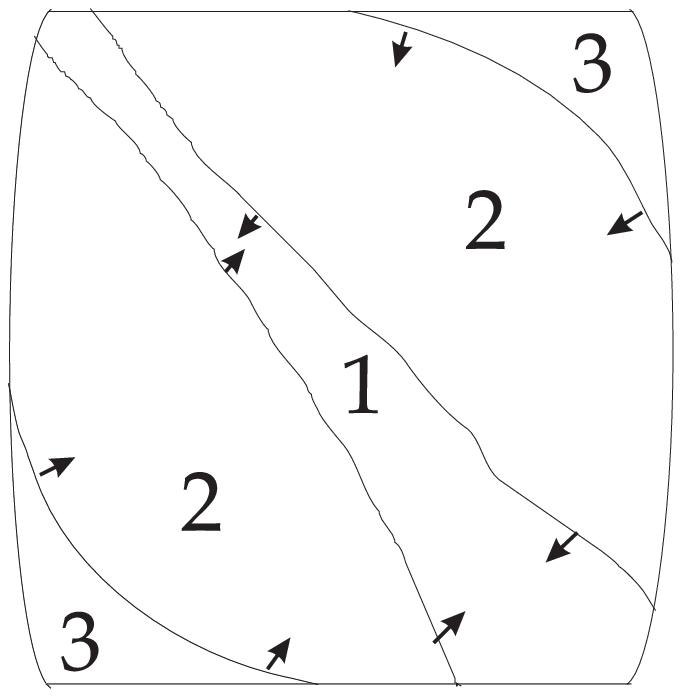}
 (b)
 \end{center}
 \end{minipage}
 \caption[Behavior of $u_{\lambda i j}$ and sones of
 $u_{\lambda}$]{\label{uproperties2}(a) Behavior of $u_{\lambda i j}$ (b) Zones of
 $u_{\lambda}$. This diagram is schematic, but is typical for a
 wide range of Hamiltonians. The arrows indicate the movement of
 the zones as $\lambda$ decreases.}
 \end{figure}

We announce our intention to get serious about our renormalization
scheme by demanding that
 \be \label{1requirement}
 H_{\lambda} = u_\lambda G_\lambda
 \ee
where $G_\lambda$ is not yet specified. Let us review the
framework we are setting up. Each matrix element $G_{\lambda i j}$
has a corresponding $u_{\lambda i j}$ which multiplies it (see
Fig. \ref{uproperties2}).

Given the diagonal elements $E_{i \lambda}$ at a certain value of
$\lambda$, and the function $\bar{u}(x)$, the $u_{\lambda i j}$
are determined and may be conveniently arranged on the graph of
$\bar{u}(x)$ as in Fig. \ref{uproperties2}(a). The diagonal
$u_{\lambda nn}$ always remain at 1 since $x_{\lambda i j} =0$
when $i=j$ from Eq. (\ref{auxfunction}). The off-diagonal terms
will be arranged in such a way that those involving a large jump
in energy are further out than those that involve a smaller jump
in energy. Since $x_{\lambda i j} = -x_{\lambda j i}$,  the
$u_{\lambda j i}$ will be positioned symmetrically opposite the
$u_{\lambda i j}$ and have the same value. The idea is that as the
flow proceeds (i.e. as $\lambda$ decreases), the $u_{\lambda i j}$
will move away from the line $x=0$, moving into zone 2 where they
decrease quickly, and finally into zone 3 where they become zero.
This may be seen in another way in Fig. \ref{uproperties2}(b),
which shows how the zones in the matrix $u_\lambda$ will evolve as
$\lambda$ decreases. As $\lambda \rightarrow 0$, zone 3 grows at
the expense of zone 1 and zone 2 until only the diagonal terms are
in zone 1, and all the off-diagonal terms are in zone 3 where they
are zero. In this limit $u_\lambda$ will be diagonal, and hence
$H_\lambda$ will be diagonalized from the requirement
(\ref{1requirement}).

In terms of $G_\lambda$, the unitary flow (\ref{arbU}) reads
 \be \label{P1}
 \frac{du_\lambda}{d\lambda}G_\lambda +
 u_\lambda\frac{dG_\lambda}{d\lambda} = [T_\lambda, H_{0\lambda}]
 + [T_\lambda, H_{I\lambda}].
 \ee
The initial Hamiltonian $H_\Lambda = H_{\lambda \rightarrow
\infty}$ is supplied at the beginning of the procedure. In order
to obtain equations for $\frac{dG_\lambda}{d\lambda}$ and
$T_\lambda$, let us regard them as unknown, and attempt to solve
for them in terms of $u_\lambda$, $H_{0 \lambda}$, $H_{I \lambda}$
and $G_\lambda$. So far there is one equation and two unknowns. In
this way the requirement (\ref{1requirement}) has not uniquely
specified the flow we are to undertake through unitary space. To
produce another equation, we rearrange Eq. (\ref{P1}) as
 \be \label{P2}
 [H_{0\lambda}, T_\lambda] + u_\lambda\frac{dG_\lambda}{d\lambda}
 =  [T_\lambda, H_{I\lambda}] -
 \frac{du_\lambda}{d\lambda}G_\lambda \equiv Q_\lambda.
 \ee
Now we make the rather arbitrary requirement that the manner in
which the two terms on the left equal $Q_\lambda$ must somehow be
in harmony with the zone structure, term for term. We decide that
 \be \label{P3}
 [H_{0 \lambda}, T_\lambda] = r_\lambda Q_\lambda,
 \ee
which means that
 \be \label{P4}
 u_\lambda \frac{dG_\lambda}{d\lambda} = Q_\lambda - r_\lambda
 Q_\lambda = u_\lambda Q_\lambda.
 \ee

We now have two equations for the two unknowns $T_\lambda$ and
$\frac{dG_\lambda}{d\lambda}$ or for $T_\lambda$ and
$\frac{dH_{\lambda i j}}{d \lambda}$ through the requirement
(\ref{1requirement}). These equations must be evaluated for the
matrix elements in the three different zones. Specifically, in
zone 1($0 \leq |x| \leq x_1$, $r_{\lambda i j} = 0$, $u_{\lambda i
j} = 1$) we have
 \begin{eqnarray}
 T_{\lambda i j} &=& 0 \label{RG1} \\
 \frac{dH_{\lambda i j}}{d \lambda} &=& [T_\lambda, H_{I
 \lambda}]_{ij}.
 \end{eqnarray}
In zone 2($x_1 < |x| < x_2$, both $r_{\lambda i j}$ and
$u_{\lambda i j}$ between 0 and 1),
 \begin{eqnarray}
 T_{\lambda i j} &=& \frac{r_{\lambda i j}}{E_{j \lambda} - E_{i
 \lambda}} \left( [T_\lambda, H_{I \lambda}]_{ij} -
 \frac{du_{\lambda i j}}{d \lambda} \frac{H_{\lambda i
 j}}{u_{\lambda i j}} \right) \\
 \frac{dH_{\lambda i j}}{d \lambda} &=& u_{\lambda i j}[T_\lambda, H_{I
 \lambda}]_{ij} + r_{\lambda i j} \frac{du_{\lambda i j}}{d
 \lambda} \frac{H_{\lambda i j}}{u_{\lambda i j}},
 \end{eqnarray}
while in zone 3($|x| > x_2$, $r_{\lambda i j} = 1$, $u_{\lambda i
j} = 0$),
 \begin{eqnarray}
 T_{\lambda i j} &=& \frac{1}{E_{j \lambda} - E_{i
 \lambda}}[T_\lambda, H_{I \lambda}]_{ij} \\
 \frac{dH_{\lambda i j}}{d \lambda} &=& H_{\lambda i j} = 0
 \label{RGL}.
 \end{eqnarray}
Recall that $u_{\lambda i j}$ is a shorthand for
 \be
 u_{\lambda i j} = \bar{u}\left(x_{\lambda i j}(E_{i \lambda}, E_{j
 \lambda})\right),
 \ee
so that
 \be
 \frac{du_{\lambda i j}}{d \lambda} = \frac{d
 \bar{u}}{dx}\left(\frac{\partial x}{\partial E_{i
 \lambda}}\frac{dE_{i \lambda}}{d \lambda} + \frac{\partial x}{\partial
 E_{j \lambda}}\frac{dE_{j \lambda}}{d \lambda} \right).
 \ee
In this way the renormalization group equations (\ref{RG1}) -
(\ref{RGL}) may be expressed in the schematic form
 \begin{eqnarray}
 T_{\lambda i j} = f(T_{\lambda k l}, H_{\lambda q r}, E_{m \lambda}, \frac{d E_{p
 \lambda}}{d \lambda}) \\
 \frac{dH_{\lambda i j}}{d \lambda} = g(T_{\lambda k l}, H_{\lambda q r}, E_{m \lambda}, \frac{d E_{p
 \lambda}}{d \lambda}),
 \end{eqnarray}
which shows they are a complicated set of non-linear algebraic and
differential equations. Normally the only way to tackle them is
through an iterative process or through perturbation theory. The
beauty of the equations (\ref{RG1}) - (\ref{RGL}) though is that
energy denominators only arise for $|x|
> x_1$ which means, from the expression for the auxiliary function
(\ref{auxfunction}), that
 \be
 \left| \frac{1}{E_{j \lambda} - E_{i \lambda}} \right| \leq
 \frac{1}{x_1} \frac{1}{E_{i \lambda} + E_{j \lambda} + \lambda},
 \ee
so that a reciprocal of an energy difference cannot be larger than
a constant times $\lambda^{-1}$. In this way we have conquered the
problem of ordinary perturbation theory where energy denominators
can grow arbitrarily small.

For more discussion on the relevance of the similarity
renormalization group equations to renormalization problems in
quantum field theory (eg. QCD), we refer the reader to the
original papers of Glazek and Wilson \cite{GlazekWilson1,
GlazekWilson2}. Some recent contributions include the analysis of
the similarity renormalization group to the Poincar$\acute{e}$
algebra \cite{Glazek3}, bound-state dynamics of effective fermions
\cite{Glazek4}, and even to theories of gravity
\cite{GlazekWilson3}.

Finally, a comment on the name ``Similarity renormalization
group''. All the transformations involved are strictly unitary and
not just similarity transformations. Of course, since all unitary
transformations are also similarity transformations, the
designation remains true, albeit imprecise. It appears to be a
historical accident from some slightly careless terminology in the
original paper in 1993 \cite{GlazekWilson2}.

\section{Wegner's flow equation and the similarity renormalization
group}

How does Wegner's flow equation tie in with renormalization, and
in particular, the similarity renormalization group equations
(\ref{RG1})-(\ref{RGL})? Both are unitary flows on the initial
Hamiltonian, and hence can be written as
 \be
 \frac{d{\cal H}_\lambda}{d\lambda} = [ F_\lambda \{\cal{H}_\lambda\}, {\cal
 H}_\lambda].
 \ee
In the similarity renormalization approach, $\lambda$ has the
dimensions of energy and can be taken to begin at $\lambda=\Lambda
\rightarrow \infty$ and asymptotically ends at $\lambda = 0$. In
Wegner's flow equation, $\ell$ has the dimensions of
1/$\mbox{energy}^2$ and begins at $\ell =0$ and asymptotically
ends at $\ell = \infty$. If we set $\ell = 1/\lambda^2$ then
Wegner's equation
 \be
 \frac{d {\cal H}(\ell)}{d \ell} = [[\mbox{Diag}({\cal H}(\ell)),
 {\cal H}(\ell)], {\cal H}(\ell)]
 \ee
follows from a specific choice of $F_\lambda$, namely
 \be
 F_\lambda\{ {\cal H} \} = \frac{1}{\lambda^2} \frac{d\ell}{d \lambda}[\mbox{Diag}({\cal H}_\lambda), {\cal
 H}_\lambda].
 \ee
However, as mentioned in Section \ref{Sect1}, the renormalization
properties of this choice of $F_\lambda$ are not entirely clear.
In the similarity renormalization group, the equations were
derived from first specifying precisely how the Hamiltonian was to
change with $\lambda$. The requirements were that states with
large energy differences should be treated first, that energy
denominators must be bounded with respect to $\lambda$, that as
$\lambda$ decreased the far off-diagonal elements become zero (not
just very small), and other similar properties resulting from the
form of $x_{\lambda i j}$. The price we had to pay for these
strict requirements resulted in a complicated set of equations for
determining $T_\lambda$, and hence $\frac{d {\cal H}}{d\lambda}$.

In contrast, Wegner's equation is simpler because $T_\lambda$ is
completely specified, and is chosen so that the Hamiltonian flows
in steepest descent fashion either to block-diagonal form or to
diagonality. Along the path to block-diagonal form, there is no
control over how the off-diagonal elements behave. In particular,
the final Hamiltonian could still contain large off-diagonal
elements. If the generator is chosen so that the Hamiltonian flows
to diagonality, then we know that the off-diagonal matrix elements
decrease monotonically, and those with the largest energy
differences decrease the fastest. However, there are no other
renormalization restrictions (such as zone structure, $\lambda$
dependent bounds on denominators, etc.) on the flow. It should
also be noted that although the off-diagonal elements become
exponentially small, they never strictly vanish. Perhaps this will
be made more clear by remembering that, to first order in the
interaction, the off-diagonal elements flow as (from Eq.
(\ref{pv}))
 \be
 v_{ij}^{(1)}(\ell) = V_{ij}e^{-2(E_i-E_j)^2\ell},
 \ee
so that, to first order (and {\em not} in general to higher
orders), the off-diagonal elements behave as if a uniform gaussian
band function $x_{\lambda i j}$ had been chosen in the similarity
renormalization group, as opposed to the usual ``widening band''
structure of the conventional $x_{\lambda i j}$ shown in Fig.
\ref{uproperties2}.

\section{Renormalization in action}
Let us continue in the spirit of renormalization via continuous
unitary transformations and look at two examples from the
literature.

\subsection{Glazek and Wilson's discrete 2-d delta function
model}

The fathers of the similarity renormalization group, Glazek and
Wilson, introduced an extremely useful toy model as a testing
ground for renormalization type flow equations
\cite{GlazekWilson4, GlazekWilson5, Glazek1}. The model displays
many important characteristics of more complex Hamiltonians since
it is asymptotically free, contains a bound state (negative
eigenvalue) and has a large range of energy scales.

The model can be considered as a discretized version of the two-
dimensional delta function Hamiltonian:
 \be
 H = \frac{\vec{p}^{\; 2}}{2m} - g\delta^2(\vec{r}).
 \ee
Such Hamiltonians have been studied before. It is known that there
is one $s$-wave bound state with a negative eigenvalue. The other
eigenstates have positive energies and are $s$-wave scattering
states. The model can be discretized by looking at it from a
variational principle point of view (see Ref.
\cite{GlazekWilson5}); we shall not go into the details here. The
discretized version is
 \be \label{discretized}
 H_{ij} = \delta_{ij}E_i - g\sqrt{E_iE_j},
 \ee
where $E_i = b^i$ and $b>1$. The parameter $b$ serves roughly as a
measure of the severity of the discretization process - as $b
\rightarrow 1$, the discretized Hamiltonian approaches the
continuous one. For numerical calculations, we shall employ $b=2$.
The infinite matrix (\ref{discretized}) will be made finite by
employing an infrared and ultraviolet cutoff. Specifically, the
index $i$ ranges from $M$ (a large negative number) to $N$ (a
large positive number). Glazek and Wilson chose $M=-21$ and
$N=16$, which, if we (with tongue in cheek) call one unit of
energy 1 GeV, means that the energy range in the Hamiltonian goes
from 0.5 eV to 65 TeV. The coupling constant $g$ is adjusted so
that the bound state eigenvalue is precisely -1 GeV, which gives
$g=0.06060600631$. Thus the bound state energy scale is three
orders of magnitude smaller than the largest energy scale in the
problem, and we have in front of us a candidate for
renormalization. The model will be renormalized using Wegner's
flow equation which, expressed in terms of the diagonal matrix
elements $\varepsilon_i$ and the off-diagonal elements $v_{ij}$,
are (see Chapter 1)
 \begin{eqnarray}
 \dot{\varepsilon_i} &=& 2\sum_k (\varepsilon_i -
 \varepsilon_k)v_{ij}^2 \label{epsdiffe} \\
 \dot{v_{ij}} &=& -(\varepsilon_i - \varepsilon_j)^2 v_{ij} + \sum_k
 (\varepsilon_i + \varepsilon_j - 2\varepsilon_k)v_{ik}v_{kj}
 \label{vdiffs}.
 \end{eqnarray}
Language familiar to renormalization gurus will be invoked by
introducing an effective running coupling constant. Since the
initial off-diagonal terms have $H_{ij} = -g\sqrt{E_iE_j}$, the
exact running coupling constant will be defined by
 \be
 \tilde{g}(\ell) = -\frac{H_{M, M+1}(\ell)}{\sqrt{E_M
 E_{M+1}}},
 \ee
where $H(\ell)$ is the exact (obtained numerically) solution to
the flow equations. This definition is motivated by the
requirement that the coupling constant determines the strength of
the interaction at small energies ($E_M$ is the smallest energy in
the problem), as is standard practice. We can also define an
approximate running constant $\tilde{g}_a(\ell)$ by allowing the
coupling in front of the first order solutions to the flow
equation to become $\ell$-dependent:
 \begin{eqnarray}
 \varepsilon_i (\ell) &=& (1-\tilde{g}_a(\ell))E_i \label{eansatz} \\
 V_{ij}(\ell) &=& -\tilde{g}_a(\ell) \sqrt{E_i E_k} e^{-(E_i-E_k)^2
 \ell} \label{vansatz}.
 \end{eqnarray}
If we evaluate $\dot{\varepsilon}_M$ by inserting the diagonal
entries (\ref{eansatz}) into the flow equation (\ref{epsdiffe}),
and make the approximation $E_M - E_k \approx -E_k$ since $E_M$ is
very small, we obtain a differential equation for the approximate
running coupling constant:
 \be
 \frac{d \tilde{g}}{d \ell} = -\tilde{g}^2 \frac{d}{d\ell}
 \sum_k e^{-2E_k^2 \ell}.
 \ee
The integration of this equation gives
 \be
 \tilde{g}_a(\ell) = \frac{g}{1-g(N+1+0.4 + \mbox{ln}
 (\ell)/\mbox{ln}4)}.
 \ee
The question now is how good an approximation Eqs. (\ref{eansatz})
and (\ref{vansatz}) actually are.
 \begin{figure}
 \begin{center}
 \includegraphics[width=0.5\textwidth]{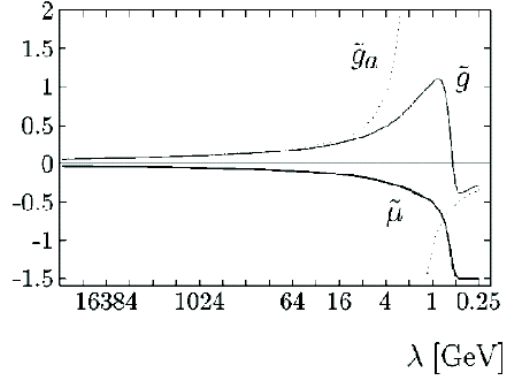}
 \end{center}
 \caption[The exact and approximate running coupling constants $\tilde{g}$ and
 $\tilde{g}_a$ as functions of the effective Hamiltonian width $\lambda = 1/\ell^2$]{\label{RenormFig1} The exact and approximate running coupling constants $\tilde{g}$ and
 $\tilde{g}_a$ as functions of the effective Hamiltonian width $\lambda = 1/\ell^2$. Source
 Glazek and Wilson, Phys Rev. D 57 3558 (1998) }
 \end{figure}
The answer is provided by Fig. \ref{RenormFig1}, which plots the
exact and approximate running coupling constants $\tilde{g}$ and
$\tilde{g}_a$ as functions of the energy width $\lambda=1/\ell^2$.
In order to compare these graphs with the bound state formation
scale (the value of $\lambda$ when the bound state eigenvalue
begins to appear on the diagonal), the matrix element
 \be
 \tilde{u}(\ell) = H_{-1, -1}(\ell) - 0.5 \mbox{ GeV}
 \ee
is also plotted (the bound state eigenvalue appears at $H_{-1,
-1}$). $\tilde{u}$ is displaced by 0.5 GeV so as not to obscure
the other graphs. The graph shows that the approximate solution
blows up in the flow before the effective Hamiltonian width is
reduced to the scale where the bound state is formed. However, the
exact effect coupling constant does not diverge. This encouraged
Glazek and Wilson to expand the solution in terms of the running
 \begin{figure}
 \begin{center}
 \includegraphics[width=0.5\textwidth]{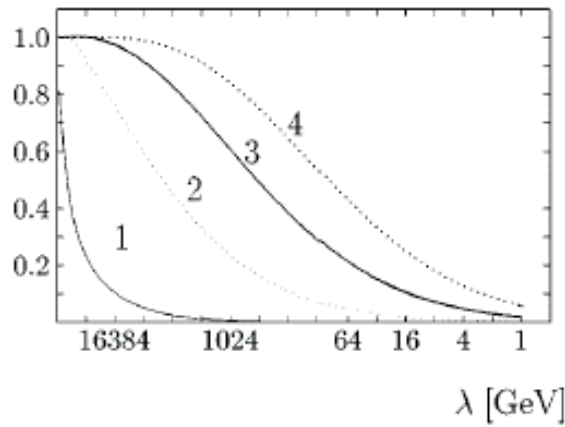}
 \end{center}
 \caption[The ratio of the bound state eigenvalue of the effective
 Hamiltonian $H^{(i)}(\ell)$ to the exact result]{\label{RenormFig2} The ratio of the bound state eigenvalue of the effective
 Hamiltonian $H^{(i)}(\ell)$ to the exact result, where $H^{(i)}(\ell)$ is the $i$th
 perturbative approximation to the exact result in terms of the bare coupling $g$.
 $\lambda = 1/\ell^2$. Source Glazek and Wilson, Phys Rev. D 57 3558 (1998) }
 \end{figure}
coupling constant $\tilde{g}(\lambda_0)$ instead of expanding it
in terms of the bare coupling $g$. This point is emphasized in
Fig. \ref{RenormFig2}, which plots the ratio of the bound state
eigenvalue of the effective Hamiltonian $H(\ell)$ to the exact
result, where $H(\ell)$ is calculated using the perturbative
solution of the flow equations (\ref{epsdiffe}) and (\ref{vdiffs})
to first, second, third and fourth order in the bare coupling $g$.
The perturbative solution $H^{(i)}(\ell)$ is diagonalized exactly
(numerically) to obtain its bound state eigenvalue. This figure
clearly demonstrates that the perturbative expansion in terms of
the canonical coupling constant in the initial Hamiltonian is not
suitable for applications in the bound state dynamics.

Glazek and Wilson went on to investigate the analogous expansion
in terms of the running coupling constant, $\tilde{g}(\lambda_0)$,
which turns out to be far more accurate. They also investigated
the effect of cutting out a small window around the bound state
eigenvalue matrix element $H_{-1,-1}(\ell)$, diagonalizing it
exactly, and comparing the result to the exact diagonalization of
the full $H(\ell)$. Their remarkable result was that accurate
results could be obtained (on the order of 10\%) by expanding the
flow to second order in the running coupling constant, and
diagonalizing a small window around the bound state for a fixed
$\ell_0$. In this way {\em the complexity of the problem has been
divided into two parts: the complexity of the exact Hamiltonian
flow, and the complexity of the eigenvalue problem for small
matrices}. Since the latter problem is tractable, we have indeed
made progress for future applications of the flow equations.

In a very recent paper \cite{Glazek1}, Glazek and Mlynik have
returned to this matrix model and investigated if changing the
flow equation slightly leads to more accurate results. In
particular, they considered multiplying the usual generator by a
matrix dependent function $\phi_{ij}$,
 \be
 \eta_{ij} = \phi_{ij} [\mbox{Diag}({\cal H}), {\cal H}].
 \ee
In their work, they found that slowing down the flow by a factor
of the type
 \be
 \phi_{ij} = \frac{1}{1 + c|i-j|},
 \ee
where $c=1$, made significant improvements. They also attempted
making $\phi_{ij}$ dependent on $\ell$, but this did not seem to
improve the accuracy. More work is needed in this regard. Also,
one is not completely free to change the generator at will, since
only the original Wegner choice is proved to lead to diagonality.
Another choice may not diagonalize the Hamiltonian. In conclusion
though, this toy model has demonstrated that Wegner's flow
equation is an exciting new renormalization tool.

 \subsection{Renormalization of the electron-phonon interaction}
 \label{RenSec}
As promised in Section \ref{LastSec}, we now briefly comment on
the problem of eliminating the electron-phonon interaction by some
kind of renormalization. In fact, the electron-phonon problem has
now been treated using no less than {\em four} renormalization
methods (this excludes the original Fr\"{o}hlich approach, which
was not a renormalization scheme). Firstly, it was treated using
Wegner's flow equation \cite{LenzWegner}, as in Section
\ref{TEPI}. Next it was treated using Glazek and Wilson's
similarity renormalization scheme \cite{Mielke1997a}. Thirdly it
was treated using a new renormalization scheme proposed by
H\"{u}bsch and Becker \cite{Becker}; this scheme works by
expanding the unitary transformation in terms of the generator and
then requiring that the resulting Hamiltonian has no energy jumps
larger than $\lambda$. It has also been briefly treated using the
algebraic Bloch-Feshbach formalism from the theory of nuclear
dynamics \cite{Rau1, Rau2}. This last method does not provide one
with an explicit form of the induced electron-electron attraction.

As a caveat, we now display the original Fr\"{o}hlich result,
followed by the results from the first three methods mentioned
above, for the effective interaction between the electrons in a
Cooper pair. The notation is taken to be the same as in Section
\ref{TEPI}. The results read
 \begin{eqnarray}
  V^{F}_{k,-k,q} &=& -|M_q|^2\frac{\omega
    _q}{\omega_q^2 - (\varepsilon _{k+q}-\varepsilon_k)^2} \quad
    \mbox{Fr\"{o}hlich's result} \\
 V^{LW}_{k,-k,q} &=& - |M_q|^2\frac{\omega _q}{\omega_q^2 + (\varepsilon
    _{k+q}-\varepsilon_k)^2} \quad \mbox{Lenz and
    Wegner's flow equation result} \\
 V^M_{k, -k, q} &=& - |M_q|^2\frac{1}{\omega_q + |\varepsilon
    _{k+q}-\varepsilon_k|}\theta(\omega_q + |\varepsilon
    _{k+q}-\varepsilon_k| - \lambda) \quad
    \mbox{Mielke's similarity renormalization} \\
 V^{HB}_{k,-k,q} &=& - |M_q|^2\frac{1}{\omega_q +|\varepsilon
    _{k+q}-\varepsilon_k|}\theta(\omega_q - |\varepsilon
    _{k+q}-\varepsilon_k|) \quad \mbox{H\"{u}bsch and Becker's
    result}.
\end{eqnarray}
The results differ since in each case an approximate solution is
found to different equations. It is not clear which result has
`proceeded the furthest' along the renormalization road.
Nevertheless, they show the vigourous nature of the field and the
prospects for future developments are highly likely.

%% file: Conclusion.tex
\chapter{Conclusion}

How valuable are flow equations as a tool for the intrepid quantum
mechanic? This thesis has attempted to provide an overview of the
subject. The method has been to present the theory in a unified
way, and then to ``teach by example'' by applying it to a few well
known problems. The overview has not been exhaustive and models
were chosen by weighing up their simplicity and pedagogical
merits.

The presentation may be summarized as follows. In Chapter
\ref{WegFE} Wegner's flow equation was introduced, and it was
shown how Wegner's choice of generator causes the off-diagonal
elements to decrease monotonically, thereby diagonalizing the
Hamiltonian. The second order perturbative solution of the
equations was presented. Thereafter the little known general
mathematical framework of flow equations, involving steepest
descent flow on the manifold of unitarily equivalent matrices, was
summarized and subsequently used as a framework for the entire
thesis. Two extensions of the machinery were presented, namely
Safonov's one step scheme and block-diagonal flow equations. This
latter method was contrasted with existing effective methods such
as that of Lee and Suzuki.

Chapter \ref{Chapter2} reviewed two instructive applications of
the flow equations program. The Dirac Hamiltonian was found to
possess a commutation structure which rendered it readily
accessible to flow equations, and the exact recursive perturbative
solution was computed, the first few steps of which reproduced the
standard Foldy-Wouthuysen transformation. The renormalization
utility of Wegner's flow equation was showcased in the
electron-phonon problem. There it was shown how the dynamical
nature of the generator allows it to eliminate high energy terms
first, and thereby proceed further up to a given order in
perturbation theory than static unitary transformation schemes,
such as that of Fr\"{o}hlich. This produced an effective
electron-electron interaction which was always attractive and less
singular. Along the way, the framework of $\cal L$ ordering was
introduced. This allows one, for purposes of comparison, to
explicitly calculate an effective static generator from the flow
equation's dynamical counterpart.

The Lipkin model was introduced in Chapter \ref{Chap3} as a
laboratory for flow equations experiments. The model was chosen
due to it being numerically solvable but still possessing a
non-trivial phase space structure. A brief numerical exercise was
undertaken to highlight explicitly the nature of the flow through
unitary space, for different values of the coupling. Three recent
ways of approaching the model using flow equations were reviewed.
Pirner and Friman closed the flow equations by linearizing newly
generated operators around their ground state expectation values,
which were assumed to be equal to their value in the zero coupling
system. Mielke constructed a new generator in order to conserve
tridiagonality, and then concentrated on the flow equations for
the matrix elements themselves. We showed that this generator was
nothing more than a slight (and unnecessary) modification of our
underlying mathematical framework. Stein employed the
Holstein-Primakoff mapping of SU(2) to cast the problem into
bosonic language. This had the noted distinction of providing a
systematic and simple method for solving the flow equations order
for order in $1/N$. Next we presented our work on the Lipkin
model, which essentially modified the Pirner and Friman method by
allowing for the dynamical alteration of the expectation value
during the flow. This was first achieved by using an external
variational calculation. A more sophisticated method utilized a
self-consistent calculation and required no external input. The
advantage of both methods was that the new equations were able to
deal with the second phase, where the others failed. Extending the
scheme, in the case of the variational calculation, proved
problematic. A similar extension of the self-consistent method has
not yet been attempted, and is an avenue for future development.
The chapter was concluded by briefly presenting other possible
methods of a type not seen before in the literature.

Chapter \ref{RenormChapter} focused on the utility of flow
equations as a renormalization framework. Glazek and Wilson's
similarity renormalization group was explained, and compared with
Wegner's flow equation. Renormalization using flow equations was
further explored by reviewing numerical results from a discretized
two dimensional delta function toy model. Finally we returned to
the renormalization of the electron-phonon interaction, and
highlighted the recent interest in this area by displaying results
from four different renormalization schemes.

Finally we consider the open questions and possibilities for
advancement related to this work. The basic problem is how to
close the flow equations by making a reasonable and consistent
approximation. Normally post hoc reasoning must be employed to
ascertain just how accurate a certain procedure is. A systematic
framework for computing the errors involved in these types of
approximations is necessary.

Wegner's flow equation has proved very useful as a renormalization
tool, and much has been learned from Glazek and Wilson's toy
model. The task now remains to apply this method to realistic
Hamiltonians to investigate its true potential. This work has
already begun \cite{Gubankova}.

The technique of tracking the ground state during the flow, by
employing a self-consistent calculation, proved very useful.
Further refinements of the approximations can obviously be made,
such as linearizing around the expectation value of $J_z$ in
excited states in order to describe the excited spectrum, and thus
the gap, more accurately. This method can then also be tested on
more complex Hamiltonians.